\documentclass[preprint]{JFM-FLM_Au}

\usepackage{amsmath}
\usepackage{booktabs}
\usepackage{multirow}
\usepackage{array}
\usepackage{makecell}
\usepackage{booktabs,threeparttable}
\usepackage{subcaption}
\usepackage{cleveref} 
\usepackage{url}
\usepackage{orcidlink}
\usepackage{adjustbox}
\usepackage{graphicx}

\usepackage{etoolbox}
\makeatletter

\patchcmd{\NAT@citex}
  {\@citea\NAT@hyper@{%
     \NAT@nmfmt{\NAT@nm}%
     \hyper@natlinkbreak{\NAT@aysep\NAT@spacechar}{\@citeb\@extra@b@citeb}%
     \NAT@date}}
  {\@citea\NAT@nmfmt{\NAT@nm}%
   \NAT@aysep\NAT@spacechar\NAT@hyper@{\NAT@date}}{}{}

\patchcmd{\NAT@citex}
  {\@citea\NAT@hyper@{%
     \NAT@nmfmt{\NAT@nm}%
     \hyper@natlinkbreak{\NAT@spacechar\NAT@@open\if*#1*\else#1\NAT@spacechar\fi}%
       {\@citeb\@extra@b@citeb}%
     \NAT@date}}
  {\@citea\NAT@nmfmt{\NAT@nm}%
   \NAT@spacechar\NAT@@open\if*#1*\else#1\NAT@spacechar\fi\NAT@hyper@{\NAT@date}}
  {}{}

\makeatother

\lefttitle{Kanehira et al.}
\righttitle{Journal of Fluid Mechanics}

\title{Dynamics of jet formation and collapse for axisymmetric surface gravity waves: coupled 3D potential flow and SPH simulations}

\author{Taiga Kanehira\aff{1,$\dagger$}, Peter K. Stansby\aff{1}, Benedict D. Rogers\aff{1}, Mark McAllister\aff{2,3}, T. S. van den Bremer\aff{4}, \and Samuel Draycott\aff{1}}

\affiliation{
\aff{1}School of Engineering, University of Manchester, Manchester, UK
\aff{2}Wood Thilsted Partners, London, UK 
\aff{3}Department of Engineering Science, University of Oxford, Oxford, UK
\aff{4}Faculty of Civil Engineering and Geosciences, Delft University of Technology, Delft, The Netherlands
}

\corresau{$\dagger$ Taiga Kanehira, taiga.kanehira@manchester.ac.uk}

\begin{document}
\maketitle

\begin{abstract}
Axisymmetric waves occur across a wide range of scales. This study analyses large-scale (with a jet height of up to 6 m) gravity-dominated axisymmetric waves (Bond number $B_o=\textit{O}(10^{5})$ and Weber number $We=\textit{O}(10^{4}\text{--}10^{6})$), for which surface-tension effects are negligible. 
Our aim is to clarify the dynamics of highly nonlinear axisymmetric jet formation, cavity collapse and the consequent generation of secondary jets. 
The newly developed three-dimensional framework \textit{OceanSPHysics3D}, combining unsteady potential flow with Smoothed Particle Hydrodynamics (SPH), enables full simulation of jet initiation and collapse. 
The computed free surface elevations and jet evolution agree well with the experiments of McAllister et al. (J. Fluid Mech., vol.~935, 2022) and an analytical jet-tip-angle formulation of Longuet–Higgins (J. Fluid Mech., vol.~127, 1983). The simulations elucidate how the falling primary jet induces a secondary jet. 
The mechanisms forming the pre-jet trough and post-jet cavity are fundamentally different.
The pre-jet trough arises geometrically from directional focusing of the constituent waves, yielding a self-similar shape when appropriately scaled. In contrast, the post-jet cavity is formed inertially by the falling continuous jet, and lacks both spatial and temporal self-similarity.
Its collapse also differs: the cavity pinches off at the neck to generate upward and downward secondary jets, with local accelerations reaching $150g$.
The primary jet scale governs the ensuing secondary-jet dynamics, including vortex-ring formation and strong vertical mixing.
These findings illustrate the complexity of axisymmetric jet dynamics, and demonstrate the ability of the present framework to reproduce the key coupled processes in such extreme free surface events.
\end{abstract}


\section{Introduction}
\label{Sec:Intro}
In the ocean, surface gravity waves are inherently three-dimensional, composed of interacting frequency and directional components. 
Owing to this three-dimensionality, ocean waves can exceed the classical breaking threshold steepness for two-dimensional long-crested waves---$ak=0.44$ (with $a$ and $k$ denoting wave amplitude and wavenumber) in  \citet{Stokes1847}, and $H_D/L_F=0.142$ (with $H_D$ and $L_F$ denoting the deep-water wave height and the finite-amplitude deep-water wavelength) in \cite{Miche1944}. These three-dimensional effects lead to distinct breaking patterns \citep{McAllister19} and existing wave breaking criteria are not valid in highly spread conditions \citep[e.g.,][]{Kanehira21}.
In \cite{McAllister24}, it was shown that for highly directionally spread wave conditions, breaking onset does not occur until the wave steepness is over double the unidirectional counterpart.
Accounting for this three dimensionality is therefore essential for understanding wave-energy dissipation, air--sea momentum exchange and the loading or response of offshore systems in realistic conditions. 
Wave breaking behaviour in nature lies between two limiting cases: unidirectional waves with infinitely long crests (no directional spreading) and axisymmetric waves (maximum directional spreading) as described in \cite{McAllister22} (hereafter referred to as MC22). The axisymmetric focused wave thus provides a simplified yet physically relevant model for studying the essential hydrodynamics of three-dimensional breaking behaviour in its most extreme form.
Axisymmetric focusing and the ensuing jet formation occur across a wide range of scales and regimes \citep{Basak_2021}---from Faraday jets generated by parametric resonance in vibrating containers \citep{Miles_1984} to Worthington jets formed by droplet or disc impacts \citep{Worthington1897,Eggers_2008,Truscott_2014}. More recently, similar axisymmetric focusing has been experimentally reproduced for gravity waves through linear dispersive focusing by MC22.
Despite differences in the relative importance of gravity, surface tension, and viscosity, these processes display similar dynamics across scales, and such jet formation may represent a fundamental feature of free surface flows \citep{Longuet-Higgins_1983}. 

The theoretical understanding of axisymmetric standing waves and jet formation has evolved over the past decades. Following the third-order theory for axisymmetric standing waves derived by \citet{Mack1962}, the first analytical model for jet formation was proposed by \citet{Longuet-Higgins_1983} (hereafter also referred to as LH83), who obtained a time-dependent inviscid solution of the Laplace equation with a free surface in the form of a Dirichlet hyperboloid.
Subsequent developments, including the asymptotic analysis of \citet{Ockendon24} and the non-perturbative potential-flow theory of \citet{Wilkinson_2025}, further refined the theoretical framework, elucidating pressure-driven jet initiation and the smooth hyperbolic evolution of the axisymmetric free surface. Since axisymmetric focusing concentrates flow toward the axis from all radial directions, the maximum impact pressure is expected to be significantly higher than in two-dimensional flip-through events \citep{Lugni06,MARTINMEDINA18}. However, despite indications from large-scale (up to 6 m) vertical jet experiments supporting this expectation, no direct measurements or fully three-dimensional characterisations of the pressure field at jet onset are available.  Under inertia-dominated conditions, the experiments of MC22 showed that the jet height follows a pre-jet cavity scaling closely related to the $H(H/L)^2$ law of \cite{Ghabache_2014}, where $H$ and $L$ denote the cavity depth and lateral extent, respectively. Recent laboratory experiments by \cite{Fillette_2022} also examined axisymmetric gravity–capillary standing waves and reported that the maximum wave steepness saturates at $5/(2\pi)$ under large forcing.

Recent studies have shown that the limiting waveform and critical crest angle of two-dimensional standing waves remain an open question. \citet{Penney_1952} argued that the downward acceleration of the free surface cannot exceed $-g$, implying a limiting crest angle of $90^\circ$, a conclusion later supported by the experiments of \citet{Taylor_1953}. In contrast, \citet{Longuet-Higgins_1994} proposed a different dynamical mechanism, suggesting that a self-similar superposition of short-wave components could lead to a blow-up of curvature and Lagrangian particle acceleration near the crest. High-resolution computations by \citet{Wilkening_2011} further indicated that large-amplitude standing waves do not converge to a unique limiting form; instead, high-frequency oscillations emerge near the crest, breaking the expected self-similarity. These results collectively suggest that even in two dimensions the existence of a universal limiting waveform remains unresolved, and that the fundamental criteria for the breaking threshold are still unclear.

Regarding axisymmetric waves, LH83 predicted that once a jet is initiated its shape evolves hyperbolically, and that axisymmetric standing waves should break when the jet angle reaches $2\arctan\sqrt2$ (i.e. $109.47^\circ$). However, as noted in MC22, identifying the onset of breaking in axisymmetric standing waves is particularly challenging. Unlike progressive waves, overturning cannot be used as an unambiguous indicator, because both wave motion and jet formation involve vertical displacements of the free surface, making the two processes difficult to distinguish. As a consequence, a clear criterion for the onset of breaking in strongly focused axisymmetric waves has not yet been established. Indeed, while MC22 confirmed that the jet angles in axisymmetric focusing waves follow the LH83 model, experimental limitations prevented verification near the predicted critical angle. Thus, significant uncertainty remains regarding the onset of breaking.

The formation and collapse of cavities generated by the impact of disks, droplets and other bodies have been extensively studied \citep{Truscott_2014}. The resulting cavity shape depends strongly on the impactor geometry, wettability and impact velocity. It has also been shown by \citet{Glasheen_1996} that the non-dimensional pinch-off time $\tau=t\sqrt{2g/D}$ where $t$ is the physical pinch-off time, $g$ is gravitational acceleration and $D$ is a characteristic length scale, remains nearly constant over a broad range of conditions, suggesting the possibility of temporally self-similar collapse \citep{TRUSCOTT_TECHET_2009}. Moreover, \citet{GEKLE09_PRL} demonstrated that the emergence of the high-speed Worthington jet does not originate from a purely local singularity at the pinch-off point. Rather, it is driven by the global collapse of the cavity wall and the associated focusing of kinetic energy, implying that a full understanding of jet formation requires analysing the dynamics of the entire collapsing cavity. Despite these advances, the cavity shape and collapse induced by a continuously falling primary jet remain poorly understood. While studies on cavities formed by a liquid column freely falling onto a quiescent free surface exist \citep{Kersten2003}, the cavity dynamics driven by a continuously falling primary jet have not been directly studied to date.

High-resolution computational fluid dynamics (CFD) models have further advanced our understanding of jet formation. \citet{Duchemin2002} directly solved the Navier--Stokes equations and reproduced bubble-bursting jets for the first time. More recent axisymmetric simulations employing interface-capturing techniques such as the coupled level-set and volume-of-fluid (CLSVOF) method have systematically examined the roles of viscosity, inertia and surface tension in determining jet shape, speed and scaling \citep{Orozco2015,Farsoiya_2017,Ismail2018,Rodríguez_2020}. However, the dynamics leading to a secondary jet remain poorly understood. Most existing theoretical and numerical studies terminate their analysis once the primary jet has been launched, and, to the authors' knowledge, experimental observations of secondary jets are absent. From the large-scale measurements of MC22, it is evident that the primary jet generates a secondary cavity whose pinch-off produces a strong acoustic impulse not observed during the formation of the initial jet, (\href{https://static.cambridge.org/content/id/urn:cambridge.org:id:article:S0022112021010235/resource/name/S0022112021010235sup007.mp4}{e.g., supplementary movie 7}). 
\citet{Grumstrup_2007,Yu_2023} also showed that cavity pinch-off generates a sharp acoustic burst, manifested as surface ripples and strong acoustic emissions during rapid cavity collapse.
Although the present study does not resolve the gas phase or acoustic propagation, \citet{Francoeur_2025} and \citet{Nelli2025} has demonstrated the potential of acoustic emission to infer breaking intensity, highlighting the importance of understanding gas entrainment and strength of cavity-collapse \citep{Deike2022}. Furthermore, the primary cavity collapse can strongly modify the subsequent free surface evolution. Experiments by \citet{Rabbi_2021} showed that the impact force on the trailing sphere is determined by the timing between the first cavity’s pinch-off and the second sphere’s arrival at the free surface. These findings indicate that it is insufficient to consider only the jet generated at the primary collapse; rather, the entire sequence of post-collapse free surface dynamics must be evaluated to understand the full progression toward secondary-jet formation. 
Therefore, accurately characterising the four-stage post-primary collapse sequence --- the primary jet impact, the subsequent cavity formation, its pinch-off and collapse, and the re-emergence of a secondary jet --- is essential.  Such analysis not only advances the fundamental understanding of axisymmetric breaking but also provides crucial insight into the underlying fluid mechanics governing impact pressures, air entrainment and associated acoustic phenomena.

\begin{figure}[htbp]
    \centering
    \includegraphics[width=1.0\textwidth]{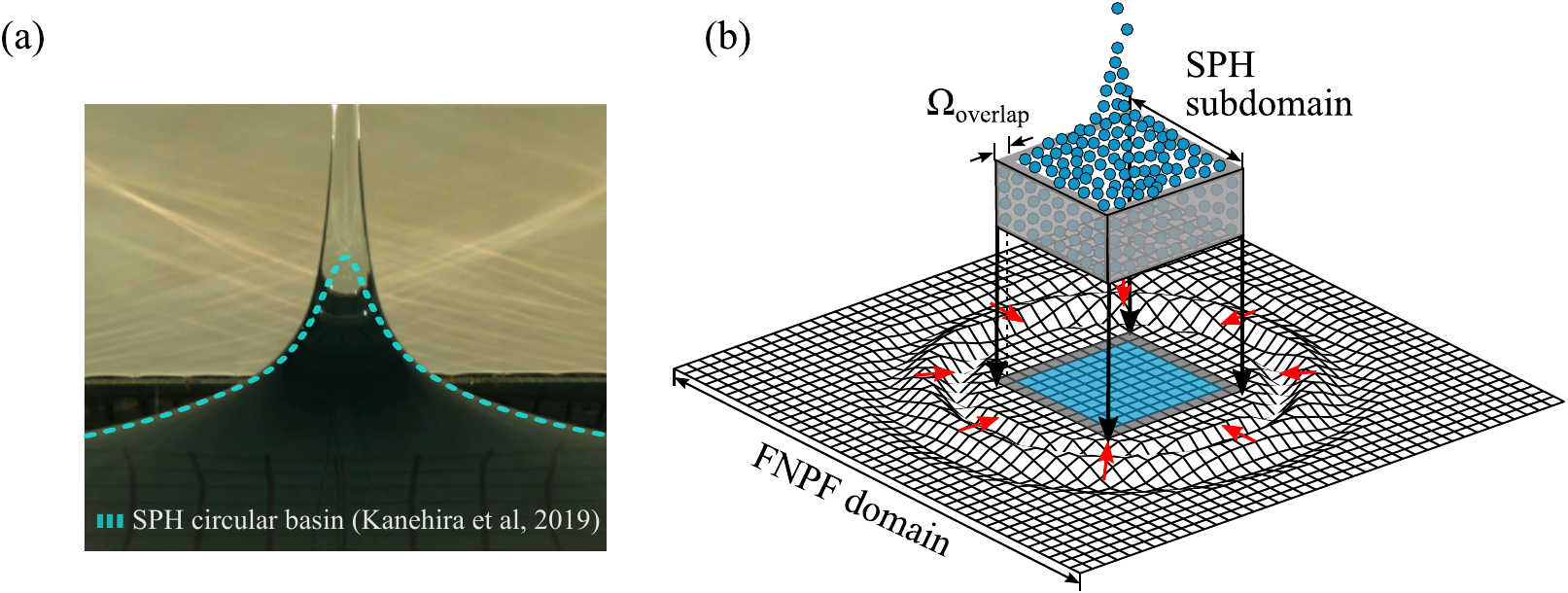}
    \caption{(a) Comparison between the experimental axisymmetric jet profile and the circular-basin SPH simulation \citep{Kanehira19}, showing that SPH underestimates the jet height and fails to reproduce the sharp jet tip. (b) Conceptual diagram of the \textit{OceanSPHysics3D} framework, in which a fully nonlinear potential-flow (FNPF) domain solves the large-scale wave propagation, while an embedded local SPH subdomain resolves wave breaking. The two solvers exchange information through the overlap region $\Omega_{\mathrm{overlap}}$.}
    \label{Fig:SPH_basin}
\end{figure}

To address this gap, we develop a fully three-dimensional numerical framework capable of resolving three-dimensional wave breaking problems, including, as studied here, axisymmetric jet formation and post-collapse dynamics. Mesh-free methods such as smoothed particle hydrodynamics (SPH) \citep{Lucy77,Gingold_and_Monaghan_77} and moving particle semi-implicit (MPS) \citep{Koshizuka96}, which naturally accommodate strongly deformed free surfaces, have therefore been widely applied to breaking-wave problems \citep{Corrado2023,King_2023}. However, as illustrated in figure~\ref{Fig:SPH_basin}a, resolving highly localised breaking without excessive dissipation or cost is still difficult in a 3D domain \citep[see also][]{Kanehira19,Kanehira21}. 
Coupled frameworks \citep{Narayanaswamy10,Sriram14,ALTOMARE18,VERBRUGGHE18,Fourtakas2017} offer a promising route forward, but existing SPH-based schemes often rely on particle creation/deletion for free surface tracking, limiting their applicability to multidirectional seas. No current framework addresses fully three-dimensional coupling with arbitrary incident wave directions \citep{Liu23}. 
These challenges motivate the present development of \textit{OceanSPHysics3D}, a fully three-dimensional coupling system that integrates high-fidelity wave information from the fully nonlinear potential-flow solver OceanWave3D \citep{ENGSIGKARUP09} with the GPU-accelerated SPH solver DualSPHysics \citep{Dominguez22} as illustrated in figure \ref{Fig:SPH_basin}b. 
In this work, we employ \textit{OceanSPHysics3D} to resolve, for the first time in fully three dimensions, the entire sequence of axisymmetric wave breaking—including primary-jet formation, free-fall and cavity evolution, violent pinch-off, and secondary-jet formation—thereby bridging long-standing gaps in the the physical understanding of this phenomenon.

In this study, our aim is to develop a unified physical description of the entire sequence of axisymmetric wave breaking—from primary-jet formation to free-fall, secondary cavity formation and collapse and the emergence of the secondary jet—by analysing the associated pressure forcing, jet kinematics, cavity self-similarity and secondary-jet morphology. To guide our investigation, we focus on the following questions: (i) what is the three-dimensional pressure field responsible for primary-jet formation, and what is its peak magnitude? (ii) can the critical crest-angle criterion of \cite{Longuet-Higgins_1983} reliably indicate breaking onset when assessed through jet kinematics and acceleration? (iii) how does the cavity collapse evolve after jet impact, and does it follow or deviate from classical self-similar scaling?, and (iv) what mechanisms govern the formation and bifurcation behaviour of the secondary jet? The remainder of this paper is organised as follows. \S\ref{Sec:Numerical Methods} introduces the \textit{OceanSPHysics3D} framework and the implementation of the open-boundary coupling strategy. \S\ref{Sec:AxisymmetricWave} presents the validation against axisymmetric focused wave experiments, together with the numerical setup and convergence tests. \S\ref{Sec:Analysis} addresses the questions posed above through a detailed analysis of primary and secondary jet formation, cavity dynamics and the applicability of the critical crest-angle criterion. The main physical insights are summarised in \S\ref{Sec:Conclusions}.

\section{Numerical Methods}
\label{Sec:Numerical Methods}
\subsection{Non-linear Potential Flow} 
The present study employs OceanWave3D \citep{ENGSIGKARUP09}, a fully nonlinear potential flow solver, to simulate wave evolution away from breaking. The potential-based formulation offers a computationally efficient alternative to solving the full Navier–Stokes equations while retaining the essential nonlinear wave dynamics including wave dispersion and nonlinear wave–wave interactions with high accuracy, but excluding viscous effects.

\subsubsection{Governing Equations}
We consider a Cartesian coordinate system in which the $xy$-plane coincides with the still water surface, and the vertical $z$-axis is oriented upwards. The free surface elevation at time $t$ is given by $z = \eta(\boldsymbol{x}, t)$, where $\boldsymbol{x} = (x, y)$ denotes the horizontal position vector. Assuming that the fluid is incompressible and irrotational, a velocity potential
$\phi(\boldsymbol{x},z,t)$ exists such that
\begin{equation}
    \boldsymbol{u} = \nabla\phi 
    = (\phi_x,\phi_y,\phi_z),
    \label{eq:u_from_phi}
\end{equation}
and the incompressibility condition $\nabla\cdot\boldsymbol{u}=0$ reduces to the Laplace equation:
\begin{equation}
    \nabla^2 \phi = 0, \quad \text{for } -h \leq z < \eta,
    \label{eq:laplace}
\end{equation}
where $h(\boldsymbol{x})$ denotes the still water depth.
Equations~\eqref{eq:u_from_phi}--\eqref{eq:laplace} constitute the governing equations for the potential-flow formulation used in OceanWave3D.

\subsubsection{Boundary Conditions and Coordinate Transformation}
The kinematic and dynamic boundary conditions for potential flow are expressed as follows:
\begin{equation}
    \partial_t \eta = -\nabla \eta \cdot \nabla \tilde{\phi} + \tilde{w} (1 + \nabla \eta \cdot \nabla \eta),
    \label{eq:kinematic_bc}
\end{equation}
\begin{equation}
   \partial_t \tilde{\phi} = -g \eta - \frac{1}{2} \left( \nabla \tilde{\phi} \cdot \nabla \tilde{\phi} - \tilde{w}^2 (1 + \nabla \eta \cdot \nabla \eta) \right),
   \label{eq:dynamic_bc}
\end{equation}
where $\tilde{\phi} = \phi(\boldsymbol{x}, z = \eta, t)$ denotes the velocity potential, and $\tilde{w} = \partial_z \phi|_{z = \eta}$ represents the vertical velocity, both evaluated at the free surface. The horizontal gradient operator is defined as $\nabla = (\partial_x, \partial_y)$, and $g$ is the gravitational acceleration.
To numerically integrate equations~\eqref{eq:kinematic_bc} and \eqref{eq:dynamic_bc}, initial conditions for both $\eta(\boldsymbol{x}, 0)$ and $\tilde{\phi}(\boldsymbol{x}, 0)$ must be specified. In this study, the initial conditions are obtained from linear wave theory, specified at a time $t_0=-11$ s when the wave group is sufficiently dispersed and remains approximately linear as in MC22.
To advance equations~\eqref{eq:kinematic_bc} and \eqref{eq:dynamic_bc} in time, the Laplace equation \eqref{eq:laplace} is solved within the fluid domain subject to the following boundary conditions:
\begin{equation}
    \phi = \tilde{\phi}, \quad \text{at } z = \eta,
    \label{eq:bc_surface}
\end{equation}
\begin{equation}
    \partial_z \phi + \nabla h \cdot \nabla \phi = 0, \quad \text{at } z = -h.
    \label{eq:bc_bottom}
\end{equation}

After solving equation~\eqref{eq:laplace}, the vertical velocity $\tilde{w}$ at the free surface is obtained, and the time evolution of $\eta$ and $\tilde{\phi}$ is computed using equations~\eqref{eq:kinematic_bc} and \eqref{eq:dynamic_bc}. 

To transform the physical domain with a moving free surface into a fixed computational domain, the $\sigma$ coordinate system is introduced as
\begin{equation}
    \sigma \equiv \frac{z + h(\boldsymbol{x})}{\eta(\boldsymbol{x}, t) + h(\boldsymbol{x})} \equiv \frac{z + h(\boldsymbol{x})}{d(\boldsymbol{x}, t)},
    \label{eq:sigma_def}
\end{equation}
where $d(\boldsymbol{x}, t) = \eta(\boldsymbol{x}, t) + h(\boldsymbol{x})$ is the total instantaneous water depth. This transformation maps the time-dependent vertical domain $[-h(\boldsymbol{x}), \eta(\boldsymbol{x}, t)]$ to the fixed interval $[0, 1]$. The governing equations are discretised using sixth-order central finite differences in all spatial directions and integrated in time using a classical fourth-order Runge–Kutta scheme. For further details on the numerical solution, the reader is referred to \citet{ENGSIGKARUP09}. 

\subsection{SPH}
\subsubsection{SPH Formulation}
The particle-based SPH solver, DualSPHysics version 5.2 \citep{Dominguez22}, used in this study follows the SPH formulation \citep{Lucy77,Gingold_and_Monaghan_77}. Physical quantities are approximated through kernel interpolation over neighbouring particles.
For comprehensive reviews of the SPH methodology and its applications in fluid dynamics, readers are referred to \cite{Violeau16,Touzé25,GotohKhayyer25}.
In the SPH framework, the physical quantities ($\phi(\boldsymbol{r})$) at a calculation point $\boldsymbol{r}$ are first approximated using an integral interpolation:

\begin{equation}
    \phi(\boldsymbol{r}) = \int_\Omega \phi(\boldsymbol{r}') W(\|\boldsymbol{r} - \boldsymbol{r}'\|, h_{SPH}) \, d\boldsymbol{r}',
\end{equation}
where $W$ is the smoothing kernel with compact support, meaning that its value is non-zero only within a finite radius proportional to the smoothing length $h_{SPH}$. Here, $\boldsymbol{r}'$ denotes a neighbouring position vector in the kernel support and $\Omega$ is the integration region. This integral is then discretised into a summation over particles as:
\begin{equation}
    \phi(\boldsymbol{r}_i) \approx \sum_j^N \phi(\boldsymbol{r}_j) W(\|\boldsymbol{r}_i - \boldsymbol{r}_j\|, h_{SPH})V_j ,
\end{equation}
where $V_{j}={m_j}/{\rho_j}$ is the volume of neighbouring particle $j$, with $m_j$ its mass and $\rho_j$ its density, and $N$ denotes the total number of particles. 
Using this SPH interpolation, the gradient of physical quantities ($\phi(\boldsymbol{r_i})$) can be expressed as:
\begin{equation}
    \nabla_i \phi(\boldsymbol{r}_i) = \sum_j^N \phi(\boldsymbol{r}_j) \nabla_i W_{ij} {V_j},
\end{equation}
where $W_{ij}= W(\|\boldsymbol{r}_i - \boldsymbol{r}_j\|, h_{SPH})$. Other expressions for gradients are also used \citep[see][]{Violeau16}. Using this numerical operator, the governing equations of the fluid are discretised as SPH particles. The kernel function used in this paper is the Wendland kernel \citep{Wendland95} which is express as follow:
\begin{equation}
  W(q,h)=
  \begin{cases}
    \alpha_D\,\left(1-\dfrac{q}{2}\right)^{4}\,(2q+1), & 0 \le q \le 2,\\[6pt]
    0, & 2<q,
  \end{cases}
\end{equation}
where $q={r}/{h}={\| \boldsymbol{r}_i-\boldsymbol{r}_j\|}/{h}$, $\alpha_{D}$ is the normalisation constant, equal to ${7}/{4\pi h^{2}}$ in 2D and ${21}/{16\upi h^{3}}$ in 3D.

\subsubsection{Governing Equations}
The SPH discretised form of the governing equations for fluid flow is given below. The continuity equation, including a density diffusion term is written as:
\begin{equation}
    \frac{d \rho_i}{dt} = \sum_j^N m_j (\boldsymbol{u}_j - \boldsymbol{u}_i) \cdot \nabla_i W_{ij} + \delta h_{SPH} c_0 \sum_j^N \psi_{ij} \cdot \nabla_i W_{ij}  V_j,
\end{equation}
where the diffusion term is introduced to suppress spurious density oscillations \citep{MOLTENI09}. Here, $c_0$ is the speed of sound, the $\delta$-SPH coefficient $\delta=0.1$ and the second term on the right-hand side (RHS) is defined according to \cite{FOURTAKAS19} in which the term $\psi_{ij}$ can be written as follows:
\begin{equation}
    \psi_{ij} = 2(\rho^T_{ji}-\rho^H_{ij})\frac{\boldsymbol{r}_{ij}}{\|\boldsymbol{r}_{ij}\|^2},
\end{equation}
where the superscripts $T$ and $H$ denote the total and hydrostatic components, respectively, whose calculation is described in \cite{FOURTAKAS19}. This approach smooths only the dynamic component of the density while excluding the hydrostatic part, and does not require the calculation of density gradients \citep{ANTUONO12}. As a result, the correct hydrostatic pressure gradient is preserved in still-water conditions. This makes the method computationally efficient and particularly effective for gravity-dominated flows as considered in this study.

The momentum equation is expressed in an extended SPH formulation as:
\begin{equation}
\label{eq:momentum_sph}
\begin{aligned}
\frac{d \boldsymbol{u}_i}{dt} =
& - \sum_j m_j \left( \frac{p_i+p_j}{\rho_i\rho_j} \right) \nabla_i W_{ij} + \boldsymbol{g} \\
& + \sum_j m_j \left( \frac{4 \nu_0 \boldsymbol{r}_{ij} \cdot \nabla_i W_{ij}}{(\rho_i + \rho_j)(|\boldsymbol{r}_{ij}|^2 + \zeta^2)} \right) (\boldsymbol{u}_i - \boldsymbol{u}_j) \\
& + \sum_j m_j \left(\frac{\boldsymbol{\tau}_i+\boldsymbol{\tau}_j}{\rho_i\rho_j} \right) \cdot \nabla_i W_{ij},
\end{aligned}
\end{equation}
where $\boldsymbol{r}_{ij} = \boldsymbol{r}_i - \boldsymbol{r}_j$, 
$\nu_{0}$ is the kinematic viscosity (typically $1.0 \times 10^{-6}~\mathrm{m^{2}/s}$ for water), 
and $\zeta^{2} = 0.01 h^{2}$ is a small regularisation parameter introduced to avoid singularities. 
The second term on the RHS corresponds to the laminar viscous stresses as formulated in \citet{LO02}. 
The last term involves $\boldsymbol{\tau}$, which denotes the Sub-Particle Scale (SPS) stress tensor presented in \citet{DALRYMPLE06} which is employed to model the effect of the SPS turbulence associated with wave breaking.

\subsubsection{Pressure and Time Stepping Schemes}
Assuming the fluid is weakly compressible, the pressure can be calculated using Tait's equation of state, which relates pressure to the density:

\begin{equation}
  p = b \left[\left(\frac{\rho}{\rho_0}\right)^\gamma - 1\right],
\end{equation}
where $\gamma = 7$, $b = c_0^2 \rho_0 / \gamma$, $\rho_0 = 1000~\text{kg/m}^3$ is the reference density. This formulation allows explicit computation of pressure, significantly reducing computational costs compared to solving Poisson's equation with an implicit method. In this study, the speed of sound was defined as $c_0=20\sqrt{g/k_0}$ based on the phase speed of deep water, where $k_0$ denotes the characteristic wavenumber of an axisymmetric standing wave discussed in \S \ref{Sec:Spike3D}.  An explicit second-order time integration method (the symplectic scheme) was employed. The size of time stamps is determined according to \cite{Monaghan92} with a CFL (Courant–Friedrichs–Lewy) number of 0.2.


\subsubsection{Boundary conditions}
\label{Sec:mOBC}
Numerous versions of boundary conditions have been used where SPH applies to the whole domain, effectively as a numerical wave basin \citep{NI18,TAFUNI18,VERBRUGGHE19,Kanehira19,TSURUTA21,yang2023numerical}.
Here the SPH domain is embedded within the OceanWave3D domain which covers the whole domain as in figure\ref{Fig:SPH_basin}b; this is equivalent to Chimera meshing for mesh-based methods \citep{BENEK_1983}, sometimes known as overset meshing in OpenFOAM \citep[][]{OpenFOAM_v1706}.
Buffer zones are initially set up on the boundaries of the SPH domain (figure~\ref{Fig:Sim_Domain_Spike3D}b) and particles specified as boundary particles are given velocities from the OceanWave3D computation through trilinear interpolation.
Unlike conventional schemes that restrict buffer particles to horizontal motion, the present method directly tracks their full three-dimensional motion, allowing accurate free surface representation and avoiding spurious vortices near the boundary \citep{NI18,Kanehira25}.
The refilling mode used in previous studies \citep{TAFUNI18,VERBRUGGHE19,yang2023numerical} is disabled, as buffer particles are directly advected by interpolated velocities.
Although boundary particles may be gradually lost due to Stokes drift, the buffer region is sufficiently wide for the simulation duration.
Pressure is not interpolated, since the weakly compressible SPH formulation is inconsistent with the incompressible OceanWave3D field, which can cause instability.
Instead, the horizontal pressure gradient into the inner SPH domain is set to zero, providing stable simulations and close agreement between the overlapping SPH and OceanWave3D results for waves up to moderate steepness.
Further details of the DualSPHysics implementation and the coupling procedure are provided in \Cref{App:Implementaion}.
We use the modified dynamic boundary condition (mDBC) by \citet{English22} for the solid boundaries.

\section{Simulated Axisymmetric wave and its validation}
\label{Sec:AxisymmetricWave}

\subsection{Numerical set-up}
To demonstrate the proposed framework \textit{OceanSPHysics3D} in \S\ref{Sec:Numerical Methods}, we demonstrate an axisymmetric standing wave in the three-dimensional numerical domain. A temporally and spatially localised standing wave is generated at the centre of the computational domain. The numerical setup used here corresponds to the experiments in MC22.

\begin{figure}[htbp]
\centering
\includegraphics[width=0.8\textwidth] {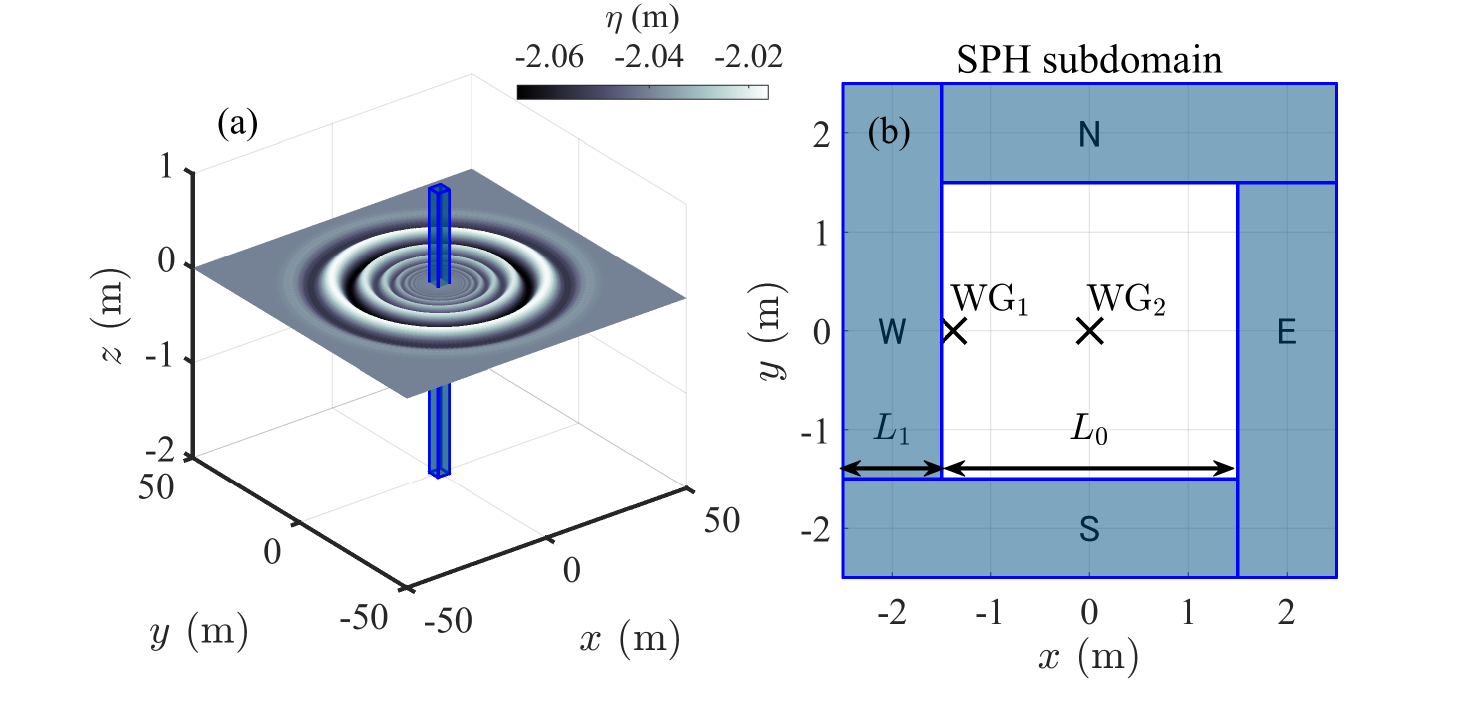}
\caption{Numerical domain in OceanWave3D (a) and the SPH subdomain (b). The SPH subdomain is highlighted by the blue rectangle in panel (a).}\label{Fig:Sim_Domain_Spike3D}
\end{figure}

The axisymmetric wave is generated through linear dispersive focusing, in which a broad-banded spectrum of axisymmetric components is phase-aligned to concentrate energy at the tank centre. As discussed in MC22, the focusing mechanism is predominantly linear, with nonlinearity only influencing the crest height at focus in the evolution. The limiting steepness of such axisymmetric waves can be characterised by the critical crest angle $2\gamma = 2\arctan\sqrt{2}  \,\, \rm{(i.e.\,\,}109.47^{\circ}\rm{)}$ proposed by \cite{Longuet-Higgins_1983}, beyond which jet formation occurs.

Figure~\ref{Fig:Sim_Domain_Spike3D}a illustrates the initial water free surface in OceanWave3D ($t_0=-11$ s) and the SPH subdomain represented by the centrally-located vertical rectangular volume. To initialise OceanWave3D, two input fields were prescribed: the velocity potential at the free surface and the free surface elevation, both generated from linear wave theory. The directional spectrum was set identical to the input spectrum of MC22, consisting of a uniform directional distribution and the ITTC frequency spectrum \citep{Matthews72}, with mean period $\overline{T} = 2.3$~s. 
The linear amplitude sum $A_0$ was taken directly from MC22 (table~2). In particular, $A_0 = 0.407$~m was used for Exp.~20, $A_0 = 1.018$~m for Exp.~50, and $A_0 = 1.526$~m for Exp.~75.
The frequency spectrum was discretised with $\delta f = 1/64 = 0.0156$~Hz, with maximum frequency 2~Hz, consistent with the wavemaker limit in the experiments.

The numerical setup used in OceanWave3D is listed in table \ref{Table:Setting_OCW3_3D}. The OceanWave3D domain is $100 \times 100$~m with water depth $h=2$~m, consistent with simulations presented in MC22. 
A coarser horizontal grid ($N_x=N_y=513$) was used instead of $N_x=N_y=1024$ in MC22, since finer grids captured steep jets that caused instability. 
As our aim is to provide boundary conditions for the SPH subdomain, this resolution ensures stability beyond the focusing event. 
In OceanWave3D, when the vertical acceleration exceeds a specified threshold of $0.4g$, wave breaking is assumed to occur, and a spatial filter is applied to smooth the free surface and maintain numerical stability.
The time step was set to $\Delta t_{\mathrm{FNPF}}=0.04$~s so that the Courant number $C_g \Delta t_{\mathrm{FNPF}} / \Delta x = 0.6$. 
The group velocity in finite depth is then given by
\begin{equation}
C_g = n \frac{\omega_0}{k_0}, \quad 
n = \tfrac{1}{2}\left(1+\tfrac{2k_0 h}{\sinh(2k_0 h)}\right),
\end{equation}
where $\omega_0=\sqrt{gk_0 \tanh(k_0 h)}$ is the angular frequency. The characteristic wavelength was $\lambda_0 = 2r_0 = 12.26$~m with the radial position of the wave trough $r_0=6.13$~m at linear focus. 
The simulations started at $t_0=-11$~s, ensuring all frequency components had time to propagate across the domain and constructively interfere at $t=0$.

Figure~\ref{Fig:Sim_Domain_Spike3D}b presents an enlarged view of the SPH subdomain. 
The domain length $L_0$ was chosen to ensure that reflected waves would not interfere with the breaking phenomena at the centre of the domain. 
The thickness of the boundary zone $L_1$ was selected such that open-boundary particles remain within the zone even under nonlinear effects such as Stokes drift, since particle-refilling was not enabled.
The four open boundaries (N, E, S, W) were introduced following the configuration of the existing InOut boundary condition \citep{TAFUNI18}, for which the code was modified accordingly in this study. In the coupling zone, boundary particles are driven solely by the predefined velocity and their velocity is not updated by the equations of motion as described in \S \ref{Sec:mOBC}.

\begin{table}
  \centering
  \begin{threeparttable}
    \begin{tabular}{cccccc}
      \toprule
      \multicolumn{2}{c}{No. Grid} & Grid Size (m) &  Time stamp (s)& CFL &\\
      $N_x$, $N_y$ & $N_z$ & $\Delta x, \Delta y$ & $\Delta t_{\mathrm{FNPF}}$ &
      $\bigl(c_g \Delta t_{\mathrm{FNPF}} / \Delta x\bigr)$ & $t_0 $ (s) \\
      513 & 9 & 0.195  & 0.04 & 0.60 & -11 \\
      \bottomrule
    \end{tabular}
  \end{threeparttable}
  \caption{Numerical parameter used in OceanWave3D.}
    \label{Table:Setting_OCW3_3D}
\end{table}

Table~\ref{Table:Test cases_3D} summarises the test cases in this study. To demonstrate applicability, three $A_0$ values were used corresponding to Exp.~20, 50, and 75. Following MC22, crest-angle and velocity analysis shows that breaking occurs between Exp.~30 and 40. Thus, Case~1 (Exp.~20) represents a non-breaking wave, while Cases~2--7 (Exp.~50) and Cases~8--12 (Exp.~75) correspond to vertically breaking waves with jet formation, Exp.~75 being the largest condition in MC22. For Exp.~50 (Cases~3--5) and Exp.~75 (Cases~8--10), the effect of smoothing was tested by varying $h/dp=1.4, 1.7,$ and $2.0$. Convergence was also checked by varying $A_0/d_p$ (Cases~4, 6, 7 for Exp.~50 and Cases~9, 11, 12 for Exp.~75). The results of these sensitivity and convergence tests are provided in Appendix \ref{App:Convergence} and Appendix \ref{App:hdp}. Higher resolution improves agreement with experiments in jet evolution, while insufficient particles ($D/d_p<10$) cause jet instability ($D$ is a jet diameter). Smaller $h_{\mathrm{SPH}}/d_p$ enhances jet height.
In table \ref{Table:Test cases_3D}, the maximum jet amplitudes ($A$) and peak crest velocities $\dot{\eta}_C$ obtained from the MC22 experiments and SPH simulations are also presented.
In SPH, $A$ represents the maximum elevation of the main jet, excluding detached droplets as shown later in figure \ref{Fig:Gamma_Surf_Exp_75}. 
For the SPH simulations, the crest velocity $\dot{\eta}_C$ is evaluated using the forward finite difference,
$\dot{\eta}_C(t^n)=(\eta_C^{n+1}-\eta_C^{n})/\Delta t,$
where $n$ denotes the discrete output time index. 

\begin{table}
\centering
\begin{tabular}{cccccccccccc}
\toprule
\multirow{2}{*}{Case} & \multirow{2}{*}{Exp.} & \multirow{2}{*}{$L_0,L_1$(m)} & \multirow{2}{*}{$A_0/d_p$} &  \multirow{2}{*}{$h_{\rm{SPH}}/d_p$} & \multirow{2}{*}{$R$ (hr)} & \multirow{2}{*}{$M_{\rm GPU}$ (GB)} & &
\multicolumn{2}{c}{$A$ (m)} &
\multicolumn{2}{c}{$\dot{\eta_C}$ (m/s)} \\
\cmidrule{8-12}
 &  &  &  &  &  &  & & MC22 & SPH & MC22 & SPH \vspace{2mm} \\

1  & 20 & 3 & 20 & 1.7 & 6.17 & 1.57 & & $0.338^{\dagger}$ & 0.456    & $0.965^I$ & 1.02  \vspace{2mm}\\
\midrule
2  & \multirow{6}{*}{50}  & \underline{5}, 1 & 51 & 1.7 & 16.7 & 4.31 &  & \multirow{6}{*}{$1.680^{\ddagger}$}      & 1.95 &  \multirow{6}{*}{$4.83^{\star}$}     & 4.65 \\
3  &  & 3, 1 & 51 & \underline{1.4} & 9.6 & 3.20 & &       & 2.15 &       & 5.35 \\
4  &  & \underline{3}, 1 & \underline{51} & \underline{1.7} & 13.1 & 3.20 & &  & 1.86 &   & 4.65 \\
5  &  & 3, 1 & 51 & \underline{2.0} & 16.4 & 3.20 & &       & 1.93 &       & 5.09 \\
6  &  & 3, 1 & \underline{26} & 1.7 & 0.9 & 0.45 & &       & 1.57 &       & 4.03 \\
7  &  & 3, 1 & \underline{34} & 1.7 & 2.7 & 1.01 & &       & 1.60 &       & 4.12 \vspace{2mm} \\
8  & \multirow{5}{*}{75} & 3, 1.2 & 76 & \underline{1.4} & 12.0 & 3.79 & & \multirow{5}{*}{$6.027^{\ddagger}$} & 6.19 & \multirow{5}{*}{$10.38^{\star}$} & 10.72 \\
9  &  & 3, 1.2 & \underline{76} & \underline{1.7} & 15.6 & 3.79 & &       & 5.16 &       & 10.20 \\
10 &  & 3, 1.2 & 76 & \underline{2.0} & 19.9 & 3.79 & &       & 4.81    &       & 10.12 \\
11 &  & 3, 1.2 & \underline{38} & 1.7 & 1.3 & 0.53 & &       & 3.58    &       & 7.92 \\
12 &  & 3, 1.2 & \underline{153} & 1.7 & 254.8 & 31.96 & &       & 6.55 &       & 13.76 \\
\bottomrule
\end{tabular}
\caption{%
Comparison of the crest amplitudes ($A$) and peak crest velocities ($\dot{\eta}_C$) obtained from the experiments (MC22) and SPH simulations. 
In the experimental data, superscripts ${\dagger}$ and ${\ddagger}$ denote values measured using wave gauges and those obtained from calibrated images. 
Superscripts ${I}$ and ${\star}$ in $\dot{\eta}_C$ indicate velocities derived from calibrated images and from the theoretical expression $\dot{\eta}_C = \sqrt{2g(A - A_{\mathrm{cusp}})}$, where $A_{\mathrm{cusp}}$ denotes the crest amplitude at the onset of cusp formation.
Underlined values are highlighted to aid visual comparison among selected cases; 
for instance, cases 2 and 4 illustrate the effect of $L_0$, whereas cases 4, 6, and 7 highlight the influence of particle resolution.
}
\label{Table:Test cases_3D}
\end{table}

\begin{figure}[htb]
\centering
\includegraphics[width=0.9\textwidth] {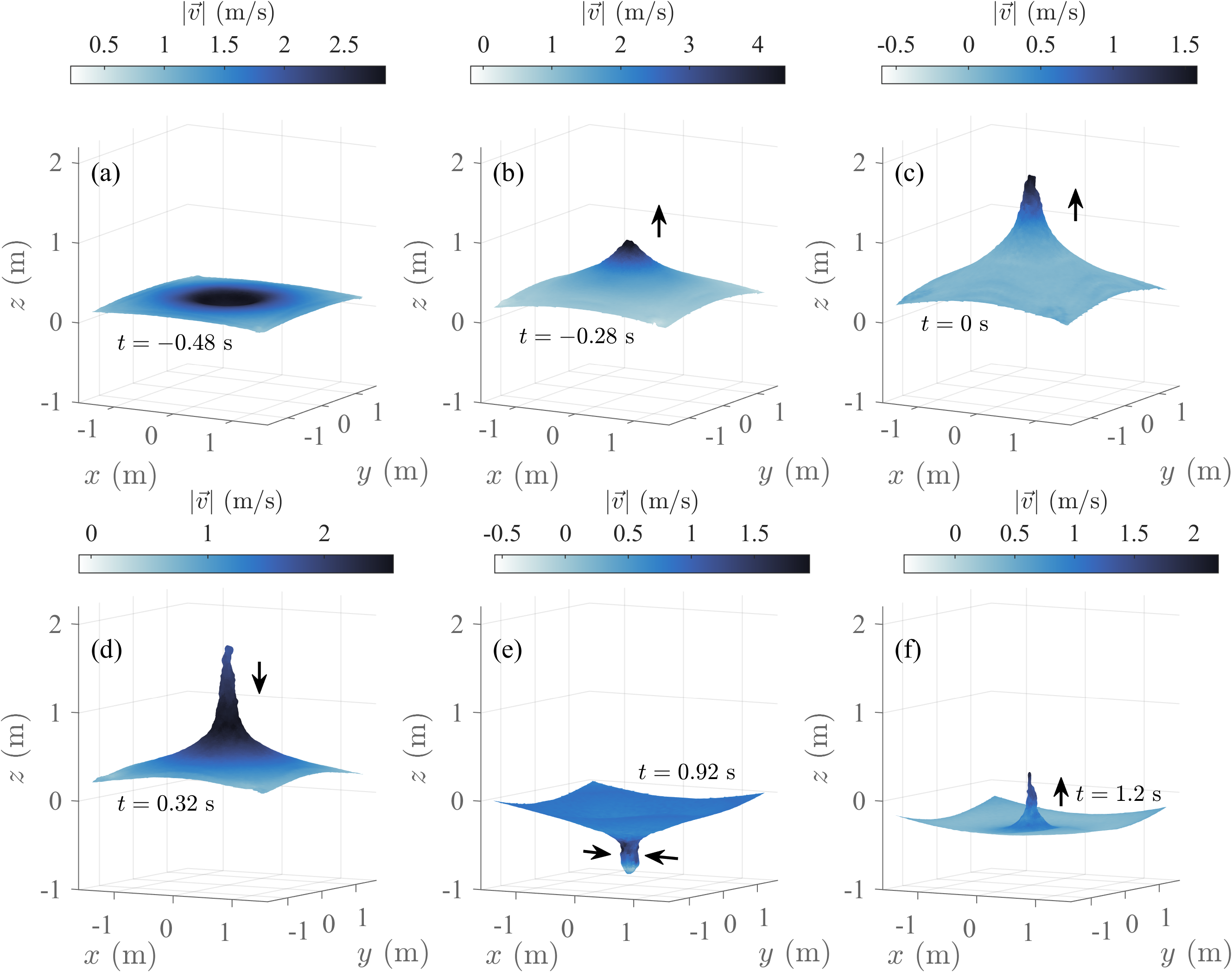}
\caption{Sequence of three-dimensional water free surface profiles simulated in the SPH for Exp. 50. Panels (a)--(c) represent curvature collapse to vertical jet evolution, whereas (d)--(f) show falling jet, secondary cavity formation and collapse.}\label{Fig:SufaceSnapshots_Exp50}
\end{figure}

\begin{figure}[htb]
\centering
\includegraphics[width=0.9\textwidth] {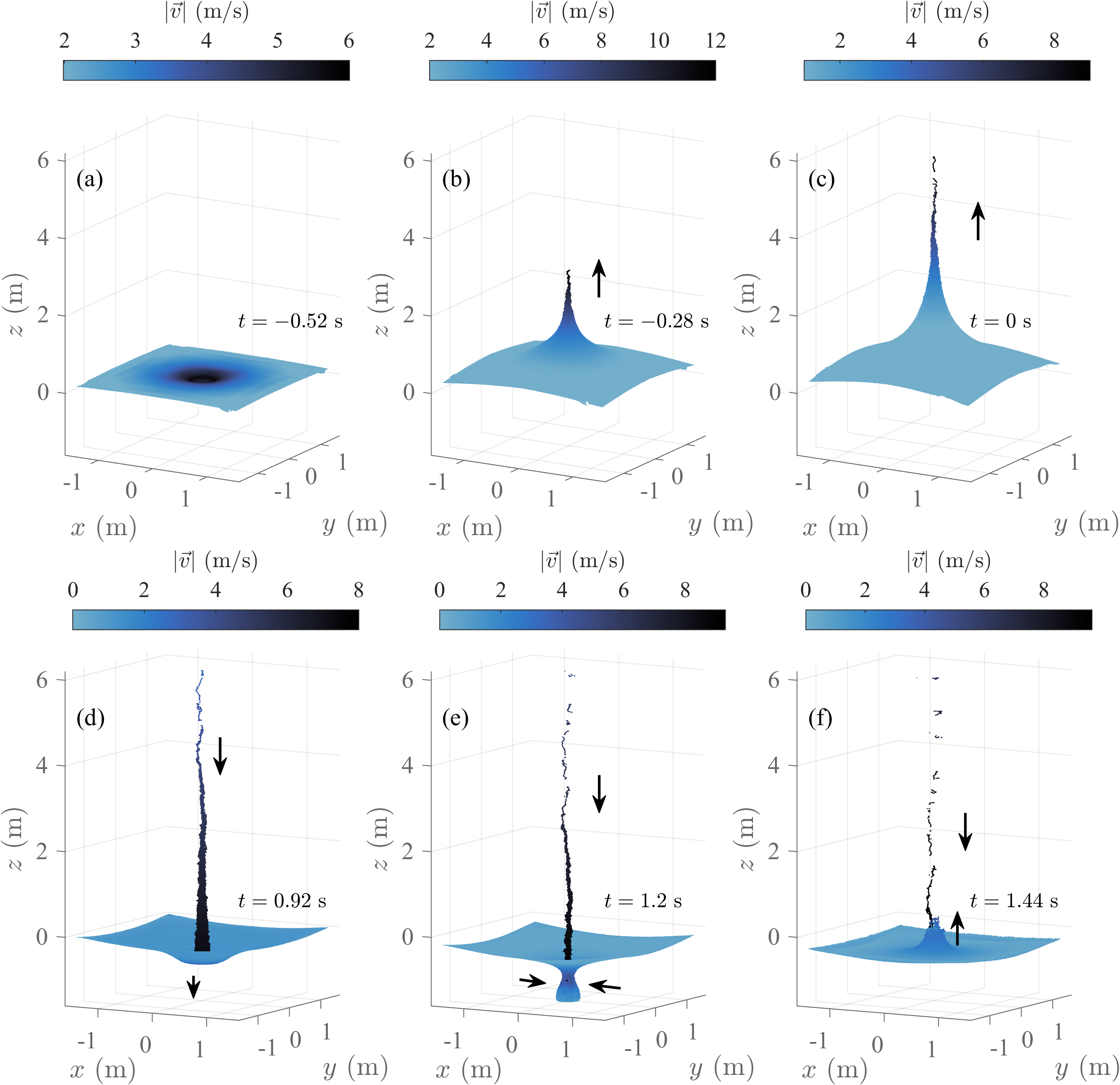}
\caption{Sequence of three-dimensional water free surface profiles simulated in the SPH for Exp. 75. Panels (a)--(c) represent curvature collapse to vertical jet evolution, whereas (d)--(f) show falling jet, secondary cavity formation and collapse.}\label{Fig:SufaceSnapshots_Exp75}
\end{figure}

Figure~\ref{Fig:SufaceSnapshots_Exp50} shows time series snapshots of Exp.~50 from the embedded SPH model, capturing the sequence from trough focusing  (a) to primary jet formation (b), jet evolution (c), falling jet (d), secondary cavity (e), and secondary jet (f).
Previous studies have used a variety of terms to describe the collapse of the free surface leading to jet formation.
\citet{Zeff2000} referred to this process as `curvature collapse' and stated that it is inertia-driven.
In contrast, \citet{Longuet-Higgins01} described a similar process that is gravity-driven as a `collapse of the trough'.
\citet{GEKLE10} examined `cavity collapse' generated by the impact of a rigid body on a free surface.
MC22 used the term `curvature collapse' for the trough focusing and subsequent cavity collapse that leads to primary jet formation in axisymmetric wave focusing.
In the present study, we follow MC22 and use `curvature collapse' for the primary jet formation. In contrast, the collapse of the deep cavity formed during the free fall of the primary jet can generate a secondary jet, and because the governing mechanism may be fundamentally different from that of the primary jet formation, we refer to this process as `cavity collapse' (see \S\ref{Sec:Cavity_Formation} for details).

Figure~\ref{Fig:SufaceSnapshots_Exp75} illustrates the same sequence for Exp.~75, but with a more slender jet that extends higher before fragmenting into droplets. In panel (e), the falling jet is entrained into the collapsing cavity, and interaction with a secondary jet is evident. For both Exp.~50 and Exp.~75, the maximum velocity occurs during the initial jet formation (panel b) and decreases as the jet develops. The details of jet formation and evolution are further discussed in \S \ref{Sec:Jet_Formation}.

\subsection{Axisymmetric wave breaking in 3D domain}
\label{Sec:Spike3D}

\begin{figure}[htbp]
\centering
\includegraphics[width=1.0\textwidth] {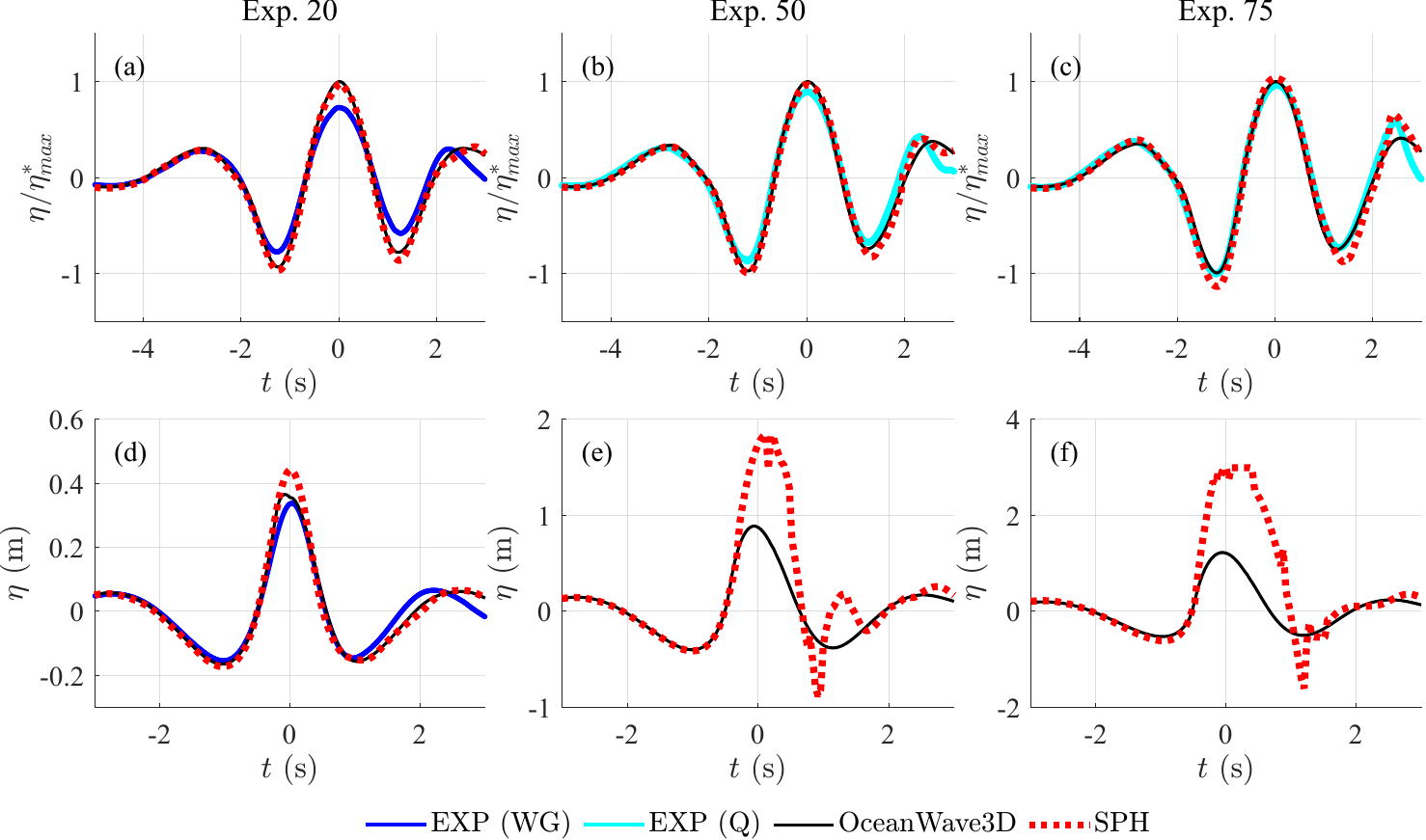}
\caption{Time series comparison of water surface elevation. Panels (a--c) show the surface elevations measured near the open boundary ($x=-1.37$), while (d--f) show the surface elevations at the centre of the simulation domain ($x,y=0$). WG denotes the wave-gauge measurements and Q denotes the Qualisys floating-marker 
measurements.}\label{Fig:TimeSeriesComp_Spike3D}
\end{figure}

The time series of surface elevations for Exp.~20, 50, and 75 are shown in figure~\ref{Fig:TimeSeriesComp_Spike3D}. Panels (a--c) display results near the open boundary (WG1), while panels (d--f) correspond to the centre of the domain (WG2). Experimental wave-gauge data (blue) are available only for Exp.~20, whereas floating-marker measurements (cyan) are used for Exp.~50 and 75. Comparisons show good agreement between OceanWave3D (black) and SPH (red) near the open boundary, although amplitudes are slightly overestimated relative to experiments (WG: wave gauges, Q: Qualisys floating markers obtained in MC22). This is partly due to imperfect wave generation, since the target amplitude $A_0$ from linear theory does not exactly match the realised crest in the laboratory.
At the domain centre, OceanWave3D and SPH agree well up to the onset of focusing and jet formation (panels e,f). After $t=0$, OceanWave3D cannot recreate wave breaking and cavity collapse, which could cause reflected waves when its output is used as boundary input. The domain was therefore chosen to be sufficiently large to avoid interference with the central region of interest. In panels (e,f), the trough deepens after $t=0$ s due to secondary cavity formation (figures~\ref{Fig:SufaceSnapshots_Exp50}e, \ref{Fig:SufaceSnapshots_Exp75}e). 
It should be noted that the value of $\eta$ in figure \ref{Fig:TimeSeriesComp_Spike3D} is not the amplitude of the jet itself.
The SPH quantity plotted here is a free surface elevation reconstructed from particle density (using a threshold 0.5$\rho_0$), and therefore reflects an SPH-based surface measure rather than the peak jet height.
Because the slender jet contains relatively few particles, this reconstructed elevation does not represent the true jet amplitude in the SPH subdomain.

\begin{figure}[htbp]
\centering
\includegraphics[width=1.0\textwidth] {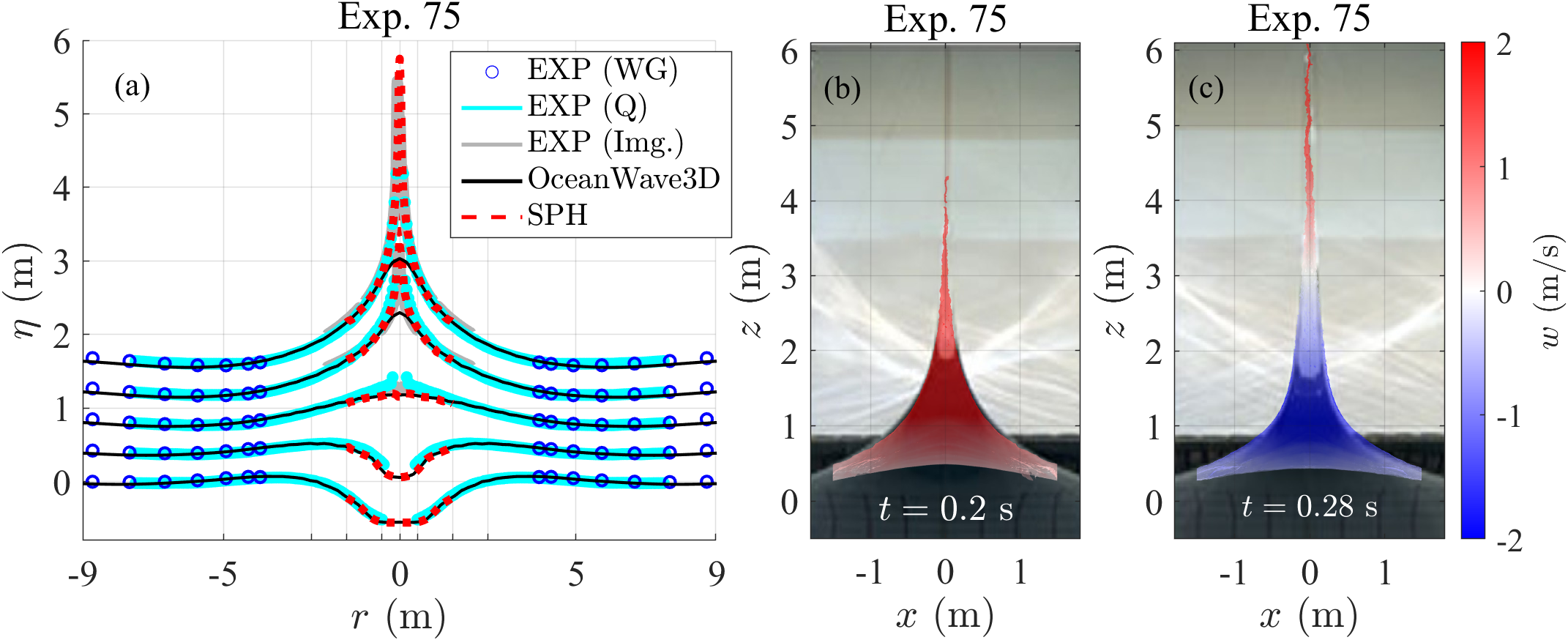}
\caption{Comparison of water surface profiles during the development of a vertical jet in Exp. 75.
Panel (a) shows the time evolution of surface elevations from the wave trough to the formation of the vertical jet, comparing experimental measurements, OceanWave3D, and SPH results at multiple time instants.
Panel (b) presents a side view of the free surface profile at $t=0.2$ and $t= 0.28$ s, with the latter corresponding to the instant when maximum jet height is observed in the experiment. WG denotes the wave-gauge measurements, Q denotes the Qualisys floating-marker 
measurements, and Img. denotes the calibrated image-based surface-elevation data.}\label{Fig:Time_Space_Comp_Exp75}
\end{figure}

Figure~\ref{Fig:Time_Space_Comp_Exp75} illustrates the development of the vertical jet in Exp.~75. 
Panel (a) presents the temporal evolution of radial surface profiles from the initial wave trough to the jet evolution. 
The results demonstrate that both OceanWave3D and SPH closely follow the experimental measurements (WG, Q and Img.: Calibrated high-speed images in MC22) during the axisymmetric focusing to vertical jet formation. 
As the vertical jet develops, OceanWave3D increasingly underestimates the peak elevation due to the application of its spatial filtering, whereas SPH continues to capture the jet growth with high accuracy. 
Panels (b,c) in figure~\ref{Fig:Time_Space_Comp_Exp75} show side views of the SPH output free surface overlaid onto calibrated image data, with the colour scale representing the vertical velocity $w$.
Colouring by vertical velocity reveals the axial stretching of the jet, with upward motion persisting only near the jet tip in panel (c), while the surrounding fluid has already reversed to downward motion. As a result, the jet is stretched into a slender, conical shape (figure \ref{Fig:SufaceSnapshots_Exp75}d).
It is notable that the presented method accurately reproduces the steep wave profile, including the vertical jet. 
These comparisons highlight the capability of the approach to capture the strongly nonlinear dynamics associated with vertical jet formation and breaking.

The comparison of maximum wave crest velocity defined by $\dot{\eta}_C$ and wave amplitude $A=max(\eta_C)$ between experiments and SPH simulations is summarised in table \ref{Table:Test cases_3D}.
A sensitivity test was first conducted with respect to the horizontal domain length $L_0$ in figure~\ref{Fig:Sim_Domain_Spike3D}b, by comparing Case~2 ($L_0=5$) and Case~4 ($L_0=3$) under otherwise identical conditions. The resulting maximum wave amplitude $A$ differed by approximately 14.9$\%$ for Case~2 and 8.9$\%$ for Case~4 relative to the experimental measurements, while the computational cost did not show a significant variation. Since a higher particle resolution was required in Exp.~75, we adopted $L_0=3$ for the remaining cases. 
Additional sensitivity tests on the particle resolution $A_0/d_p$ and the smoothing-length ratio $h/d_p$ revealed that smaller $h/d_p$ values (e.g.\ 1.4) tend to yield slightly larger peak amplitudes and velocities, as local interpolation of velocity and pressure becomes less smoothed compared with larger $h/d_p$. However, as shown in Appendix \ref{App:hdp}, excessively small $h/d_p$ values also caused segmentation of the jet, and thus $h/d_p=1.7$ was selected as the representative value.
Compared with the experimental data (MC22), the SPH results for Cases~1, 4, and~12 slightly overestimated both the crest amplitude and peak crest velocity, mainly due to differences between the target and actual incident wave conditions in the experiments. Nevertheless, the SPH simulations reproduced the overall trends with good fidelity, confirming the validity of the present numerical approach.
In terms of computational cost, the present coupling approach reduced the run time and memory usage by approximately 90\% compared with the conventional full-basin SPH simulation \citep{Kanehira19}, while maintaining comparable accuracy. Even at $d_p=1$~cm, however, the computational expense remained substantial, highlighting the difficulty of performing large numbers of high-resolution single-SPH simulations.


\section{Jet formation and post-breaking behaviour}
\label{Sec:Analysis}
By capitalising on the results of high-fidelity computations presented in \S\ref{Sec:Spike3D}, we carry out a detailed analysis of the vertical wave breaking dynamics. \S\ref{Sec:Jet_Formation} focuses on the onset and evolution of jet formation, whereas \S\ref{Sec:Cavity_Formation} investigates the post-breaking cavity formation, its subsequent collapse, and the mechanism by which the secondary jets are generated.


\subsection{Pre-jet curvature collapse and primary jet evolution}
\label{Sec:Jet_Formation}
\begin{figure}[htbp]
\centering
\includegraphics[width=0.95\textwidth] {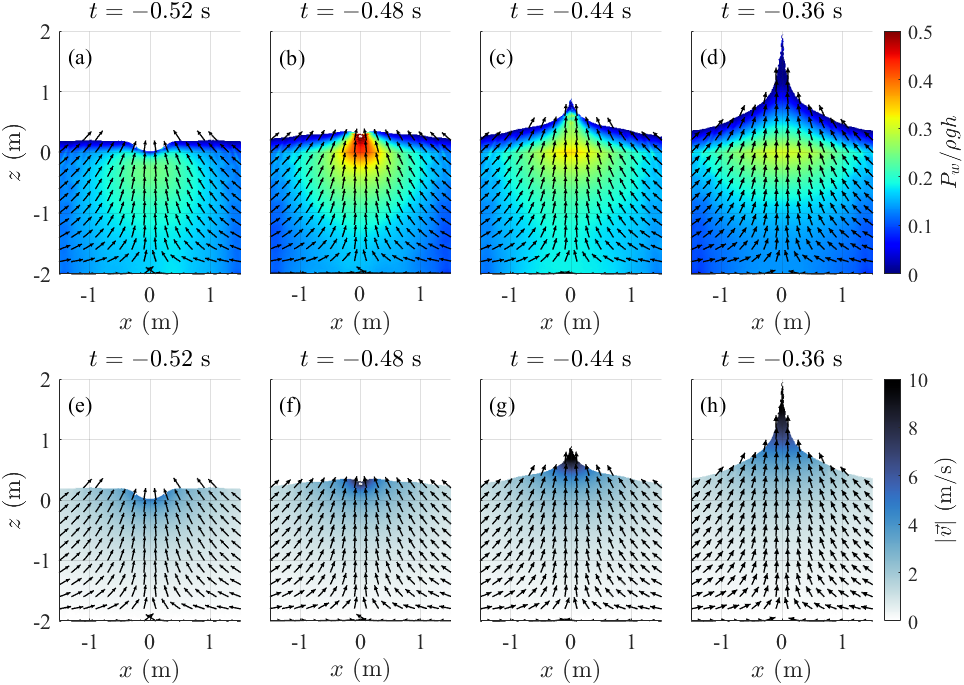}
\caption{Temporal evolution of the normalised wave-induced pressure (a–d) and the vertical velocity component (e–h) during the formation of the vertical jet in Exp. 75. The black arrows indicate instantaneous velocity vectors.}\label{Fig:Press and Vel}
\end{figure}

As reported in MC22, a highly localised jet is formed as the linear superposition of incident wave components increases in amplitude. Our simulations presented in figures \ref{Fig:SufaceSnapshots_Exp50} and \ref{Fig:SufaceSnapshots_Exp75} clearly capture the development of a vertical jet at the wave crest. Comparable jets have also been observed in standing waves and in the so-called ``flip trough'' wave breaking in front of a vertical wall \citep{Lugni06}. 

In each of these cases, rapid collapse is the key factor in jet formation, and the pressure impulse theory \citep{Cooker_Peregrine_1995} can be applied to explain the resulting localised jet. However, this theory is valid only during an exceedingly brief instant, for example, when a steep wave front impacts a vertical wall. In contrast, the present study considers a focused wave produced by the axisymmetric convergence of multiple wave frequency components. Under such conditions, the free surface evolves continuously in both time and space up to the point of jet formation; therefore, the impulsive approximation may no longer be applicable. 

Instead, we investigate jet formation by tracing how wave-induced pressure gradients drive the flow and transfer energy into fluid motion. 
Following the analysis of \citet{Wilkinson_2025} and consistent with the physical transitions described by \citet{Eggers_2008}, the evolution of a free surface jet can be conceptually divided into three successive phases: 
(I) a \textit{pressure-dominated phase}, in which strong pressure gradients accelerate the surface upward and dominate the dynamics; 
(II) an \textit{inertia-dominated phase}, where the pressure gradients rapidly vanish and the jet evolves ballistically under its own momentum; and 
(III) a \textit{gravity–capillary phase}, in which the vertically stretched jet thins under gravity and the internal velocity approaches zero, allowing surface-tension effects to become significant (if included).

Figure \ref{Fig:Press and Vel}(a--d) illustrates snapshots of the wave-induced pressure, defined as $P_w = P - \rho g (z_0 - z)$, where $z_0$ is the still-water level ($z_0 = 0$), and (e--h) show the corresponding vertical velocity magnitude. These panels capture the transition from the pressure-dominated phase (I) to the inertia-dominated phase (II). At $t=-0.52$ s (a,e), the trough is still collapsing, driving fluid toward the symmetry axis. The stored pressure beneath the free surface increases as the cavity collapses and reaches its maximum at $t=-0.44$~s (b,f), generating strong upward pressure gradients (I). Around $t=-0.40$~s (c,g), these gradients drive the formation of the initial jet, accelerating the fluid vertically (I to II). By $t=-0.36$~s (d,h), the pressure gradients have largely vanished as the motion becomes momentum-dominated, producing a jet with a velocity of about 10~m/s (II).
From this figure, the upward–acceleration phase of the emerging jet is seen to persist over $0.04$ to $0.08\,\mathrm{s}$, depending on whether the onset is taken at the pressure maximum (b,f) or at the trough–collapse stage (a,e). Thus, the effective forcing time is consistently of order $O(10^{-2})$.

Regarding Phase (I), the overall process of vertical jet formation through curvature collapse is consistent with the mechanism described by \cite{Longuet-Higgins_2001}. As the converging flow approaches the axis, the radial velocity diminishes, requiring a strong radial pressure gradient that generates an overpressure near the axis. Since the free surface is constrained to atmospheric pressure, this overpressure relaxes over a short vertical distance, producing a large negative $\partial p/\partial z$ that drives the upward jet. This mechanism has also been confirmed in two-dimensional potential flow computations \citep{Scolan22,Ockendon24}.
To the best of our knowledge, this is the first time-resolved observation of the Longuet-Higgins mechanism in three-dimensional axisymmetric curvature collapse. For Phase~(II), the analysis of MC22 revealed that the jet tip undergoes inertial motion consistent with a free-fall trajectory (see figure 15 in MC22). 
In Phase~(III), as noted in MC22, once the jet rises and approaches its maximum height, surface-tension effects become important, leading to droplet formation and jet breakup. Although our model does not include surface tension, droplet-like fragmentation is still observed. This behaviour is presumed to arise from kernel-deficiency effects in the SPH method rather than from physical surface tension, and this point is further examined in figure~\ref{Fig:Gamma_Surf_Exp_75}.

\begin{figure}[htbp]
\centering
\includegraphics[width=0.95\textwidth] {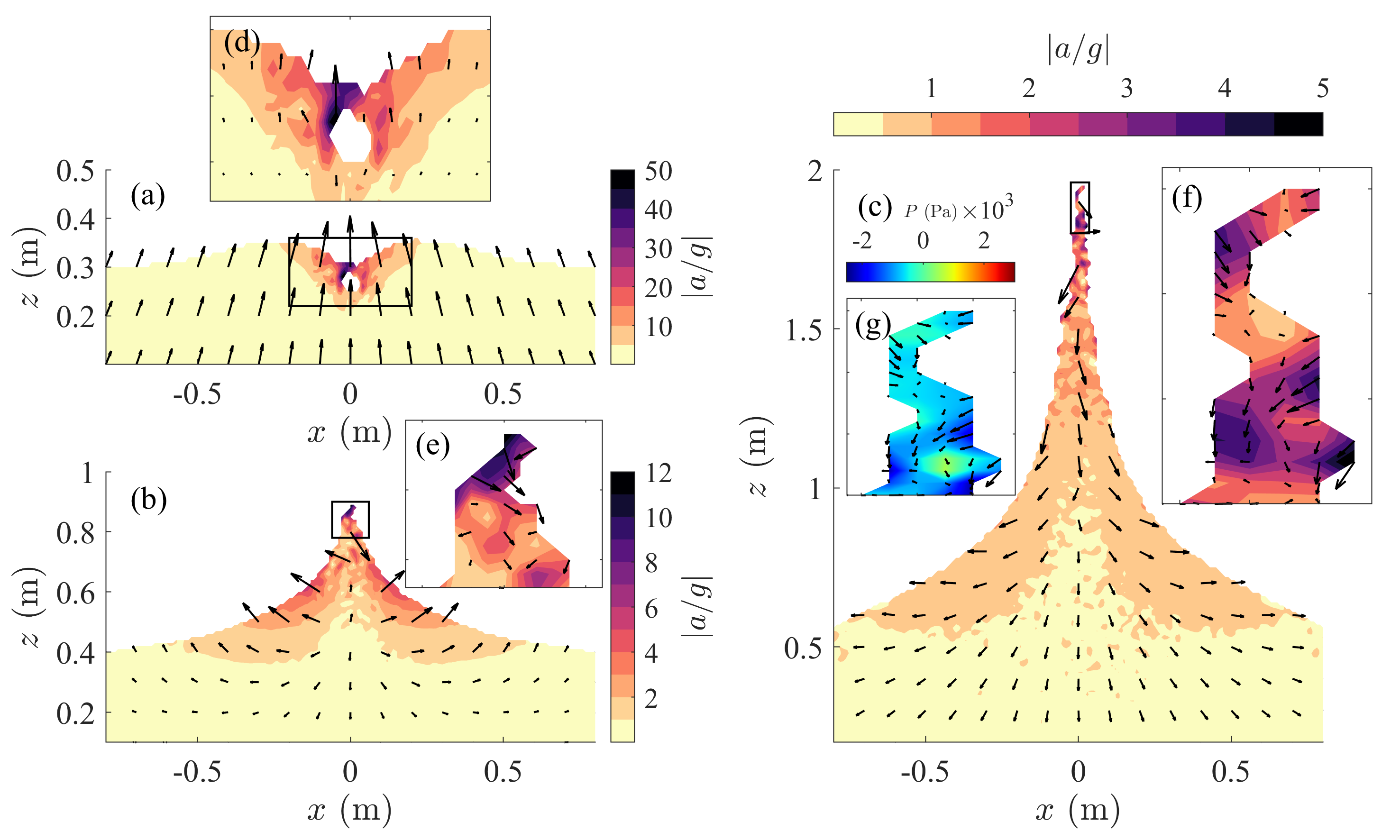}
\caption{Contours of the normalised acceleration $a/g$ in the vertical cross-section at $y=0$ during the formation of the vertical jet. (Exp. 75). Panels (a--c) show the acceleration field at $t=-0.48, -0.44, -0.36$ s. (d--f) are enlarged insets of the boxed regions in (a), (b) and (c), respectively. (g) shows the same region as (f) but with corresponding pressure field.}\label{Fig:Acc}
\end{figure}

The maximum acceleration during jet formation is of particular interest, as it characterizes the intensity of the focusing process and is relevant to impact and engineering applications  \citep[e.g.,][]{Lugni06,MARTINMEDINA18}. Figure \ref{Fig:Acc} displays contours and vectors of the normalised acceleration field, $a/g$, during jet formation, where $a$ represents acceleration calculated in numerical simulations. The largest accelerations occur at the instant of closure (a). The zoomed view in the inset panel (d) shows upward values approaching $50g$. Once the vertical jet has appeared (b,c), a contour with $a/g>5$ can be seen at the jet tip; the vectors are downward, indicating a free falling jet.
The peak acceleration of approximately $50g$ is substantially lower than the values reported for flip-through impacts, where $a>1000g$ \citep{Lugni06,MARTINMEDINA18}. 
This discrepancy is most likely due to the different time scales over which energy (or momentum flux) becomes focused in the two phenomena.
In flip-through events, the wave front impacts the wall within $\textit{O}(10^{-3})$ s, whereas in our case the time required for the trough to close and form the cavity was approximately 0.52 s ($t=-1.00$ to $t=-0.48$ s).

During the stage when the jet forms, the stability of the jet tip and the accuracy of its numerical representation become critical, as small-scale instabilities emerge. Panels (e,f) in figure \ref{Fig:Acc} show the acceleration field near the jet tip, where highly non-uniform contours appear with localised regions of $a/g>1$. The free surface profile exhibits undulations, indicating loss of axial symmetry. This arises from particle clustering and kernel deficiency: as the jet narrows, fewer particles span its diameter, and near the free surface kernel support is incomplete, reducing interpolation accuracy. Consequently, negative pressure promotes particle agglomeration, destabilising the jet tip. 
Unlike dam-break or plunging breakers, the axisymmetric case offers a simpler geometry for resolution studies. Here, destabilisation occurred when fewer than ten particles spanned the jet diameter (figure~\ref{Fig:required_resolution_jet} in Appendix \ref{App:Convergence}).

\begin{figure}[htbp]
\centering
\includegraphics[width=1.0\textwidth] {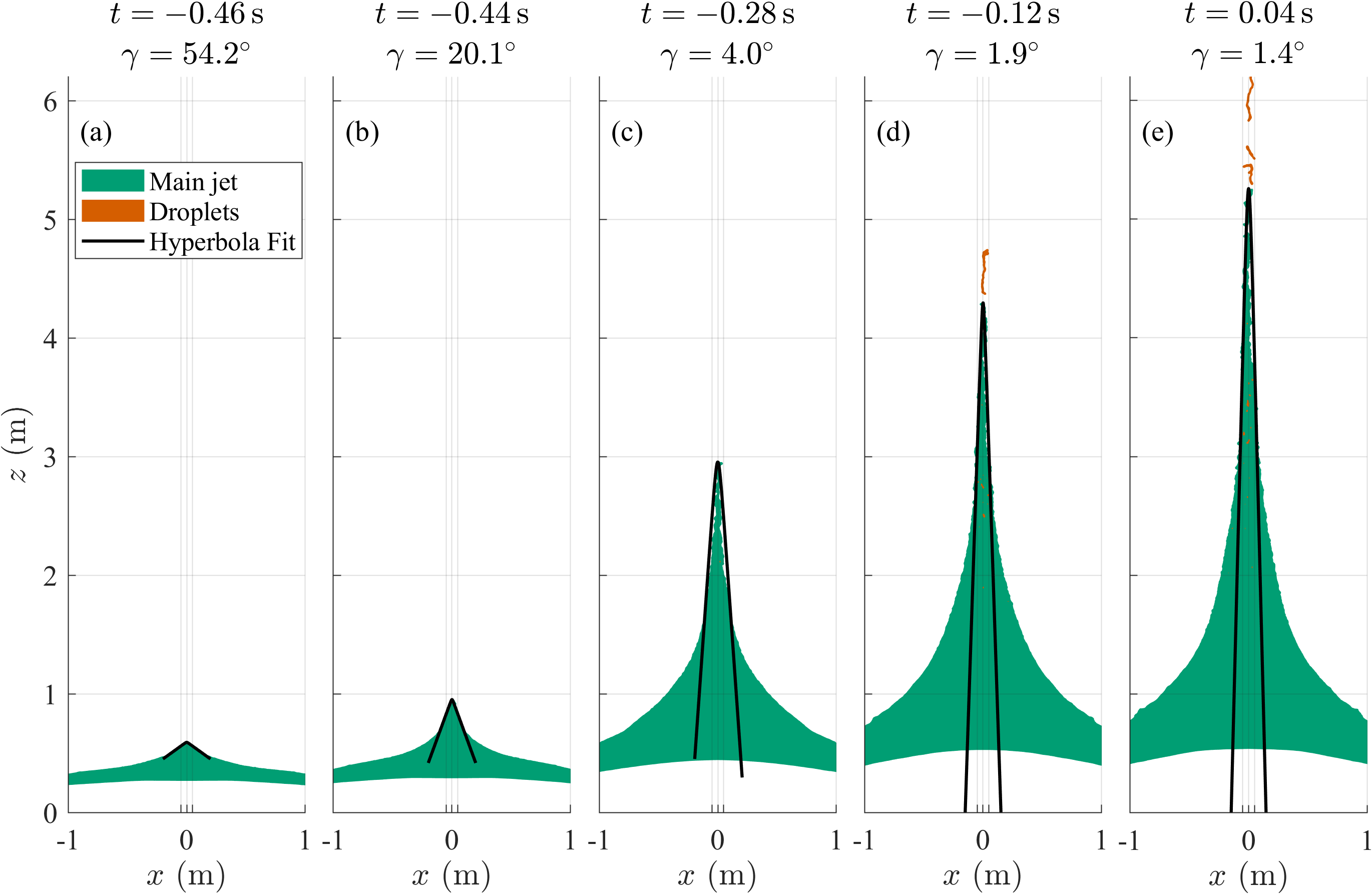}
\caption{Evolution of the SPH-computed free surface during the jet formation of an axisymmetric wave for Exp.~75. The fitted hyperbolic curves (black) capture the narrowing and vertical stretching of the jet profile. $\gamma$ denotes one-side tip angle of the jet.}\label{Fig:Gamma_Surf_Exp_75}
\end{figure}

\begin{figure}[htbp]
\centering
\begin{subfigure}[t]{0.48\textwidth}
    \centering
    \includegraphics[width=\textwidth]{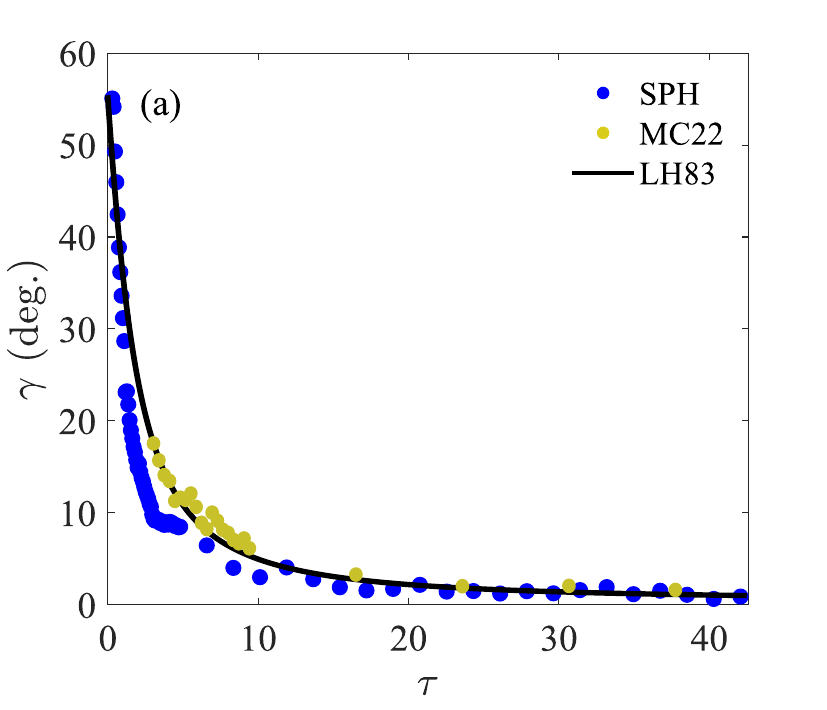}
    \caption*{}
    \label{Fig:Gamma_Exp75a}
\end{subfigure}
\hfill
\begin{subfigure}[t]{0.48\textwidth}
    \centering
    \includegraphics[width=\textwidth]{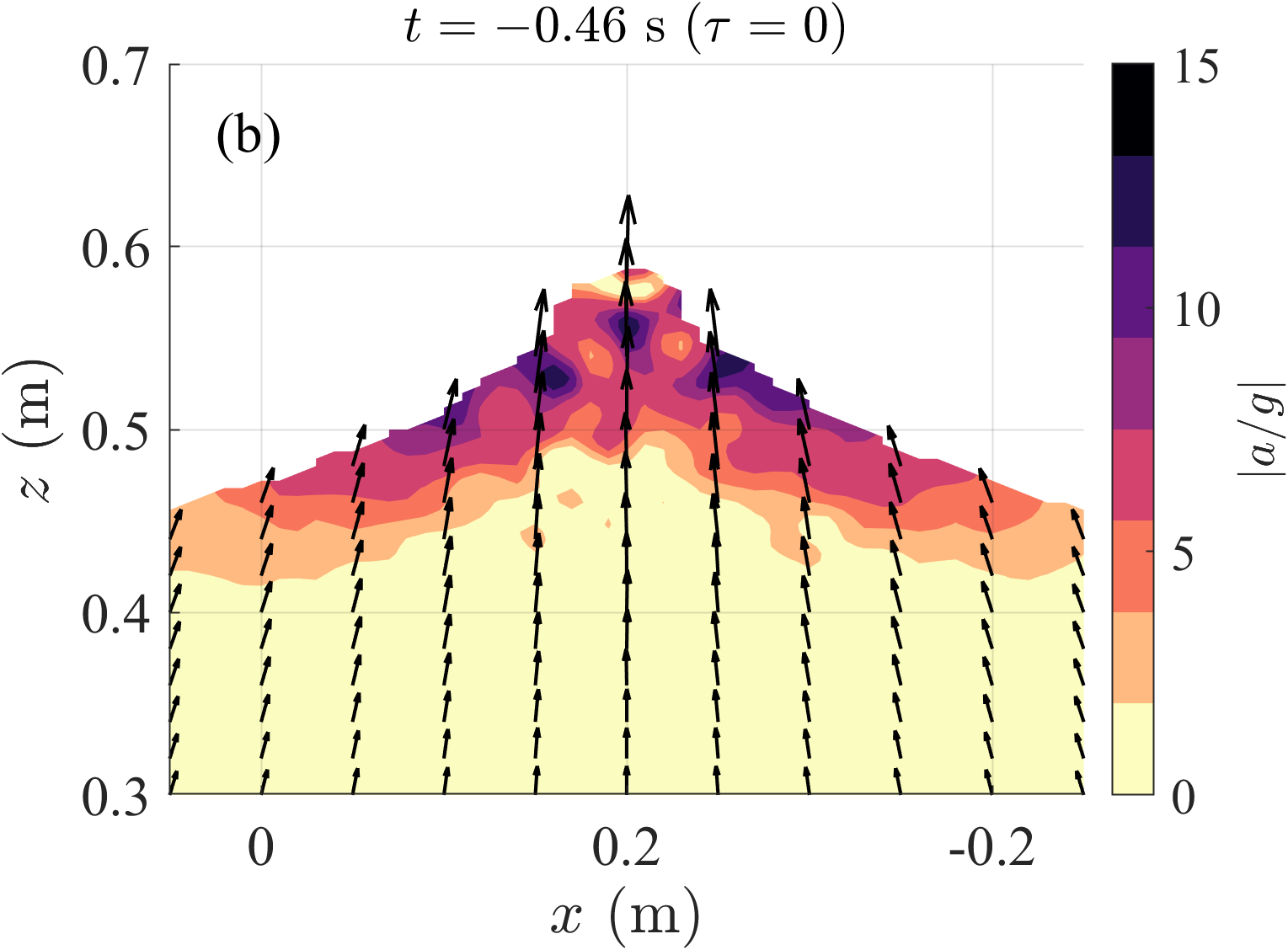}
    \caption*{}
    \label{Fig:Gamma_Exp75b}
\end{subfigure}
\caption{(a) Temporal variation of $\gamma$ with nondimensional time $\tau$, comparing SPH results (blue dots) with the Longuet–Higgins (1983) model (black line) and experimental data (MC22, yellow dots).
(b) Instantaneous acceleration magnitude $|a/g|$ and acceleration vectors from SPH at $t=-0.46$ s ($\tau=0$).}
\label{Fig:Gamma_Exp75}
\end{figure}

The growth of the tip instability observed in figure~\ref{Fig:Acc} is examined in figures. \ref{Fig:Gamma_Surf_Exp_75} and \ref{Fig:Gamma_Exp75} together with the jet tip angle.
Following the onset of vertical breaking, the jet develops upward, while its diameter decreases as it is stretched vertically (Phase III). As discussed previously, when the number of particles across the diameter falls below approximately ten, the free surface becomes unstable, promoting particle agglomeration and leading to the fragmentation of the jet (figure~\ref{Fig:Gamma_Surf_Exp_75} c, d and e). In figure~\ref{Fig:Gamma_Surf_Exp_75}, detached free surface elements separated from the main jet are shown in orange and identified as droplets.

In MC22, jet breakup into droplets occurs once the jet has sufficiently developed and the Weber number (\(\mathrm{We} = \rho U^2 D / \sigma\)) becomes small enough for surface tension to dominate, leading to Rayleigh–Plateau instability \citep{Rayleigh1878,Plateau1873}. Although our model does not include a surface tension term, the jet undergoes segmentation due to instability induced by insufficient kernel support. Nevertheless, the maximum height of the unsegmented main jet body at \(t = 0.28\ \mathrm{s}\) is \(6.08\ \mathrm{m}\), which agrees very well with the experimental value of \(6.0\ \mathrm{m}\), suggesting that the primary vertical momentum of the jet is accurately reproduced in this embedded SPH formulation, despite the absence of a surface tension model and the appearance of jet tip segmentation.


According to MC22, in axisymmetric standing waves, the formation of a free‐falling jet may be a more reliable indicator of breaking than the crest angle. This is because previous studies \citep[e.g.,][]{Longuet-Higgins01} have shown that 2D standing wave breaking may lack a universal limiting waveform, making geometric measures such as the crest angle less robust as breaking criteria. 
To assess the applicability of the crest-angle criterion in our simulation, we first estimate the jet angle from the simulated water free surface, applying the hyperbolic model of \cite{Longuet-Higgins_1983}. Subsequently, using the acceleration field inside the jet, we examine whether this angle provides a valid criterion for the onset of vertical jet breaking.

In the LH83 model, the jet is treated as an inviscid, inertia–dominated flow, neglecting viscous, capillary, and gravitational effects. Once the free‐surface crest encloses a critical angle of $2\gamma = 109.47^\circ$, the subsequent jet evolution is governed solely by inertial dynamics. The model predicts that the jet angle $\gamma$ follows:

\begin{equation}\label{Eq.gamma}
\tan\gamma \approx \frac{\sqrt{2}}{\left(1 + \tau/\sqrt{3}\right)^{3/2}}, 
\quad \tau = \frac{U(t-t_0)}{l},
\end{equation}
where $\tau$ is the non-dimensional time, $U$ is a characteristic velocity, $l$ is a characteristic length scale, and $t_0$ denotes the instant at which the critical angle is attained. 
In MC22, the jet angle $\gamma(t_i)$, obtained from a hyperbolic fit to the free surface profile at discrete times $t_i$, was fitted to the Longuet--Higgins model to determine the unknown scale ratio $U/l$. The estimated $U/l$ was then used to compute the non-dimensional time $\tau$, enabling direct comparison between the measured discrete jet angles $\gamma(t_i)$, converted to $\gamma(\tau_i)$, and those predicted by the model $\gamma(\tau)$.

To estimate the jet angle ($\gamma$), a hyperbolic curve was fitted to the free surface mesh coordinates at each time step as depicted in figure \ref{Fig:Gamma_Surf_Exp_75}. 
To accurately capture the critical angle, the output interval was reduced to 0.002~s between $t=-0.52$ and $t=-0.36$ s (corresponding approximately to $\tau = 0$–$5$).
The jet tip was defined as the crest region of the main jet surface (green), while detached free surface segments were classified as droplets (orange) and excluded from the fitting dataset. The corresponding fitted hyperbolae (black lines) are shown together with the free surface profiles in figure~\ref{Fig:Gamma_Surf_Exp_75}.
Panel (a) shows the time step at which the estimated jet angle is closest to the LH83 critical angle of 54.7$^{\circ}$.
Even after the onset of jet instability in the inset panel (c), the fitted curve provides a faithful overall description of the jet shape, as illustrated in figure~\ref{Fig:Gamma_Surf_Exp_75}(a–e).

Figure \ref{Fig:Gamma_Exp75}a compares the jet tip angle $\gamma$ as a function of the non-dimensional time $\tau$, obtained from SPH simulations (blue), MC22 experiments (yellow), and the LH83 model (black). The SPH results show an overall good agreement with both the experiments and the theoretical prediction. 
In particular, near the critical angle of $\gamma = 54.7^\circ$, reliable experimental measurements of the free surface geometry become difficult. At this stage, the jet velocity reaches approximately $10~\mathrm{m/s}$, and the free surface appears blurred in the images due to the rapid motion, making an accurate determination of the jet angle challenging.
This limitation does not arise in the SPH simulations, which allow the initial jet angle to be captured with high accuracy even under such extreme conditions.

For $\tau > 10$, as the jet narrows ($D/d_p<10$), the axial symmetry of the jet begins to break down. Nevertheless, the jet angle can still be estimated by fitting a hyperbola to the portions of the free surface where axial symmetry is preserved. The comparison between the free surface profile and the fitted curve can be seen in figure~\ref{Fig:Gamma_Surf_Exp_75}.

In figure~\ref{Fig:Gamma_Surf_Exp_75}a, the free surface profile at the instant when the numerically predicted jet angle $\gamma = 54.2^{\circ}$ is closest to the critical value of $\gamma = 54.7^{\circ}$ proposed by LH83 is shown.
Figure~\ref{Fig:Gamma_Exp75}b presents the corresponding vertical section of the acceleration field at $y=0$.
At this critical angle, the jet tip experiences a strong upward acceleration, reaching $a/g \approx 15$.
The corresponding time, $t=-0.46$~s, lies between $t=-0.48$ and $t=-0.44$~s in figure~\ref{Fig:Press and Vel}(b,c,f,g), representing the transition from Phase~(I) to Phase~(II).
The acceleration field shown in figure~\ref{Fig:Acc}(b,e) corresponds to figure~\ref{Fig:Press and Vel}(f,g), where downward-directed acceleration vectors start to appear.
Subsequently, at $t=-0.36$~s (figure~\ref{Fig:Acc}c,f), the acceleration becomes fully downward, marking Phase~(II).
These results suggest that the critical angle of LH83 represents the boundary between Phases~(I) and (II): beyond this point, the jet tip, having been accelerated upward, begins to fall freely and moves under its own inertia.
Hence, the critical angle proposed by LH83 can be interpreted as a physically consistent indicator of the onset of vertical wave breaking in an axisymmetric focused wave.

\begin{figure}[htb]
\centering
\includegraphics[width=0.9\textwidth]{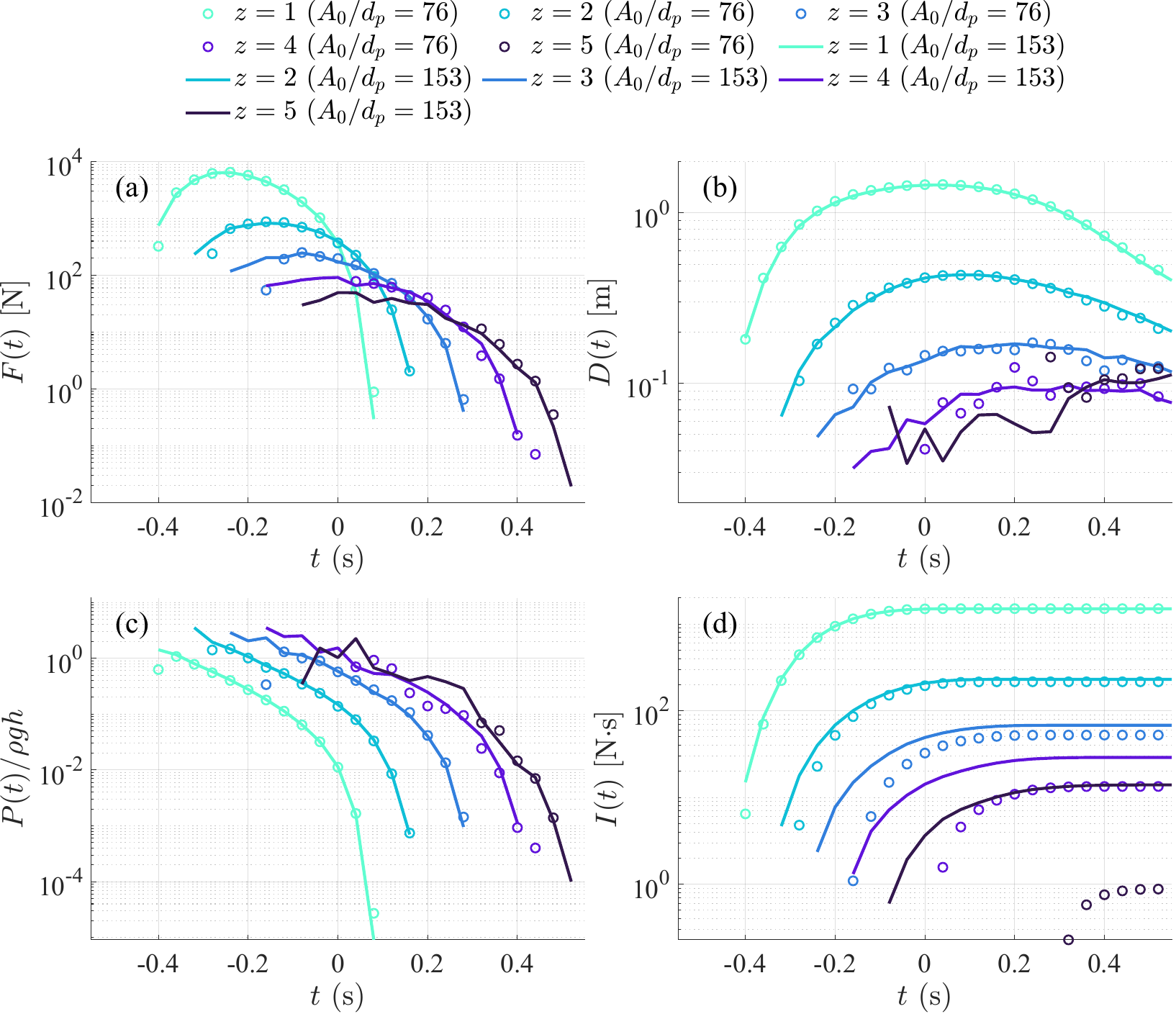}
\caption{Time histories of (a) jet force $F(t)$, (b) jet diameter $D(t)$, 
    (c) mean pressure $P(t)$, and (d) impulse $I(t)$ at various horizontal 
    $z$-sections for Exp.~75, obtained from SPH simulations with two particle 
    resolutions ($A_0/d_p = 76$ and $153$).} \label{Fig:JetConvergnece}
\end{figure}

As shown in Appendix \ref{App:Convergence} in figure~\ref{Fig:Comp_DifResol}, the particle resolution influences the maximum height reached by the jet. 
However, excluding both the maximum jet height and the detailed shape of the jet tip, the simulated jet free surface profile shows little variation between the higher resolutions and progressively approaches the experimental profile, indicating convergence of the jet-shape representation away from the tip region.
To further examine the convergence of the integrated jet-flow quantities, we evaluate the effect of particle resolution on relevant physical quantities across each horizontal z-section as depicted in figure~\ref{Fig:JetConvergnece}.
The jet force is computed as $F = \rho \, \dot{m} \, w $ [N], where $\rho$ is the fluid density [$\rm{kg/m^3}$], \(\dot{m}\) is the volumetric flux through the cross‐section, and \(w\) is the vertical velocity. 
The jet diameter \(D\) [$\rm{m}$] is obtained from the intersection of the free‐surface profile with the horizontal \(z\)‐section, and the cross‐sectional area is $ A = \upi D^2 / 4 $ [$\rm{m^2}$]. 
The mean pressure on the section is then $ P = F / A $ [$\rm{Pa}$], and the
 is defined as $ I(t) = \int_{t_1}^{t} F(t') \, \mathrm{d}t' $ [$\rm {N s}$].
Here, $t_1$ is defined as the time when the jet first passes through the section at $z = 1~\mathrm{m}$ (i.e., $t_1 = -0.4~\mathrm{s}$). The total impulse is given by $I(t_2)$, where $t_2$ corresponds to the time when the jet velocity at $z = 5~\mathrm{m}$ approaches zero (i.e., $t_2 = 0.5~\mathrm{s}$).

Firstly, the time series of $F$, $D$, and $P$ show well convergence up to $z = 3$ m as depicted in figure~\ref{Fig:JetConvergnece}(a,b,c).  Regarding the jet diameter  $D$, a clear decreasing trend is observed with increasing $z$, and for $z = 4$ and $5$ m the diameter falls below approximately $0.1\ \mathrm{m}$.  For jet shapes smaller than this, even the highest‐resolution ($A_0/d_p=153$) case does not satisfy the requirement $D/d_p > 10$, leading to free‐surface destabilisation and reduced accuracy in diameter estimation.  
Consequently, the estimated mean cross‐sectional pressure at $z = 5$ m exhibits large fluctuations.  
For the total impulse $I(t_2)$, no clear convergence is observed at higher $z$‐sections, although convergence trends are evident for $z = 1$ and $2$ m. The lack of impulse convergence at higher $z$ arises because $I(t)$ represents the time-integrated force. The instantaneous force $F(t)$ depends on the momentum flux (i.e. the volumetric flow rate) passing through the section at that moment. As shown in figure~\ref{Fig:JetConvergnece}a, the duration of the non-zero force signal at \(z=5\) m, corresponding to the time the jet passes through the section, varies substantially with resolution. This difference accumulates during the time integration and manifests as a pronounced discrepancy in $I(t_2)$ at higher $z$.

From figure~\ref{Fig:JetConvergnece}c, the maximum pressure reaches about $2\text{--}3\rho g h$ at the present $z$-section. This level is lower than flip-through impacts, which can reach a maximum pressure up to  $30\rho g h$ when the wave crest and trough collide directly against a rigid wall \citep{Lugni06}. 
In our case, the jet results from the gradual focusing of multiple frequency components rather than from the instantaneous focusing associated with a crest–trough collision against a rigid boundary. 
Consequently, the momentum builds over a longer time scale and is not concentrated instantaneously, leading to a reduced pressure peak. 
Nevertheless, the associated impulses remain on the order $10^{2}$--$10^{3}$ N\,s, and therefore still represent substantial momentum transfer.

\subsection{Post-jet cavity collapse and secondary jet formation}
\label{Sec:Cavity_Formation}
\begin{figure}[htbp]
\centering
\includegraphics[width=1.0\textwidth] {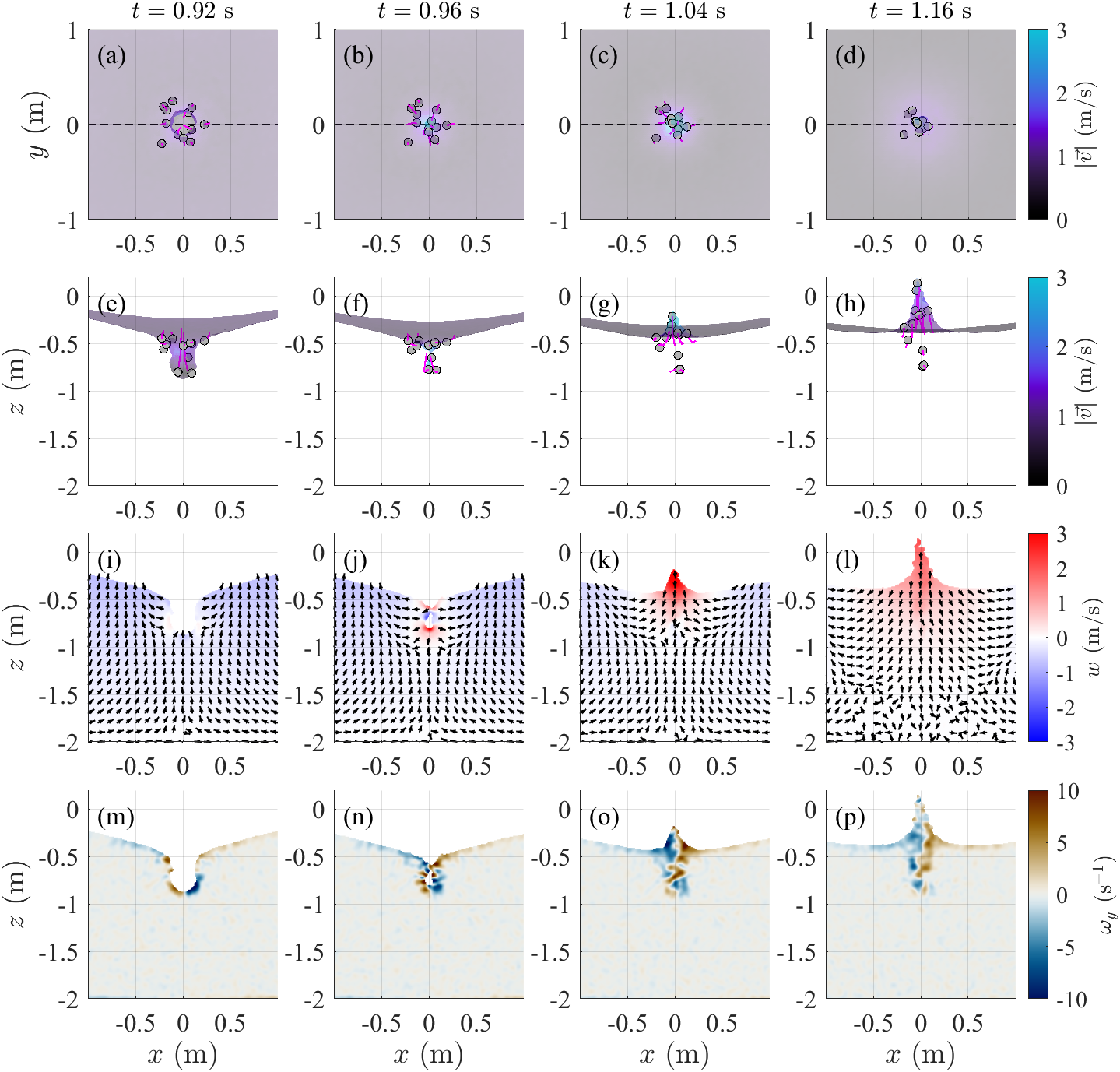}
\caption{Cavity collapse process after time of focusing and the associated vortex dynamics in the $xz$-plane at $y=0$ for Exp. 50. The upper row (a--d) shows side view of the free surface and particle trajectories during cavity collapse. Grey particles represent those entrained by vertically downward jet flow generated during the collapse. Magenta lines indicate the trajectories of these entrained particles. The lower row (e--h) presents the vorticity field $\omega_y$ and velocity vectors in the $xz$-plane at $y=0$, highlighting the development of vortical structures.}\label{Fig:Vorticy_Formation_Exp50}
\end{figure}

\begin{figure}[htbp]
\centering
\includegraphics[width=1.0\textwidth] {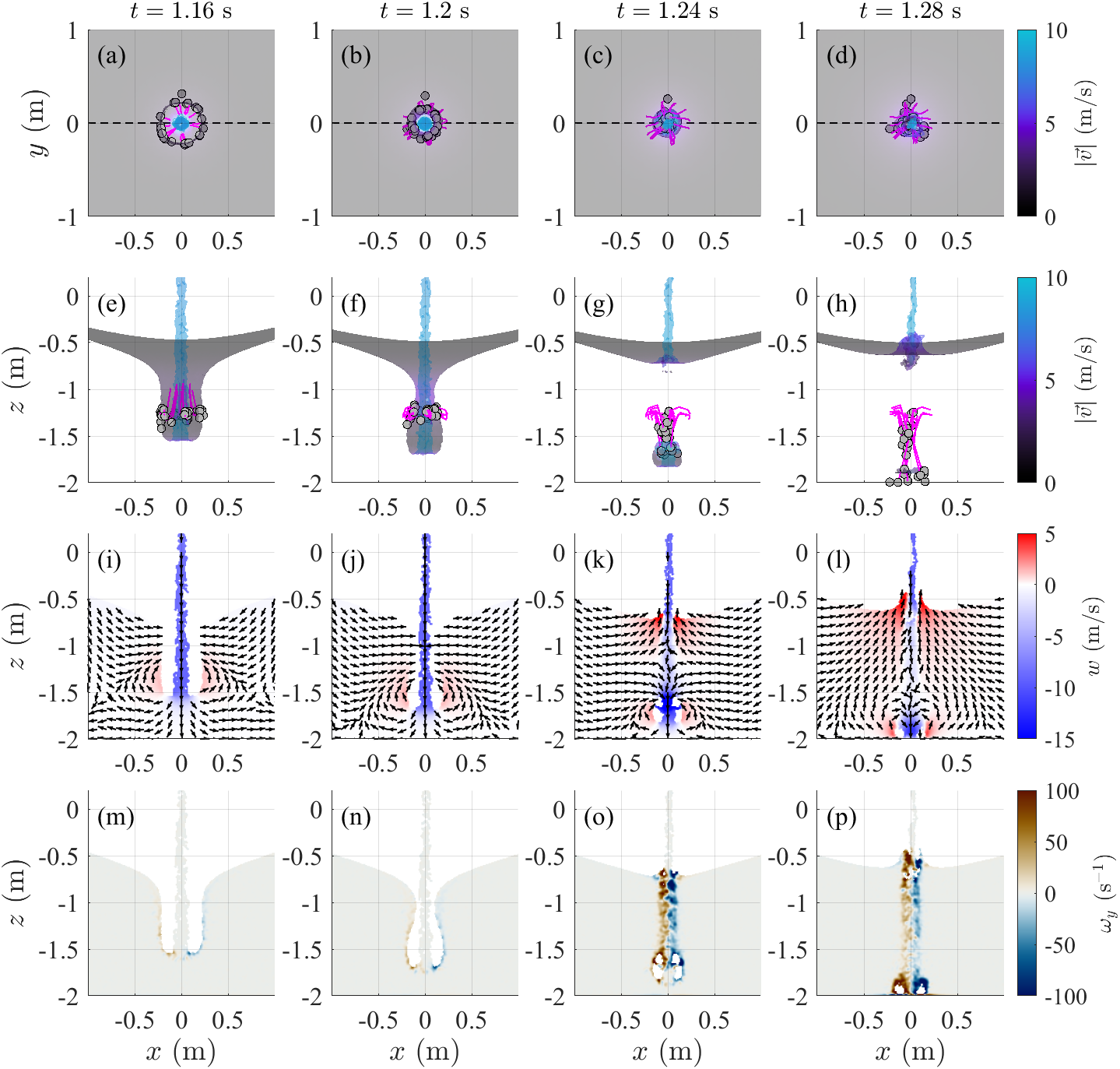}
\caption{Cavity collapse process after time of focusing and the associated vortex dynamics in the $xz$-plane at $y=0$ for Exp. 75. The upper row (a--d) shows side view of the free surface and particle trajectories during cavity collapse. Grey particles represent those entrained by vertically downward jet flow generated during the collapse. Magenta lines indicate the trajectories of these entrained particles. The lower row (e--h) presents the vorticity field $\omega_y$ and velocity vectors in the $xz$-plane at $y=0$, highlighting the development of vortical structures.}\label{Fig:Vorticy_Formation_Exp75}
\end{figure}

In the jet formation mechanism, the curvature collapse process is as shown in figure~\ref{Fig:Press and Vel}, but the present simulations also captured the formation of a cavity by a falling jet and its subsequent collapse, as observed in experiments (e.g., \href{https://www.cambridge.org/core/journals/journal-of-fluid-mechanics/article/wave-breaking-and-jet-formation-on-axisymmetric-surface-gravity-waves/487B14F30F0C90437038007C1D07FFBE#supplementary-materials}{supplementary movie 7 for Exp. 50} in MC22). This cavity, produced by the falling jet, is deeper in the vertical direction and narrower in width near the axis than the pre‐jet trough, with a stronger localisation along the central axis as shown in figures \ref{Fig:Vorticy_Formation_Exp50} and \ref{Fig:Vorticy_Formation_Exp75} for Exp. 50 and 75, respectively. Furthermore, its collapse pattern differs from that of the initial wave trough and closely resembles the collapse observed prior to the formation of Worthington jets in \citet{GEKLE10}. Here, we describe the cavity formed by the falling jet, its collapse, and the associated generation of secondary jets and vortex structures in axisymmetric focused gravity waves.

Figure~\ref{Fig:Vorticy_Formation_Exp50} and \ref{Fig:Vorticy_Formation_Exp75} illustrates the cavity collapse and associated vorticity field for Exp.50 and 75, respectively. Panels (a,e,i,m) in figures~\ref{Fig:Vorticy_Formation_Exp50} and \ref{Fig:Vorticy_Formation_Exp75} show the formation of a cavity, which is attributed to the inertia of the freely falling jet. Such inertia-driven cavity formation has also been reported in numerical simulations of disc impacts \citep{GEKLE10}, where surface tension plays a negligible role \citep{GEKLE09_PRL}. The influence of air, however, becomes significant once the cavity entrains air and approaches collapse \citep{GEKLE09_PRL}.
The cavity depth increases with the linear amplitude sum $A_0$.
In Exp.~75, the cavity depth driven by the falling jet reaches 1.8~m, close to the total water depth of 2~m as shown in figure~\ref{Fig:Vorticy_Formation_Exp75}f---about three times deeper than the 0.6~m trough preceding jet formation. 

\begin{figure}[htbp]
\centering
\includegraphics[width=0.9\textwidth] {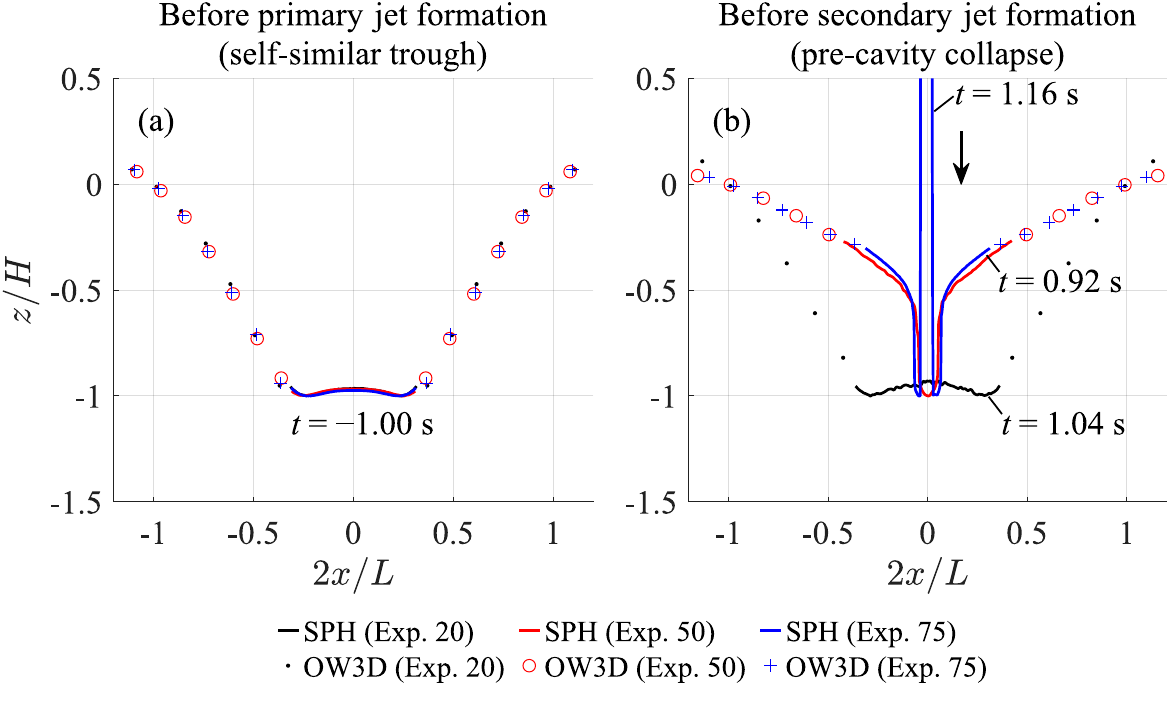}
\caption{Comparison of the trough or cavity profiles before and after jet formation. (a) The pre-jet troughs exhibit self-similarity across the three amplitudes. (b) After jet formation, this self-similarity is lost as the falling jet produces a strongly localised and deeper cavity. The gray dashed line shows the pre-jet through for Exp.~20. 
(c) Time evolution of the cavity profiles in Exp.~50. 
All profiles shown correspond to the $y=0$ cross-section. 
Profiles in (a) are taken at the time of maximum depth before the primary jet formation, and those in (b) at the time of maximum depth before pinch-off.}\label{Fig:Comp_cavity_shape}
\end{figure}

\begin{figure}[htbp]
\centering
\includegraphics[width=0.9\textwidth] {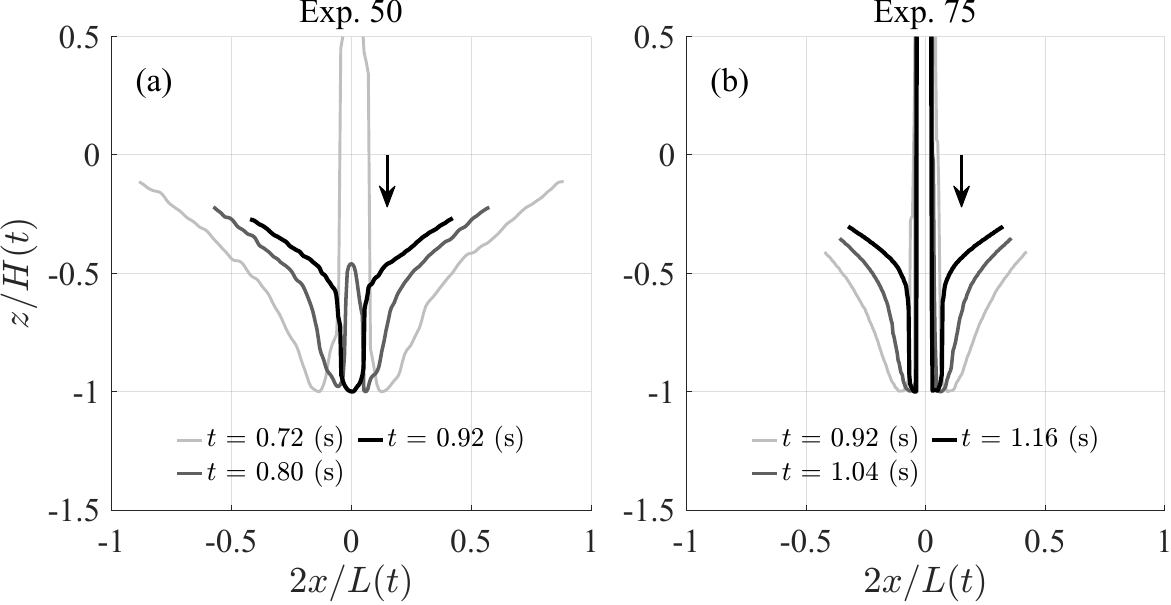}
\caption{Time evolution of the cavity profiles in Exp.~50 (a) and Exp.~75 (b). The cavity is generated by the falling primary jet produced after the axisymmetric wave focusing. All profiles correspond to the $y=0$ cross-section. Each curve is normalised by the instantaneous cavity depth $H(t)$ and lateral extent $L(t)$ for each case.}\label{Fig:Comp_cavity_evo}
\end{figure}

The cavity profiles generated by falling jets are compared with the wave troughs prior to jet formation in figure \ref{Fig:Comp_cavity_shape}. Panel (a) shows that the pre-jet troughs exhibit clear self-similarity: once scaled by their instantaneous depth $H$ and lateral extent $L$ of wave troughs, all cases collapse onto the same non-dimensional profile. Hence the pre-jet trough geometry is amplitude-independent, and the observed differences in jet height arise from the absolute scales, consistent with the $H^3$ scaling of \cite{Ghabache_2014} as reported by \cite{McAllister22}. 
In contrast, panel (b) shows that the self-similarity observed in the pre-jet trough breaks down after jet formation. For the Exp.~20 (non-breaking case), the maximum trough profiles before and after jet formation remain almost identical. However, as nonlinearity increases, the region beneath the falling jet becomes markedly deeper relative to the cavity lateral extent $L$, indicating a strongly localised cavity formation by falling jet. This marks a transition from the universal focusing dynamics to a non-universal deformation dominated by falling jet impact. Figure \ref{Fig:Comp_cavity_evo} shows the temporal evolution of the cavity generated by falling jets in Exp.~50 (a) and Exp.~75 (b). As the cavity deepens and narrows toward pinch-off, the scaled profiles at different times do not overlap, indicating that the cavity lacks temporal self-similarity and remains continuously influenced by the falling jet. This contrasts with the temporally self-similar collapse observed in an empty spherical bubble \citep[e.g.,][]{Obreschkow_2012}. 

\begin{figure}[htbp]
\centering
\includegraphics[width=0.9\textwidth] {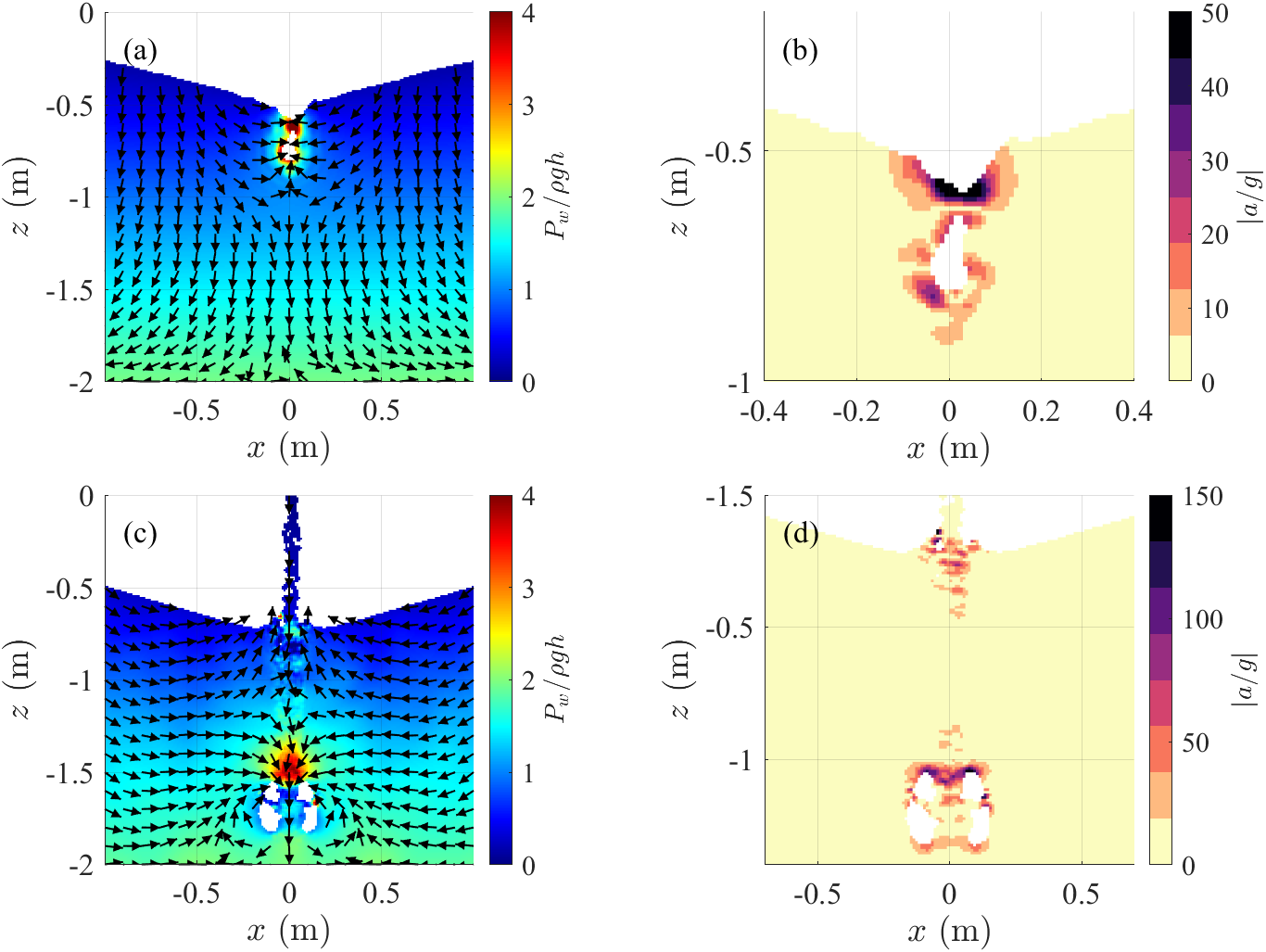}
\caption{Pressure and acceleration fields during cavity collapse.
(a,c) Normalised pressure with velocity vectors showing strong focusing toward the axis; (b,d) normalised accelerations highlighting intensified forcing near the pinch-off.
Panels (a,b)  correspond to Exp.~50 and (c,d) to Exp.~75.}\label{Fig:Cavity_Collapse_Press_Ace}
\end{figure}

During the subsequent collapse, a qualitatively different flow behaviour is observed compared with the curvature collapse at jet initiation, characterised by pinch-off at mid-depth followed by the generation of upward and downward jets as depicted in figure\ref{Fig:Vorticy_Formation_Exp75}j,k.
Similar to \citet{GEKLE10}, the cavity pinches off at mid-depth. This occurs because cavity sidewalls are forced inward by hydrostatic pressure ($\rho g z$), while vertical inertia dominates the lower part, giving weaker axial contraction. Pinch-off at mid-depth is thus a distinct feature of cavities induced by a falling continuous jet, not observed in curvature collapse. 
Figure \ref{Fig:Cavity_Collapse_Press_Ace} also shows that the collapse is far more violent than the earlier curvature collapse stage. Whereas curvature collapse in Exp. 75 reached only ~50 $g$ and pressures of $0.5\rho g h$ (see \ref{Fig:Acc}d and \ref{Fig:Press and Vel}b, respectively), the cavity collapse stage reaches accelerations up to $150g$ and pressures approaching $4\rho gh$ concentrated around the neck. 
Such strong convergence within a deep and narrow cavity is consistent with the findings of disc-impact analysis by \cite{GEKLE10}. They systematically characterised cavity collapse using a Froude number $Fr =V_D^2/(R_D g)$ based on the disc diameter $R_D$ and impact velocity $V_D$. However, in our case, the cavity grows within a continuously deepening trough, and its shape evolves smoothly up to the point where it becomes a well-defined cylindrical cavity as shown in \ref{Fig:Vorticy_Formation_Exp75}e. This continuous evolution makes it difficult to define a representative length scale, so we do not adopt such a parametrisation.

After pinch-off, upward and downward jets form as shown in panels (k,l) in figures~\ref{Fig:Vorticy_Formation_Exp50} and \ref{Fig:Vorticy_Formation_Exp75}, as also reported by \citet{GEKLE10}. 
Although Exp.~50 does not clearly capture the downward jet due to the limited temporal output interval, transient upward and downward jets are likely to occur immediately after the pinch-off (see figure~\ref{Fig:Vorticy_Formation_Exp50}j). The upward secondary jet is more pronounced in Exp.~50 (k,l), whereas in Exp.~75 it interacts strongly with the falling primary jet, preventing the formation of a distinct secondary jet.
Particle tracking shows that fluid entrained into the downward jet penetrates to the bottom (g,h), promoting vertical mixing.
The vorticity field (m–p) further demonstrates that collapse confines the downward jet, with velocity differences producing strong axial vorticity (y-direction). The downward jet also generates a vortex pair from the cavity bottom, yielding an extremely concentrated, standing-type vorticity structure distinct from conventional plunging or spilling breakers \citep[e.g.,][]{Peregrine83,WATANABE05,Giorgio22}.

\begin{figure}[htbp]
\centering
\includegraphics[width=1.0\textwidth] {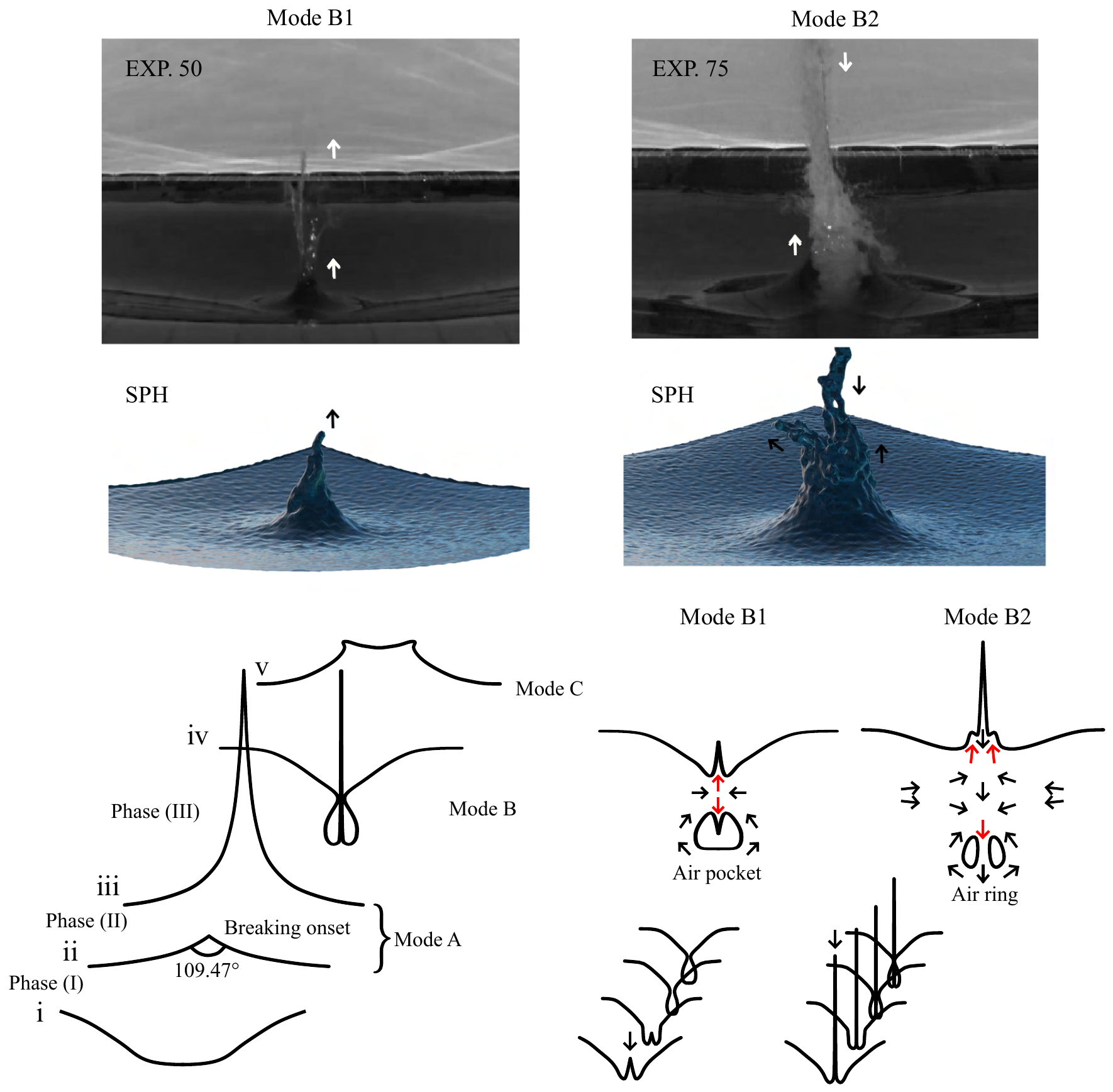}
\caption{Schematic representation of axisymmetric wave--jet evolution and the resulting breaking behaviour. Conceptual diagram of the axisymmetric breaking sequence in the lower panel, showing (i) trough formation, (ii) breaking onset with the critical angle in \cite{Longuet-Higgins_1983}, (iii) jet evolution, (iv) cavity formation and collapse and (v) radial breaking. This schematic extend the diagram of \cite{McAllister22} by incorporating the detailed secondary jet mechanism observed in stage (iv). 
The breaking is classified into Mode A (Sharp-Crested Breaking) and C (Radial Breaking) following \cite{JIANG_1998,McAllister22}, with an additional Mode B introduced here to describe the newly identified secondary jet formation. Mode~B is further subdivided into B1 and B2, depending on whether the secondary jet interacts with the falling jet (B2) or not (B1). 
The primary jet evolution is also classified into Phases (I)--(III), as defined in \S\ref{Sec:Jet_Formation}. A comparison between the experimental and SPH secondary jets is also shown in the upper panel. The experimental images are snapshots from the videos in \href{https://static.cambridge.org/content/id/urn:cambridge.org:id:article:S0022112021010235/resource/name/S0022112021010235sup004.avi}{Exp. 50} and \href{https://static.cambridge.org/content/id/urn:cambridge.org:id:article:S0022112021010235/resource/name/S0022112021010235sup001.avi}{Exp. 75}, respectively \citep[see supplemental material in][]{McAllister22}.}\label{Fig:Schematic_Jet_Breaking}
\end{figure}

Figure~\ref{Fig:Schematic_Jet_Breaking} summarises the full axisymmetric evolution by extending the schematic of \citet{McAllister22}. The figure highlights the newly identified mechanism for secondary-jet formation revealed in the present study. The lower-left panel outlines the sequential stages from trough formation to vertical jet development and finally radial breaking. The primary jet evolution is separated into Phases~(I)--(III), corresponding to the dominated mechanisms described in \S\ref{Sec:Jet_Formation}.
In terms of breaking classification, we introduce Mode~B---responsible for generating secondary jets---in addition to the previously established sharp-crested breaking (Mode~A) and radial breaking (Mode~C) in \cite{JIANG_1998,McAllister22}. The schematic shows that the deep and narrow cavity induced by the falling jet collapses inertially and can produce two distinct types of secondary jets: a non-interacting mode (B1), in which the secondary jet does not collide with the falling jet, and an interacting mode (B2), in which strong impact occurs. The directions of cavity contraction and the resulting upward and downward jets are also indicated by arrows. Our simulations capture the pocket that would entrap the air beneath the free surface, including the formation of an air pocket or ring after pinch-off, together with the associated vortex field.
The upper panel compares the newly identified secondary upward jets between the experiments and SPH simulations. Although qualitative agreement is observed, a quantitative comparison of droplet sizes produced by splash, as well as the height of the secondary jets, will require future two-phase simulations.

\begin{figure}[htbp]
\centering
\includegraphics[width=0.9\textwidth] {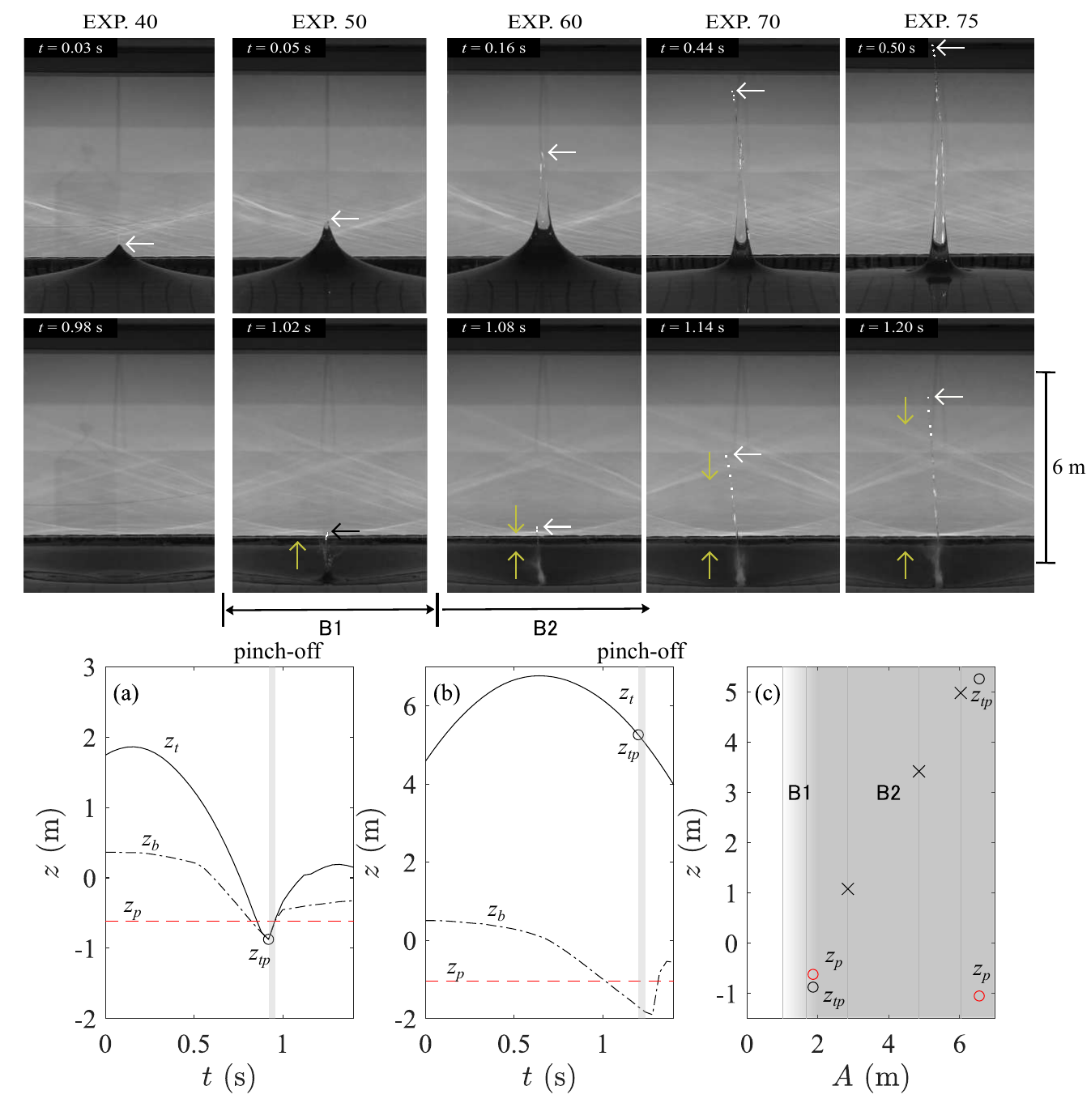}
\caption{
Upper panels: Experimental snapshots showing the maximum primary jets for Exp.~40, 50, 60, 70 and 75.
The white arrows indicate the jet-tip location.
Middle panels: Experimental snapshots around trough formation (Exp.~40) and immediately after pinch-off, when the secondary jet becomes visible (Exp.~50–75). The primary jet-tip location is indicated by the white arrows, whereas the secondary by the black arrow (Exp.~50). The yellow arrows indicate the upward secondary jet motion and the downward motion of the falling jet.
Lower panels: Time evolution of the jet-tip height $z_t$, 
cavity-bottom elevation $z_b$, and the jet-tip height at pinch-off $z_{tp}$, all extracted from the SPH subdomain for (a) Exp.~50 and (b) Exp.~75. 
The grey band marks the estimated pinch-off time.
(c) Trends of $z_p$ and $z_{tp}$ as functions of the maximum jet height $A$. 
Circles denote $z_p$ and $z_{tp}$ obtained in SPH, whereas crosses denote $z_{tp}$ estimated from the experimental snapshots in the middle panels.
A B1-type jet forms when $z_{tp} < z_p$, whereas a B2-type jet forms when $z_{tp} > z_p$.}
\label{Fig:B1B2}
\end{figure}

We further investigate how the secondary-jet morphology bifurcates into two regimes in figure~\ref{Fig:B1B2}. Jet morphology is often organised using time-scale ratios or classical non-dimensional parameters such as $We$ or $Fr$ \citep[e.g.,][]{GEKLE10,Cheng_25}, but in our continuous wave configuration, the representative cavity scales and the pinch-off time are not straightforward to determine from the initial conditions. A time-scale-based classification is therefore not applicable, and we distinguish the B1--B2 regimes solely through the geometric relation between the jet-tip position and the pinch-off depth.

We first examine the presence and morphology of the secondary jet using the experimental snapshots. The upper row of figure~\ref{Fig:B1B2} shows the free surface profiles at the time of maximum primary-jet height for Exp.~40, 50, 60, 70, and 75. The tip of the primary jet, indicated by the white arrow, subsequently undergoes free falling and, depending on the amplitude $A$, generates a secondary cavity that leads to the formation of a secondary jet. The middle row presents snapshots around the time of trough formation (Exp.~40) and immediately after pinch-off, when the secondary jet emerges (Exp. 50, 60, 70, and 75). In the weakly nonlinear breaking case Exp.~40, a trough forms but no secondary jet is observed. Secondary jet formation is first observed from Exp.~50 onward. A B1-type secondary jet (no interaction with the descending primary jet) is found only in Exp.~50. As the amplitude of the primary-jet $A$ increases, the moment at which the secondary-jet splash becomes visible is progressively delayed, suggesting that the time to start the pinch also becomes later with increasing $A$. From Exp.~60 onward, the larger amplitude causes the primary-jet tip to fall back more slowly, leading to complex B2-type breaking.

To further quantify these observations, we examine the corresponding numerical results obtained from the SPH simulations.
Panels (a) and (b) of figure~\ref{Fig:B1B2} show the time evolution of the jet-tip height $z_t$, the cavity-bottom elevation $z_b$, and the jet-tip height at pinch-off $z_{tp}$ with the pinch-off depth $z_p$ indicated as a reference level for Exp.~50 and 75 respectively. The grey band indicates the estimated timing of pinch-off, inferred from panels (f,g) of figures~\ref{Fig:Vorticy_Formation_Exp50} and \ref{Fig:Vorticy_Formation_Exp75}. In Exp.~50 (a), the jet tip descends fully before the pinch-off event and reaches the same level as the cavity bottom, so that $z_{tp} < z_p$. This configuration leads to the formation of a B1-type secondary jet. In contrast, Exp.~75 exhibits $z_{tp} > z_p$, indicating that the primary jet overshoots the cavity prior to pinch-off, thereby producing a B2-type secondary jet.

To relate this behaviour to the incident-wave amplitude, we examine how $z_p$ and $z_{tp}$ vary with the maximum primary-jet height $A$, which increases approximately with the cube of the initial trough depth $H$ \citep{Ghabache_2014,McAllister22}. Increasing nonlinearity yields a taller primary jet with stronger downward inertia, producing a deeper secondary cavity and a deeper, later pinch-off point, consistent with figures~\ref{Fig:Vorticy_Formation_Exp50} and \ref{Fig:Vorticy_Formation_Exp75}. Panel~(c) of figure~\ref{Fig:B1B2} shows the values of $z_p(A)$ and $z_{tp}(A)$ obtained from the numerical results for Exp.~50 and 75 (circles), together with the corresponding estimates of $z_{tp}(A)$ (crosses) extracted from the experimental snapshots (middle panels) for Exp.~60 70, and 75. 
It is evident that, as $A$ increases, $z_{tp}$ grows approximately linearly, while $z_p$ tends to shift to deeper values. This suggests that the condition required for a B1-type jet, $z_{tp} < z_p$, is satisfied only within a very narrow range of amplitudes (around $A \approx 2$~m).
As nonlinearity increases, the primary jet grows taller and its falling inertia intensifies, producing a deeper inertia-dominated type cavity and a deeper pinch-off point. As a result, the jet tip is more likely to lie above the pinch-off depth ($z_{tp} > z_p$), making the system increasingly prone to transition to the B2 mode.

\section{Conclusions}
\label{Sec:Conclusions}

This study aims to understand and model the dynamics of highly nonlinear axisymmetric jet formation, cavity collapse and the subsequent generation of secondary jets. This is achieved with a novel three-dimensional coupled framework, \textit{OceanSPHysics3D}, which combines the potential-flow solver OceanWave3D \citep{ENGSIGKARUP09} with DualSPHysics \citep{Dominguez22}. We have successfully reproduced the complete evolution of axisymmetric focused waves---from initial directional focusing through jet formation, and post-breaking collapse---under the experimental conditions of \citet{McAllister22}.

The simulations show that the primary jet forms through extreme directional focusing, where multiple directional components combine constructively to generate a steep crest and a strongly localised subsurface pressure maximum, consistent with the localised pressure maximum discussed by \citet{Longuet-Higgins_2001,McAllister22,Scolan22,Ockendon24}. Although the peak vertical acceleration in axisymmetric focusing is relatively modest ($\approx 50g$), compared with the extremely large values reported for two-dimensional flip-through impacts \citep[exceeding $1500g$;][]{Lugni06}, the resulting jet reaches a comparable or even greater height. This contrast indicates that the effective duration of the upward pressure forcing, set by the geometric convergence in the axisymmetric configuration and of order \textit{O}($10^{-2}$), is intrinsically longer than in flip-through and therefore imparts a larger net momentum to the jet despite the smaller peak pressure.
The geometry of the pre-jet trough exhibits clear self-similarity when scaled by its instantaneous depth $H$ and width $L$ of wave trough. Once the critical jet angle of $2\arctan\sqrt{2}$ \citep{Longuet-Higgins_1983} is reached, the leading-edge fluid transitions toward free-fall acceleration, allowing this angle to be interpreted as a physically meaningful onset point for breaking based on jet kinematics. 

A key new finding of this study is that the secondary jet forms for extreme waves form through an inertia-driven cavity collapse generated by the continuous falling jet. This process shares key features with rigid-disc–induced inertial cavity collapses that drive Worthington-type jetting \citep[e.g.][]{GEKLE10}.
The formation of this cavity is highly localised, as evidenced by the loss of self-similarity: when the cavity profiles are scaled by their depth and width, they do not collapse onto a universal shape, nor do profiles at different times align during the collapse. This inertially driven cavity collapse is accompanied by a pressure maximum almost eight times larger than that associated with the curvature-driven formation of the primary jet, together with fuid accelerations roughly three times stronger ($\approx 150g$). These values may increase further depending on the numerical configuration.
This localisation marks a departure from the universal focusing dynamics responsible for the primary jet, and highlights that a distinct physical mechanism operates in the post-jet stage. These results demonstrate, for the first time in three dimensions, that primary and secondary jets arise from two sequential but fundamentally different processes: curvature collapse for the primary jet, and an inertia-driven cavity collapse for the secondary jet. The initial maximum jet height $A$ determine whether the two jets remain separate or clash during collapse. Subsequently, the cavity dynamics generate a vortex ring and enhanced vertical mixing, distinct from the vorticity pathways known for conventional breaking waves \citep{Peregrine83,WATANABE05,Giorgio22}.

The coupled numerical framework is essential for resolving these nonlinear processes. Compared with the full-basin SPH simulations of \citet{Kanehira19}, the present method achieves an approximately 90\% reduction in computational cost while retaining the ability to capture steep free surface curvature, localised pressure fields and high-speed jet evolution in three dimensions. The high-resolution simulation of Exp.~75 ($A_0/d_p = 153$) further demonstrates the capability of \textit{OceanSPHysics3D} for modelling extreme wave events.

Overall, this study provides new physical insight into the nonlinear dynamics of axisymmetric focusing waves, showing that cavity interactions induced by the continuous falling jet give rise to secondary jets and enhanced vertical mixing. 
The ability to accurately reproduce the extreme wave in Exp.~75 further
demonstrates the reliability of the present approach, giving confidence for future simulations with more general and realistic directional spreading relevant for ocean science and engineering applications. Future work will incorporate two-phase (air--water) modelling, two-way coupling, and multi-resolution or multi-GPU approaches \citep[e.g.,][]{RICCI25}, enabling a deeper investigation of the air entrainment, cavity dynamics and jet--air interactions that underlie three-dimensional breaking.

\section*{Funding.}
This work was supported by the Japan Society for the Promotion of Science (JSPS) Overseas Research Fellowship.

\section*{Declaration of interests.}
The authors report no conflict of interest.

\renewcommand{\UrlFont}{\rmfamily}
\section*{Author ORCIDs.}
\orcidlink{0000-0002-1547-8344}T. Kanehira \url{https://orcid.org/0000-0002-1547-8344};\newline
\orcidlink{0000-0002-3552-0810}P.\,K. Stansby \url{https://orcid.org/0000-0002-3552-0810};\newline
\orcidlink{0000-0002-3269-7979}B.\,D. Rogers \url{https://orcid.org/0000-0002-3269-7979};\newline
\orcidlink{0000-0002-5142-3172}M.\,L. McAllister \url{https://orcid.org/0000-0002-5142-3172};\newline
\orcidlink{0000-0001-6154-3357}T.\,S. van den Bremer \url{https://orcid.org/0000-0001-6154-3357};\newline
\orcidlink{0000-0002-7372-980X}S. Draycott \url{https://orcid.org/0000-0002-7372-980X}.

\appendix
\renewcommand{\theHsection}{\Alph{section}}
\renewcommand{\theHsubsection}{\Alph{section}.\arabic{subsection}} 


\section{Numerical implementation in DualSPHysics and the coupling procedure}
\label{App:Implementaion}

\begin{figure}[htbp]
\centering
\includegraphics[width=0.6\textwidth] {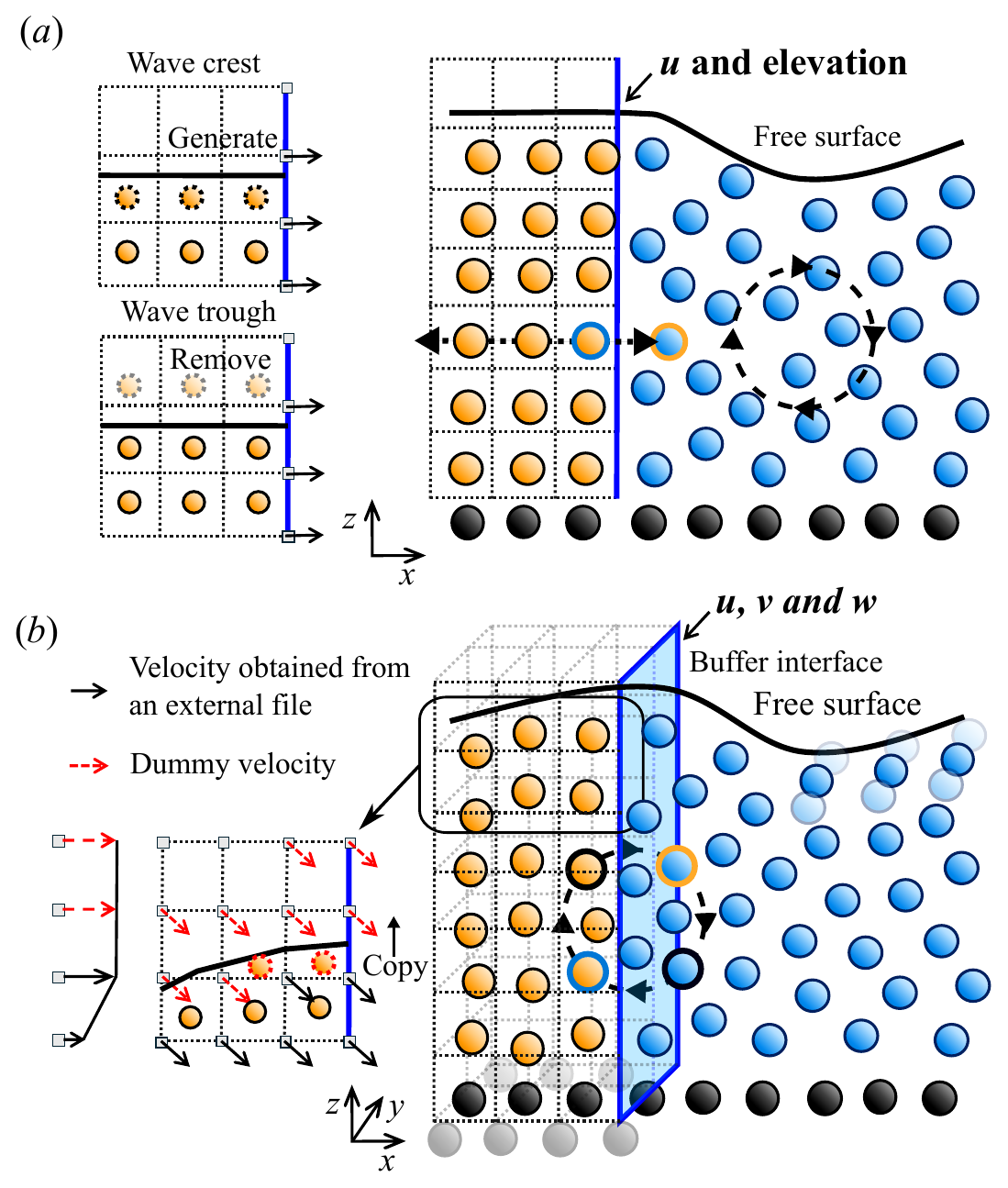}
\caption{Illustration of the improved open boundary for wave generation. (a) Conventional method in \cite{TAFUNI18}, where buffer particles move horizontally and particle creation/deletion is required for free surface tracking.
(b) Present method, which maintains orbital motion across the open boundary.
Blue, orange, and black particles denote fluid, buffer, and solid boundaries, respectively.}\label{Fig:NewOB}
\end{figure}

Following \citet{TAFUNI18}, the open boundary comprises a buffer region and the inner fluid domain (figure~\ref{Fig:NewOB}a). Buffer particles are advected with prescribed velocities obtained from analytical solutions, interpolated values, or external data. Conventional schemes \citep{NI18,TAFUNI18,VERBRUGGHE19,yang2023numerical} restrict the buffer particles to horizontal motion, as vertical displacement complicates the refilling process by making the insertion of new particles ambiguous. However, this constraint can generate spurious anticlockwise vortices near the generation boundary \citep[see figure~8 of][]{NI18}.

To extend the method to fully three-dimensional, multidirectional waves, we removed the refilling algorithm and adopted a direct-tracking approach (figure~\ref{Fig:NewOB}b). Buffer particles are explicitly updated using all three velocity components \citep{Sriram14,Liu23}, allowing accurate free surface tracking while retaining the Lagrangian consistency of SPH. The buffer zone is discretised into a 3D grid ($n_x,n_y,n_z>1$), where velocity components are defined at each grid point (the original GPU version in DualSPHysics used only $n_x=1$). This extension enables stable coupling with multidirectional wave fields and effectively suppresses unphysical circulation at the interface between OceanWave3D and SPH \citep{Kanehira25}.

A limitation is the gradual loss of boundary particles due to Stokes drift \citep{Stokes1847}. Although mass-flux replenishment \citep{LEROY16} can mitigate this, it often leads to irregular distributions that require corrective shifting \citep{Bouscasse13,Liu23}. Therefore, the replenishment is omitted here, which is acceptable since the buffer region used in this study is sufficiently larger than the expected Stokes drift over the simulation duration.

\section{Convergence Study}
\label{App:Convergence}

\begin{figure}[htbp]
\centering
\includegraphics[width=0.9\textwidth]{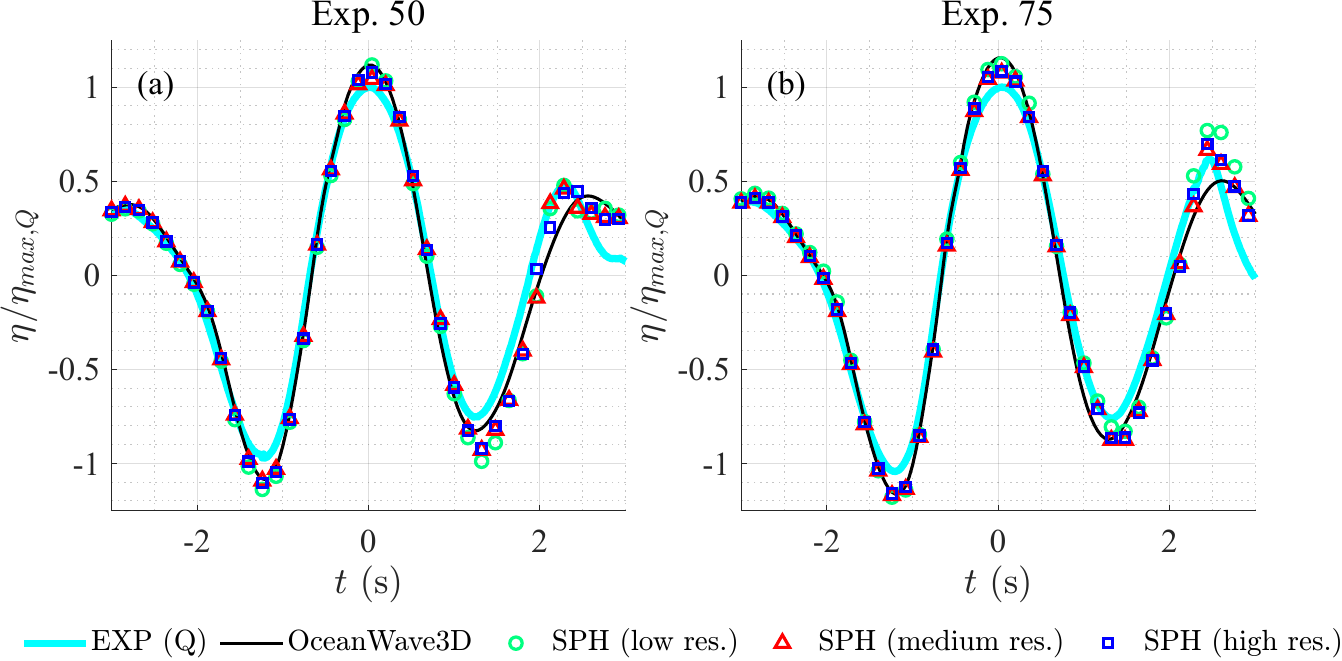}
\caption{Time series comparison of the water free surface elevation near the open boundary (WG1) between experiments (EXP Q: the Qualisys floating-marker 
measurements), OceanWave3D, and SPH with different particle resolutions ($A_0/d_p$).}
\label{fig:spike3d_convergence_wg1}
\end{figure}

Figure \ref{fig:spike3d_convergence_wg1} show a comparison of the free surface elevation at WG1 near the open boundary for Exp. 50 and Exp. 75 between laboratory measurements, OceanWave3D, and SPH at different particle resolutions. For Exp. 50, the particle resolutions used are $A_0/d_p = 26$, 34, and 51, which correspond to coarse, medium, and fine, respectively. For Exp. 75, the corresponding resolutions are $A_0/d_p = 38$, 78, and 153.
The vertical axis is normalised by the maximum experimental amplitude ($\eta_{max,Q}$). For both cases, the results up to the focusing time ($t=0$) exhibit very good agreement on particle resolution, showing numerical convergence in the vicinity of the open boundary.
The two numerical approaches are in close agreement with each other. Particularly good agreement is observed in the largest input amplitude case (Exp. 75), thus confirming the robustness and effectiveness of the present coupling methodology.

\begin{figure}[htbp]
\centering
\includegraphics[width=0.75\textwidth] {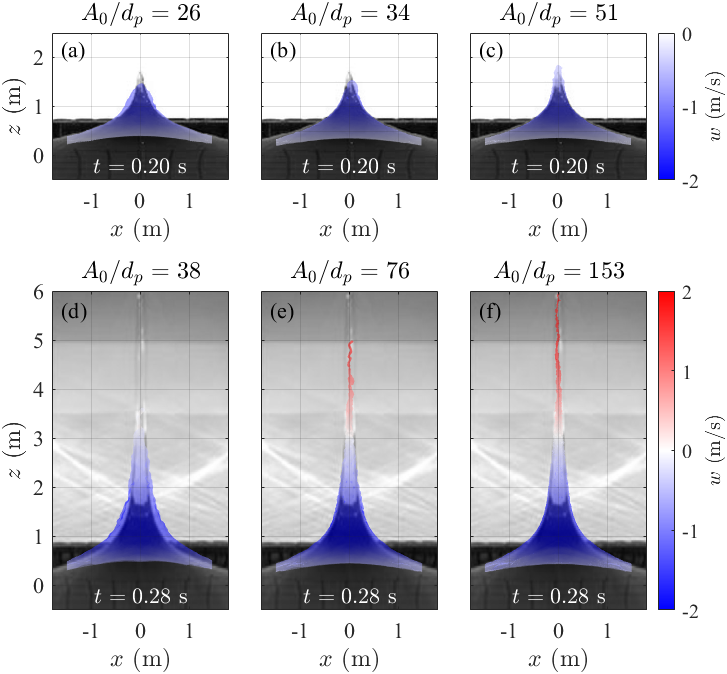}
\caption{Comparison between experimental images and SPH results of the free surface profile for different dimensional particle resolutions ($A_0/d_p$). The upper panels (a--c) corresponds to Exp.50, while the lower (d--f) shows Exp. 75.}\label{Fig:Comp_DifResol}
\end{figure}

\begin{figure}[htbp]
\centering
\includegraphics[width=0.8\textwidth] {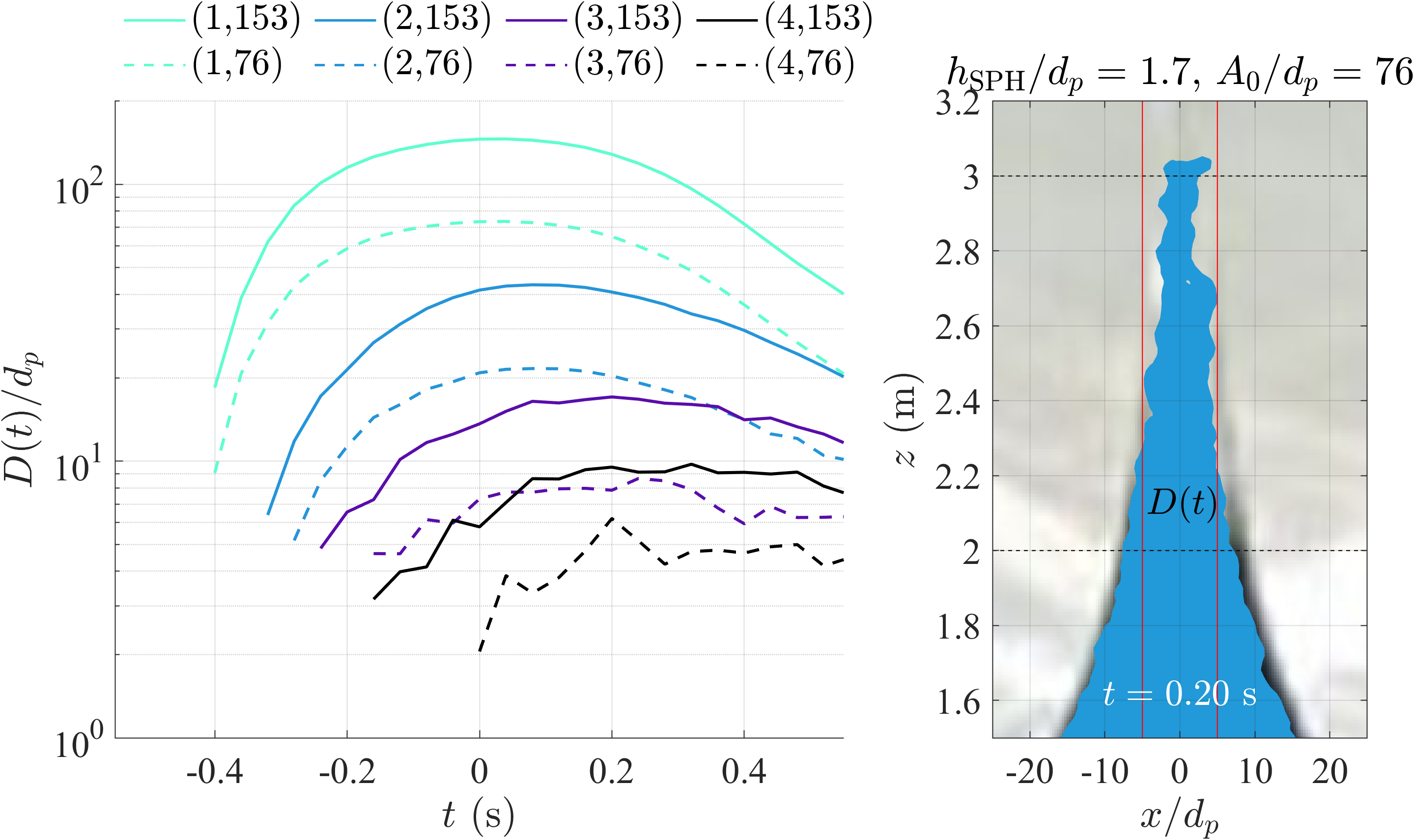}
\caption{Time evolution of the normalised jet diameter $D(t)/d_p$ at various vertical positions $z$, where $d_p$ is the initial particle distance. The legend entries ($z$, $A_0/d_p$) indicate the measurement height (m) and the non-dimensional particle resolution used in the SPH simulation, respectively.}\label{Fig:required_resolution_jet}
\end{figure}

Figure \ref{Fig:Comp_DifResol} illustrates the effect of particle resolution on the free surface profile for Exp. 50 and Exp. 75. As the resolution increases, steeper wave profiles are better captured, and in Exp. 75 the maximum jet height is also increased. As discussed in the main text, the vertical jet evolution is well reproduced, with the jet becoming progressively narrower; however, once the jet becomes too thin, surface undulations and fragmentation start to appear.

Figure \ref{Fig:required_resolution_jet} further examines the particle requirement within the jet. The left panel shows the time evolution of the jet diameter $D(t)/d_p$ at four vertical cross-sections ($z=1$, 2, 3, and 4 m). When the number of particles across the diameter falls below approximately ten, the time series of $D(t)$ becomes unstable and exhibits fluctuations. The right snapshot confirms this behaviour: the region inside the red vertical line corresponds to $D/d_p<10$, where the jet free surface displays pronounced waviness. This behaviour is consistent with the kernel-support deficiency mechanism described in the main text.

\section{Simulated velocity field}
\label{App:velocity_field}

\begin{figure}[htb]
\centering
\includegraphics[width=0.8\textwidth] {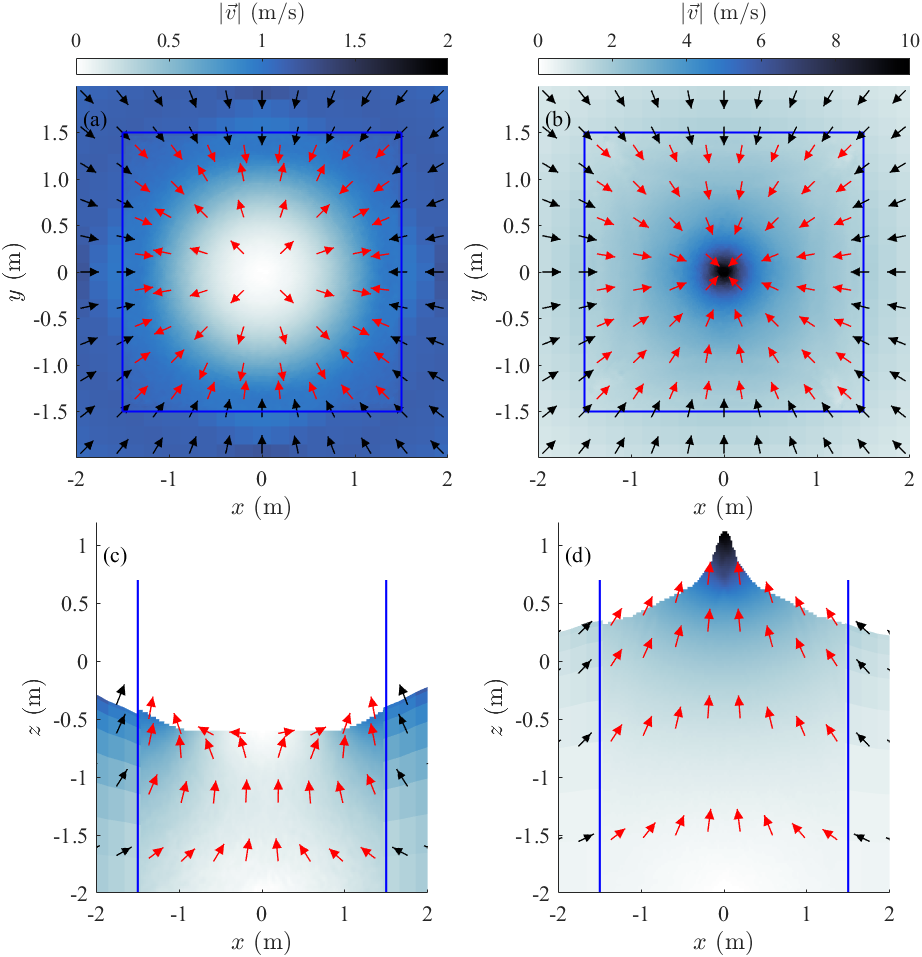}
\caption{Top views (a,b) and side views at $y=0$ (c,d) of the velocity fields during wave focusing in Exp. 75.
The outer region displays the velocity field obtained from OceanWave3D, while the inner region enclosed by the blue frame shows the embedded SPH results.
Black arrows represent velocity vectors from OceanWave3D, whereas red arrows indicate the SPH vectors. 
Panels (a) and (c) show the flow field at the instant when the water surface reaches its minimum level, while panels (b) and (d) correspond to the onset of vertical jet formation.}\label{Fig:Vel_Fields_Exp75}
\end{figure}

Figure~\ref{Fig:Vel_Fields_Exp75} shows velocity fields during focusing in Exp.~75. In the top views (a,b), velocity vectors indicate axisymmetric focusing and jet initiation, well reproduced within the SPH subdomain. Although the square-shaped coupling boundary might be expected to disturb the symmetry, no inconsistency is observed. The kernel-based interpolation of SPH preserves smooth velocity fields even at corners where vectors are not aligned with the surface normal. In the $x$–$z$ sections at $y=0$ (c,d), both the free surface and velocity field remain continuous across the coupling interface (blue lines), confirming the accurate coupling between SPH and OceanWave3D.

\section{Jet formation with different smoothing lengths}
\label{App:hdp}

\begin{figure}[htbp]
\centering
\includegraphics[width=0.8\textwidth] {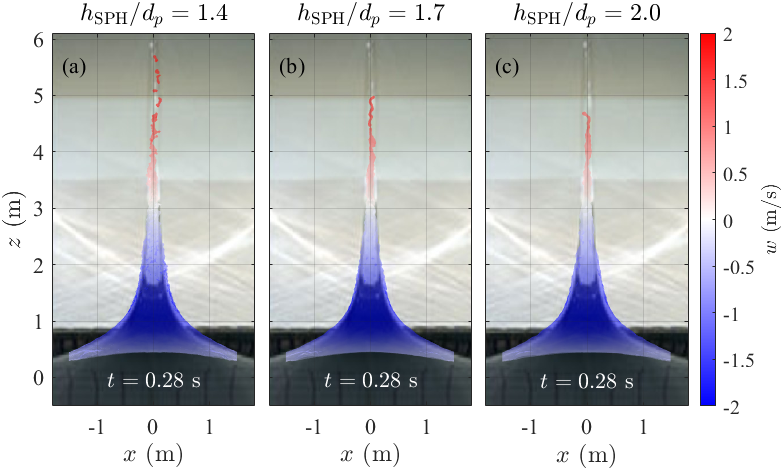}
\caption{Comparison between experimental images and SPH results of jet formation for different $h_{\rm{SPH}}/d_p$ values: (a) $h_{\rm{SPH}}/d_p=1.4$, (b) $h_{\rm{SPH}}/d_p=1.7$, and (c) $h_{\rm{SPH}}/d_p=2.0$. The non-dimensional particle resolution ($A_0/d_p$) used here is 76.}\label{Fig:Comp_Difhdp_Exp75}
\end{figure}

Figure \ref{Fig:Comp_Difhdp_Exp75} compares experimental images and SPH results of jet formation for different $h_{\mathrm{SPH}}/d_p$ values. A clear trend is observed whereby reducing $h_{\mathrm{SPH}}/d_p$ leads to an increase in the maximum jet height. This behaviour is attributed to the fact that a larger smoothing length tends to over-smooth the velocity and pressure fields during jet formation, thereby suppressing the jet development. It is also evident that the surface undulations and segmentation near the jet tip are smoothed out when a larger influence radius is used, resulting in a smoother jet tip appearance.


\bibliographystyle{jfm}
\bibliography{cas-refs}

@article{TAFUNI18,
title = {A versatile algorithm for the treatment of open boundary conditions in Smoothed particle hydrodynamics GPU models},
journal = {Computer Methods in Applied Mechanics and Engineering},
volume = {342},
pages = {604-624},
year = {2018},
issn = {0045-7825},
doi = {https://doi.org/10.1016/j.cma.2018.08.004},
url = {https://www.sciencedirect.com/science/article/pii/S0045782518303906},
author = {A. Tafuni and J.M. Domínguez and R. Vacondio and A.J.C. Crespo}
}

@article{TSURUTA21,
title = {Development of Wavy Interface model for wave generation by the projection-based particle methods},
journal = {Coastal Engineering},
volume = {165},
pages = {103861},
year = {2021},
issn = {0378-3839},
doi = {https://doi.org/10.1016/j.coastaleng.2021.103861},
url = {https://www.sciencedirect.com/science/article/pii/S0378383921000211},
author = {Naoki Tsuruta and Abbas Khayyer and Hitoshi Gotoh and Kojiro Suzuki},
}

@article{VERBRUGGHE18,
title = {Coupling methodology for smoothed particle hydrodynamics modelling of non-linear wave-structure interactions},
journal = {Coastal Engineering},
volume = {138},
pages = {184-198},
year = {2018},
issn = {0378-3839},
doi = {https://doi.org/10.1016/j.coastaleng.2018.04.021},
url = {https://www.sciencedirect.com/science/article/pii/S0378383917304349},
author = {Tim Verbrugghe and José Manuel Domínguez and Alejandro J.C. Crespo and Corrado Altomare and Vicky Stratigaki and Peter Troch and Andreas Kortenhaus},
}

@article{ALTOMARE18,
title = {Improved relaxation zone method in SPH-based model for coastal engineering applications},
journal = {Applied Ocean Research},
volume = {81},
pages = {15-33},
year = {2018},
issn = {0141-1187},
doi = {https://doi.org/10.1016/j.apor.2018.09.013},
url = {https://www.sciencedirect.com/science/article/pii/S0141118718303705},
author = {C. Altomare and B. Tagliafierro and J.M. Dominguez and T. Suzuki and G. Viccione},
}

@article{McAllister22, 
title={Wave breaking and jet formation on axisymmetric surface gravity waves}, 
volume={935}, 
DOI={10.1017/jfm.2021.1023}, 
journal={Journal of Fluid Mechanics}, 
author={McAllister, M.L. and Draycott, S. and Davey, T. and Yang, Y. and Adcock, T.A.A. and Liao, S. and van den Bremer, T.S.}, year={2022}, pages={A5}}

@article{Kanehira19,
author = {Kanehira, T. and Mutsuda, H. and Doi, Y. and Taniguchi, N. and Draycott, S. and Ingram, D.},
title = {Development and experimental validation of a multidirectional circular wave basin using smoothed particle hydrodynamics},
journal = {Coastal Engineering Journal},
volume = {61},
number = {1},
pages = {109--120},
year = {2019},
publisher = {Taylor \& Francis},
doi = {10.1080/21664250.2018.1560922},
URL = { https://doi.org/10.1080/21664250.2018.1560922},
eprint = {https://doi.org/10.1080/21664250.2018.1560922}
}

@article{RICCI25,
title = {Three-dimensional variable resolution for multi-scale modeling in Smoothed Particle Hydrodynamics},
journal = {Computer Physics Communications},
volume = {313},
pages = {109609},
year = {2025},
issn = {0010-4655},
doi = {https://doi.org/10.1016/j.cpc.2025.109609},
url = {https://www.sciencedirect.com/science/article/pii/S0010465525001067},
author = {Francesco Ricci and Renato Vacondio and José M. Domínguez and Angelantonio Tafuni},
}

@article{Longuet-Higgins_1983, title={Bubbles, breaking waves and hyperbolic jets at a free surface}, volume={127}, DOI={10.1017/S0022112083002645}, journal={Journal of Fluid Mechanics}, author={Longuet-Higgins, M. S.}, year={1983}, pages={103–121}}

@article{Longuet-Higgins_2001,
    author = {Longuet-Higgins, Michael S. and Dommermuth, Douglas G.},
    title = {On the breaking of standing waves by falling jets},
    journal = {Physics of Fluids},
    volume = {13},
    number = {6},
    pages = {1652-1659},
    year = {2001},
    month = {06},
    issn = {1070-6631},
    doi = {10.1063/1.1369141},
    url = {https://doi.org/10.1063/1.1369141},
}

@article{Ockendon24, title={Inviscid jets driven by pressure maxima}, volume={996}, DOI={10.1017/jfm.2024.625}, journal={Journal of Fluid Mechanics}, author={Ockendon, J.R. and Ockendon, H.}, year={2024}, pages={A17}}

@article{Scolan22,
  author  = {Scolan, Yves-Marie},
  title   = {Some aspects of the pressure field preceding the onset of critical jets in a breaking wave},
  journal = {Journal of Engineering Mathematics},
  year    = {2022},
  volume  = {138},
  number  = {1},
  pages   = {1},
  month   = nov,
  doi     = {10.1007/s10665-022-10246-3},
  url     = {https://doi.org/10.1007/s10665-022-10246-3},
  issn    = {1573-2703}
}

@inproceedings{Matthews72,
  author    = {Matthews, S. T.},
  title     = {A critical review of the 12th {ITTC} Wave Spectrum Recommendations},
  booktitle = {Proceedings of the 13th ITTC},
  year      = {1972},
  address   = {Berlin, Germany},
  pages     = {973--986},
  note      = {Report of the Seakeeping Committee, Appendix 9}
}

@article{GEKLE10, title={Generation and breakup of Worthington jets after cavity collapse. Part 1. Jet formation}, volume={663}, DOI={10.1017/S0022112010003526}, journal={Journal of Fluid Mechanics}, author={Gekle, Stephan and Gordillo, J. M.}, year={2010}, pages={293–330}}

@article{GEKLE09_PRL,
  title = {High-Speed Jet Formation after Solid Object Impact},
  author = {Gekle, Stephan and Gordillo, Jos\'e Manuel and van der Meer, Devaraj and Lohse, Detlef},
  journal = {Phys. Rev. Lett.},
  volume = {102},
  issue = {3},
  pages = {034502},
  numpages = {4},
  year = {2009},
  month = {Jan},
  publisher = {American Physical Society},
  doi = {10.1103/PhysRevLett.102.034502},
  url = {https://link.aps.org/doi/10.1103/PhysRevLett.102.034502}
}

@article{Rayleigh1878,
author = {Rayleigh, Lord},
title = {On The Instability Of Jets},
journal = {Proceedings of the London Mathematical Society},
volume = {s1-10},
number = {1},
pages = {4-13},
doi = {https://doi.org/10.1112/plms/s1-10.1.4},
year = {1878}
}

@book{Plateau1873,
  author    = {Plateau, J. A. F.},
  title     = {Statique exp{\'e}rimentale et th{\'e}orique des liquides soumis aux seules forces mol{\'e}culaires: Tome premier},
  volume    = {2},
  year      = {1873},
  publisher = {Gauthier-Villars},
  address   = {Paris}
}

@article{WATANABE05, title={Three-dimensional vortex structures under breaking waves}, volume={545}, DOI={10.1017/S0022112005006774}, journal={Journal of Fluid Mechanics}, author={Watanabe, YASUNORI and Saeki, HIROSHI and Hosking, ROGER J.}, year={2005}, pages={291–328}}

@article{Peregrine83,
   author = "Peregrine, D H",
   title = "Breaking Waves on Beaches", 
   journal= "Annual Review of Fluid Mechanics",
   year = "1983",
   volume = "15",
   number = "Volume 15, 1983",
   pages = "149-178",
   doi = "https://doi.org/10.1146/annurev.fl.15.010183.001053",
   url = "https://www.annualreviews.org/content/journals/10.1146/annurev.fl.15.010183.001053",
   publisher = "Annual Reviews",
   issn = "1545-4479",
   type = "Journal Article",
  }

@article{Giorgio22, title={On coherent vortical structures in wave breaking}, volume={947}, DOI={10.1017/jfm.2022.674}, journal={Journal of Fluid Mechanics}, author={Di Giorgio, Simone and Pirozzoli, Sergio and Iafrati, Alessandro}, year={2022}, pages={A44}}

@article{McAllister24,
  author    = {McAllister, M. L. and Draycott, S. and Calvert, R. and Davey, T. and Dias, F. and van den Bremer, T. S.},
  title     = {Three-dimensional wave breaking},
  journal   = {Nature},
  year      = {2024},
  volume    = {633},
  number    = {8030},
  pages     = {601--607},
  doi       = {10.1038/s41586-024-07886-z},
  url       = {https://doi.org/10.1038/s41586-024-07886-z}
}

@article{McAllister19, title={Laboratory recreation of the Draupner wave and the role of breaking in crossing seas}, volume={860}, DOI={10.1017/jfm.2018.886}, journal={Journal of Fluid Mechanics}, author={McAllister, M. L. and Draycott, S. and Adcock, T. A. A. and Taylor, P. H. and van den Bremer, T. S.}, year={2019}, pages={767–786}}

@article{Kanehira21,
title = {Highly directionally spread, overturning breaking waves modelled with Smoothed Particle Hydrodynamics: A case study involving the Draupner wave},
journal = {Ocean Modelling},
volume = {164},
pages = {101822},
year = {2021},
issn = {1463-5003},
doi = {https://doi.org/10.1016/j.ocemod.2021.101822},
url = {https://www.sciencedirect.com/science/article/pii/S1463500321000731},
author = {T. Kanehira and M.L. McAllister and S. Draycott and T. Nakashima and N. Taniguchi and D.M. Ingram and T.S. {van den Bremer} and H. Mutsuda},
}

@article{Narayanaswamy10,
author = {Muthukumar Narayanaswamy and Alejandro Jacobo Cabrera Crespo and Moncho Gómez-Gesteira and Robert Anthony Dalrymple},
title = {SPHysics-FUNWAVE hybrid model for coastal wave propagation},
journal = {Journal of Hydraulic Research},
volume = {48},
number = {sup1},
pages = {85--93},
year = {2010},
publisher = {IAHR Website},
doi = {10.1080/00221686.2010.9641249},
URL = { https://doi.org/10.1080/00221686.2010.9641249},
eprint = { https://doi.org/10.1080/00221686.2010.9641249}
}

@article{Liu23,
title = {Coupling SPH with a mesh-based Eulerian approach for simulation of incompressible free-surface flows},
journal = {Applied Ocean Research},
volume = {138},
pages = {103673},
year = {2023},
issn = {0141-1187},
doi = {https://doi.org/10.1016/j.apor.2023.103673},
url = {https://www.sciencedirect.com/science/article/pii/S0141118723002146},
author = {Kun Liu and Ye Liu and Shaowu Li and Hanbao Chen and Songgui Chen and Taro Arikawa and Yang Shi},
}

@article{Sriram14,
title = {A hybrid method for modelling two dimensional non-breaking and breaking waves},
journal = {Journal of Computational Physics},
volume = {272},
pages = {429-454},
year = {2014},
issn = {0021-9991},
doi = {https://doi.org/10.1016/j.jcp.2014.04.030},
url = {https://www.sciencedirect.com/science/article/pii/S0021999114002939},
author = {V. Sriram and Q.W. Ma and T. Schlurmann},
}

@article{NI18,
title = {A SPH numerical wave flume with non-reflective open boundary conditions},
journal = {Ocean Engineering},
volume = {163},
pages = {483-501},
year = {2018},
issn = {0029-8018},
doi = {https://doi.org/10.1016/j.oceaneng.2018.06.034},
url = {https://www.sciencedirect.com/science/article/pii/S0029801818310588},
author = {Xingye Ni and Weibing Feng and Shichang Huang and Yu Zhang and Xi Feng},
}

@article{ENGSIGKARUP09,
title = {An efficient flexible-order model for 3D nonlinear water waves},
journal = {Journal of Computational Physics},
volume = {228},
number = {6},
pages = {2100-2118},
year = {2009},
issn = {0021-9991},
doi = {https://doi.org/10.1016/j.jcp.2008.11.028},
url = {https://www.sciencedirect.com/science/article/pii/S0021999108006190},
author = {A.P. Engsig-Karup and H.B. Bingham and O. Lindberg},
}

@article{Gingold_and_Monaghan_77,
    author = {Gingold, R. A. and Monaghan, J. J.},
    title = {Smoothed particle hydrodynamics: theory and application to non-spherical stars},
    journal = {Monthly Notices of the Royal Astronomical Society},
    volume = {181},
    number = {3},
    pages = {375-389},
    year = {1977},
    month = {12},
    issn = {0035-8711},
    doi = {10.1093/mnras/181.3.375},
    url = {https://doi.org/10.1093/mnras/181.3.375},
    eprint = {https://academic.oup.com/mnras/article-pdf/181/3/375/3104055/mnras181-0375.pdf},
}

@article{Lucy77,
  author    = {Lucy, L. B.},
  title     = {A numerical approach to the testing of the fission hypothesis},
  journal   = {The Astronomical Journal},
  year      = {1977},
  volume    = {82},
  number    = {6},
  pages     = {1013--1024},
  doi       = {10.1086/112164},
  issn      = {0004-6256},
  url       = {https://ui.adsabs.harvard.edu/abs/1977AJ.....82.1013L},
}

@article{Koshizuka96,
author = {S. Koshizuka and Y. Oka},
title = {Moving-Particle Semi-Implicit Method for Fragmentation of Incompressible Fluid},
journal = {Nuclear Science and Engineering},
volume = {123},
number = {3},
pages = {421--434},
year = {1996},
publisher = {Taylor \& Francis},
doi = {10.13182/NSE96-A24205},
URL = {https://doi.org/10.13182/NSE96-A24205},
eprint = {https://doi.org/10.13182/NSE96-A24205}
}

@article{Dominguez22,
  author    = {Domínguez, J. M. and Fourtakas, G. and Altomare, C. and Canelas, R. B. and Tafuni, A. and García-Feal, O. and Martínez-Estévez, I. and Mokos, A. and Vacondio, R. and Crespo, A. J. C. and Rogers, B. D. and Stansby, P. K. and Gómez-Gesteira, M.},
  title     = {DualSPHysics: from fluid dynamics to multiphysics problems},
  journal   = {Computational Particle Mechanics},
  year      = {2022},
  volume    = {9},
  number    = {5},
  pages     = {867--895},
  doi       = {10.1007/s40571-021-00404-2},
  url       = {https://doi.org/10.1007/s40571-021-00404-2},
  issn      = {2196-4386},
}

@article{DALRYMPLE06, 
title = {Numerical modeling of water waves with the SPH method},
journal = {Coastal Engineering},
volume = {53},
number = {2},
pages = {141-147},
year = {2006},
note = {Coastal Hydrodynamics and Morphodynamics},
issn = {0378-3839},
doi = {https://doi.org/10.1016/j.coastaleng.2005.10.004},
url = {https://www.sciencedirect.com/science/article/pii/S0378383905001304},
author = {R.A. Dalrymple and B.D. Rogers},
}

@article{FOURTAKAS19,
title = {Local uniform stencil (LUST) boundary condition for arbitrary 3-D boundaries in parallel smoothed particle hydrodynamics (SPH) models},
journal = {Computers \& Fluids},
volume = {190},
pages = {346-361},
year = {2019},
issn = {0045-7930},
doi = {https://doi.org/10.1016/j.compfluid.2019.06.009},
url = {https://www.sciencedirect.com/science/article/pii/S0045793019301859},
author = {Georgios Fourtakas and Jose M. Dominguez and Renato Vacondio and Benedict D. Rogers},
}

@article{English22,
  author    = {English, A. and Domínguez, J. M. and Vacondio, R. and Crespo, A. J. C. and Stansby, P. K. and Lind, S. J. and Chiapponi, L. and Gómez-Gesteira, M.},
  title     = {Modified dynamic boundary conditions (mDBC) for general-purpose smoothed particle hydrodynamics (SPH): application to tank sloshing, dam break and fish pass problems},
  journal   = {Computational Particle Mechanics},
  year      = {2022},
  volume    = {9},
  number    = {5},
  pages     = {1--15},
  doi       = {10.1007/s40571-021-00403-3},
  url       = {https://doi.org/10.1007/s40571-021-00403-3},
  issn      = {2196-4386},
}

@article{Wendland95,
  author    = {Holger Wendland},
  title     = {Piecewise polynomial, positive definite and compactly supported radial functions of minimal degree},
  journal   = {Advances in Computational Mathematics},
  year      = {1995},
  volume    = {4},
  number    = {1},
  pages     = {389--396},
  doi       = {10.1007/BF02123482},
  url       = {https://doi.org/10.1007/BF02123482},
  issn      = {1572-9044},
}

@article{MOLTENI09,
title = {A simple procedure to improve the pressure evaluation in hydrodynamic context using the SPH},
journal = {Computer Physics Communications},
volume = {180},
number = {6},
pages = {861-872},
year = {2009},
issn = {0010-4655},
doi = {https://doi.org/10.1016/j.cpc.2008.12.004},
url = {https://www.sciencedirect.com/science/article/pii/S0010465508004219},
author = {Diego Molteni and Andrea Colagrossi},
}

@article{LO02,
title = {Simulation of near-shore solitary wave mechanics by an incompressible SPH method},
journal = {Applied Ocean Research},
volume = {24},
number = {5},
pages = {275-286},
year = {2002},
issn = {0141-1187},
doi = {https://doi.org/10.1016/S0141-1187(03)00002-6},
url = {https://www.sciencedirect.com/science/article/pii/S0141118703000026},
author = {Edmond Lo and Songdong Shao},
}

@article{Monaghan92,
   author = "Monaghan, J. J.",
   title = "Smoothed Particle Hydrodynamics", 
   journal= "Annual Review of Astronomy and Astrophysics",
   year = "1992",
   volume = "30",
   number = "Volume 30, 1992",
   pages = "543-574",
   doi = "https://doi.org/10.1146/annurev.aa.30.090192.002551",
   url = "https://www.annualreviews.org/content/journals/10.1146/annurev.aa.30.090192.002551",
   publisher = "Annual Reviews",
   issn = "1545-4282",
   type = "Journal Article",
  }

@article{VERBRUGGHE19,
title = {Non-linear wave generation and absorption using open boundaries within DualSPHysics},
journal = {Computer Physics Communications},
volume = {240},
pages = {46-59},
year = {2019},
issn = {0010-4655},
doi = {https://doi.org/10.1016/j.cpc.2019.02.003},
url = {https://www.sciencedirect.com/science/article/pii/S0010465519300463},
author = {Tim Verbrugghe and J.M. Domínguez and Corrado Altomare and Angelantonio Tafuni and Renato Vacondio and Peter Troch and Andreas Kortenhaus},
}

@inproceedings{Kanehira25,
  author    = {Kanehira, T. and Draycott, S. and Rogers, B.D. and Stansby, P.K.},
  title     = {An Enhanced Inlet Boundary Condition for Stable and Accurate Nonlinear Wave Simulations},
  booktitle = {Proceedings of the 19th SPHERIC International Workshop},
  year      = {2025},
  address   = {Barcelona, Spain},
  month     = {jun},
  note      = {Talk 11.3A, Session: SPH Application in Coastal Engineering (SS3-I)},
  url       = {https://spheric2025.upc.edu/}
}

@article{LEROY16,
title = {A new open boundary formulation for incompressible SPH},
journal = {Computers $\&$ Mathematics with Applications},
volume = {72},
number = {9},
pages = {2417-2432},
year = {2016},
issn = {0898-1221},
doi = {https://doi.org/10.1016/j.camwa.2016.09.008},
url = {https://www.sciencedirect.com/science/article/pii/S0898122116305107},
author = {A. Leroy and D. Violeau and M. Ferrand and L. Fratter and A. Joly},
}

@inproceedings{Bouscasse13,
  TITLE = {{Multi-purpose interfaces for coupling SPH with other solvers}},
  AUTHOR = {Bouscasse, B. and Marrone, S. and Colagrossi, A. and Di Mascio, A.},
  URL = {https://hal.science/hal-01158872},
  BOOKTITLE = {{SPHERIC 2013}},
  ADDRESS = {Trondheim, Norway},
  YEAR = {2013},
  HAL_ID = {hal-01158872},
  HAL_VERSION = {v1},
}

@article{Violeau16,
author = {Damien Violeau and Benedict D. Rogers},
title = {Smoothed particle hydrodynamics (SPH) for free-surface flows: past, present and future},
journal = {Journal of Hydraulic Research},
volume = {54},
number = {1},
pages = {1--26},
year = {2016},
publisher = {IAHR Website},
doi = {10.1080/00221686.2015.1119209},
URL = {https://doi.org/10.1080/00221686.2015.1119209},
eprint = {https://doi.org/10.1080/00221686.2015.1119209}
}

@article{Touzé25,
doi = {10.1088/1361-6633/ada80f},
url = {https://dx.doi.org/10.1088/1361-6633/ada80f},
year = {2025},
month = {feb},
publisher = {IOP Publishing},
volume = {88},
number = {3},
pages = {037001},
author = {Le Touzé, David and Colagrossi, Andrea},
title = {Smoothed particle hydrodynamics for free-surface and multiphase flows: a review},
journal = {Reports on Progress in Physics},
}

@book{GotohKhayyer25,
  author    = {Hitoshi Gotoh and Abbas Khayyer},
  title     = {Advanced Particle Methods},
  publisher = {Springer Singapore},
  year      = {2025},
  edition   = {1},
  isbn      = {978-981-97-7932-1},
  doi       = {10.1007/978-981-97-7933-8},
  url       = {https://doi.org/10.1007/978-981-97-7933-8},
}

@article{Lugni06,
    author = {Lugni, C. and Brocchini, M. and Faltinsen, O. M.},
    title = {Wave impact loads: The role of the flip-through},
    journal = {Physics of Fluids},
    volume = {18},
    number = {12},
    pages = {122101},
    year = {2006},
    month = {12},
    issn = {1070-6631},
    doi = {10.1063/1.2399077},
    url = {https://doi.org/10.1063/1.2399077},
}

@article{MARTINMEDINA18,
title = {Numerical simulation of flip-through impacts of variable steepness on a vertical breakwater},
journal = {Applied Ocean Research},
volume = {75},
pages = {117-131},
year = {2018},
issn = {0141-1187},
doi = {https://doi.org/10.1016/j.apor.2018.03.013},
url = {https://www.sciencedirect.com/science/article/pii/S0141118717304248},
author = {M. Martin-Medina and S. Abadie and C. Mokrani and D. Morichon},
}

@article{Cooker_Peregrine_1995, title={Pressure-impulse theory for liquid impact problems}, volume={297}, DOI={10.1017/S0022112095003053}, journal={Journal of Fluid Mechanics}, author={Cooker, Mark J. and Peregrine, D. H.}, year={1995}, pages={193–214}}

@article{Stokes1847,
  author  = {Stokes, G. G.},
  title   = {On the Theory of Oscillatory Waves},
  journal = {Transactions of the Cambridge Philosophical Society},
  volume  = {8},
  pages   = {441--455},
  year    = {1847}
}

@article{Longuet-Higgins01,
author = {Longuet-Higgins, Michael S },
title = {Vertical jets from standing waves},
journal = {Proceedings of the Royal Society of London. Series A: Mathematical, Physical and Engineering Sciences},
volume = {457},
number = {2006},
pages = {495-510},
year = {2001},
doi = {10.1098/rspa.2000.0678},
URL = {https://royalsocietypublishing.org/doi/abs/10.1098/rspa.2000.0678},
eprint = {https://royalsocietypublishing.org/doi/pdf/10.1098/rspa.2000.0678}
,
}

@article{yang2023numerical,
  title={A numerical flume for waves on variable sheared currents using smoothed particle hydrodynamics (SPH) with open boundaries},
  author={Yang, Yong and Draycott, Samuel and Stansby, Peter K and Rogers, Benedict D},
  journal={Applied Ocean Research},
  volume={135},
  pages={103527},
  year={2023},
  publisher={Elsevier}
}

@article{ANTUONO12,
title = {Numerical diffusive terms in weakly-compressible SPH schemes},
journal = {Computer Physics Communications},
volume = {183},
number = {12},
pages = {2570-2580},
year = {2012},
issn = {0010-4655},
doi = {https://doi.org/10.1016/j.cpc.2012.07.006},
url = {https://www.sciencedirect.com/science/article/pii/S0010465512002342},
author = {M. Antuono and A. Colagrossi and S. Marrone},
}

@article{Wilkinson_2025, title={Hyperbolic profile of free-surface jets}, volume={1020}, DOI={10.1017/jfm.2025.10677}, journal={Journal of Fluid Mechanics}, author={Wilkinson, Andrew and Morgan, Michael A. and Wilkinson, Michael}, year={2025}, pages={A41}}

@Article{Zeff2000,
author={Zeff, Benjamin W.
and Kleber, Benjamin
and Fineberg, Jay
and Lathrop, Daniel P.},
title={Singularity dynamics in curvature collapse and jet eruption on a fluid surface},
journal={Nature},
year={2000},
month={Jan},
day={01},
volume={403},
number={6768},
pages={401-404},
issn={1476-4687},
doi={10.1038/35000151},
url={https://doi.org/10.1038/35000151}
}

@article{Mack1962,
author = {Mack, Lawrence R.},
title = {Periodic, finite-amplitude, axisymmetric gravity waves},
journal = {Journal of Geophysical Research (1896-1977)},
volume = {67},
number = {2},
pages = {829-843},
doi = {https://doi.org/10.1029/JZ067i002p00829},
url = {https://agupubs.onlinelibrary.wiley.com/doi/abs/10.1029/JZ067i002p00829},
year = {1962}
}

@article{Duchemin2002,
    author = {Duchemin, Laurent and Popinet, Stéphane and Josserand, Christophe and Zaleski, Stéphane},
    title = {Jet formation in bubbles bursting at a free surface},
    journal = {Physics of Fluids},
    volume = {14},
    number = {9},
    pages = {3000-3008},
    year = {2002},
    month = {09},
    issn = {1070-6631},
    doi = {10.1063/1.1494072},
    url = {https://doi.org/10.1063/1.1494072},
}

@article{Farsoiya_2017, title={Axisymmetric viscous interfacial oscillations – theory and simulations}, volume={826}, DOI={10.1017/jfm.2017.443}, journal={Journal of Fluid Mechanics}, author={Farsoiya, Palas Kumar and Mayya, Y. S. and Dasgupta, Ratul}, year={2017}, pages={797–818}}

@Article{Ismail2018,
author ={Ismail, A. Said and Gañán-Calvo, Alfonso M. and Castrejón-Pita, J. Rafael and Herrada, Miguel A. and Castrejón-Pita, Alfonso A.},
title  ={Controlled cavity collapse: scaling laws of drop formation},
journal  ={Soft Matter},
year  ={2018},
volume  ={14},
issue  ={37},
pages  ={7671-7679},
publisher  ={The Royal Society of Chemistry},
doi  ={10.1039/C8SM00114F},
url  ={http://dx.doi.org/10.1039/C8SM00114F},
}

@article{Rodríguez_2020, title={On the sea spray aerosol originated from bubble bursting jets}, volume={886}, DOI={10.1017/jfm.2019.1061}, journal={Journal of Fluid Mechanics}, author={Blanco–Rodríguez, Francisco J. and Gordillo, J. M.}, year={2020}, pages={R2}}

@article{Orozco2015,
  title = {Droplet impact on deep liquid pools: Rayleigh jet to formation of secondary droplets},
  author = {Castillo-Orozco, Eduardo and Davanlou, Ashkan and Choudhury, Pretam K. and Kumar, Ranganathan},
  journal = {Phys. Rev. E},
  volume = {92},
  issue = {5},
  pages = {053022},
  numpages = {12},
  year = {2015},
  month = {Nov},
  publisher = {American Physical Society},
  doi = {10.1103/PhysRevE.92.053022},
  url = {https://link.aps.org/doi/10.1103/PhysRevE.92.053022}
}

@article{Basak_2021, title={Jetting in finite-amplitude, free, capillary-gravity waves}, volume={909}, DOI={10.1017/jfm.2020.851}, journal={Journal of Fluid Mechanics}, author={Basak, Saswata and Farsoiya, Palas Kumar and Dasgupta, Ratul}, year={2021}, pages={A3}}

@article{Worthington1897,
author = {Worthington, Arthur Mason  and Cole, R. S. },
title = {V. Impact with a liquid surface, studied by the aid of instantaneous photography},
journal = {Philosophical Transactions of the Royal Society of London. Series A, Containing Papers of a Mathematical or Physical Character},
volume = {189},
number = {},
pages = {137-148},
year = {1897},
doi = {10.1098/rsta.1897.0005},
URL = {https://royalsocietypublishing.org/doi/abs/10.1098/rsta.1897.0005},
eprint = {https://royalsocietypublishing.org/doi/pdf/10.1098/rsta.1897.0005},
}

@article{Eggers_2008,
doi = {10.1088/0034-4885/71/3/036601},
url = {https://doi.org/10.1088/0034-4885/71/3/036601},
year = {2008},
month = {feb},
publisher = {},
volume = {71},
number = {3},
pages = {036601},
author = {Eggers, Jens and Villermaux, Emmanuel},
title = {Physics of liquid jets},
journal = {Reports on Progress in Physics}
}

@article{Miles_1984, title={Nonlinear Faraday resonance}, volume={146}, DOI={10.1017/S0022112084001865}, journal={Journal of Fluid Mechanics}, author={Miles, John W.}, year={1984}, pages={285–302}}

@manual{OpenFOAM_v1706,
  title        = {OpenFOAM\,v1706: New and improved numerics},
  author       = {{OpenCFD Ltd}},
  year         = {2017},
  month        = {June},
  url          = {https://www.openfoam.com/news/main-news/openfoam-v1706/numerics},
  note         = {Accessed: 5 November 2025}
}

@inbook{BENEK_1983,
author = {J. Benek and J. Steger and F.C. Dougherty},
title = {A flexible grid embedding technique with application to the Euler equations},
booktitle = {6th Computational Fluid Dynamics Conference Danvers},
year    = {1983},
chapter = {},
pages = {},
doi = {10.2514/6.1983-1944},
URL = {https://arc.aiaa.org/doi/abs/10.2514/6.1983-1944},
eprint = {https://arc.aiaa.org/doi/pdf/10.2514/6.1983-1944}
}

@article{JIANG_1998, title={Period tripling and energy dissipation of breaking standing waves}, volume={369}, DOI={10.1017/S0022112098001785}, journal={Journal of Fluid Mechanics}, author={Jiang, L. and Perlin, M. and Schultz,W. W.}, year={1998}, pages={273–299}}

@article{Ghabache_2014, title={Liquid jet eruption from hollow relaxation}, volume={761}, DOI={10.1017/jfm.2014.629}, journal={Journal of Fluid Mechanics}, author={Ghabache, Élisabeth and Séon, Thomas and Antkowiak, Arnaud}, year={2014}, pages={206–219}}

@article{Obreschkow_2012,
  title = {Analytical approximations for the collapse of an empty spherical bubble},
  author = {Obreschkow, D. and Bruderer, M. and Farhat, M.},
  journal = {Phys. Rev. E},
  volume = {85},
  issue = {6},
  pages = {066303},
  numpages = {4},
  year = {2012},
  month = {Jun},
  publisher = {American Physical Society},
  doi = {10.1103/PhysRevE.85.066303},
  url = {https://link.aps.org/doi/10.1103/PhysRevE.85.066303}
}

@article{King_2023, title={Large eddy simulations of bubbly flows and breaking waves with smoothed particle hydrodynamics}, volume={972}, DOI={10.1017/jfm.2023.649}, journal={Journal of Fluid Mechanics}, author={King, J.R.C. and Lind, S.J. and Rogers, B.D. and Stansby, P.K. and Vacondio, R.}, year={2023}, pages={A24}}

@article{Fillette_2022,
  title = {Axisymmetric gravity-capillary standing waves on the surface of a fluid},
  author = {Fillette, Jules and Fauve, St\'ephan and Falcon, Eric},
  journal = {Phys. Rev. Fluids},
  volume = {7},
  issue = {12},
  pages = {124801},
  numpages = {12},
  year = {2022},
  month = {Dec},
  publisher = {American Physical Society},
  doi = {10.1103/PhysRevFluids.7.124801},
  url = {https://link.aps.org/doi/10.1103/PhysRevFluids.7.124801}
}

@article{Miche1944,
  author    = {Robert Miche},
  title     = {Mouvements ondulatoires de la mer en profondeur constante ou décroissante},
  journal   = {Annales des Ponts et Chaussées},
  year      = {1944},
  volume    = {114},
  number    = {1},
  pages     = {25--78},
  language  = {French}
}

@article{Cheng_25,
  author    = {Xianggang Cheng and Xiao-Peng Chen and Zhi-Ming Yuan and Laibing Jia},
  title     = {Particulate reshapes surface jet dynamics induced by a cavitation bubble},
  journal   = {Nature Communications},
  year      = {2025},
  volume    = {16},
  number    = {1},
  pages     = {7562},
  doi       = {10.1038/s41467-025-62936-y},
  url       = {https://doi.org/10.1038/s41467-025-62936-y},
}

@article{Rabbi_2021, title={Impact force reduction by consecutive water entry of spheres}, volume={915}, DOI={10.1017/jfm.2020.1165}, journal={Journal of Fluid Mechanics}, author={Rabbi, Rafsan and Speirs, Nathan B. and Kiyama, Akihito and Belden, Jesse and Truscott, Tadd T.}, year={2021}, pages={A55}}

@article{Penney_1952,
author = {Penney, William George  and Price, A. T.},
title = {Part II. finite periodic stationary gravity waves in a perfect liquid},
journal = {Philosophical Transactions of the Royal Society of London. Series A, Mathematical and Physical Sciences},
volume = {244},
number = {882},
pages = {254-284},
year = {1952},
doi = {10.1098/rsta.1952.0004},
URL = {https://royalsocietypublishing.org/doi/abs/10.1098/rsta.1952.0004},
eprint = {https://royalsocietypublishing.org/doi/pdf/10.1098/rsta.1952.0004},
}

@article{Taylor_1953,
author = {Taylor, Geoffrey Ingram },
title = {An experimental study of standing waves},
journal = {Proceedings of the Royal Society of London. Series A. Mathematical and Physical Sciences},
volume = {218},
number = {1132},
pages = {44-59},
year = {1953},
doi = {10.1098/rspa.1953.0086},
URL = {https://royalsocietypublishing.org/doi/abs/10.1098/rspa.1953.0086},
eprint = {https://royalsocietypublishing.org/doi/pdf/10.1098/rspa.1953.0086},
}

@article {Longuet-Higgins_1994,
      author = "Michael S.  Longuet-Higgins",
      title = "A Fractal Approach to Breaking Waves",
      journal = "Journal of Physical Oceanography",
      year = "1994",
      publisher = "American Meteorological Society",
      address = "Boston MA, USA",
      volume = "24",
      number = "8",
      doi = "10.1175/1520-0485(1994)024<1834:AFATBW>2.0.CO;2",
      pages=      "1834 - 1838",
      url = "https://journals.ametsoc.org/view/journals/phoc/24/8/1520-0485_1994_024_1834_afatbw_2_0_co_2.xml"
}

@article{Wilkening_2011,
  title = {Breakdown of Self-Similarity at the Crests of Large-Amplitude Standing Water Waves},
  author = {Wilkening, Jon},
  journal = {Phys. Rev. Lett.},
  volume = {107},
  issue = {18},
  pages = {184501},
  numpages = {5},
  year = {2011},
  month = {Oct},
  publisher = {American Physical Society},
  doi = {10.1103/PhysRevLett.107.184501},
  url = {https://link.aps.org/doi/10.1103/PhysRevLett.107.184501}
}

@article{Truscott_2014,
   author = "Truscott, Tadd T. and Epps, Brenden P. and Belden, Jesse",
   title = "Water Entry of Projectiles", 
   journal= "Annual Review of Fluid Mechanics",
   year = "2014",
   volume = "46",
   number = "Volume 46, 2014",
   pages = "355-378",
   doi = "https://doi.org/10.1146/annurev-fluid-011212-140753",
   url = "https://www.annualreviews.org/content/journals/10.1146/annurev-fluid-011212-140753",
   publisher = "Annual Reviews",
   issn = "1545-4479",
   type = "Journal Article",
  }

@article{Glasheen_1996,
    author = {Glasheen, J. W. and McMahon, T. A.},
    title = {Vertical water entry of disks at low Froude numbers},
    journal = {Physics of Fluids},
    volume = {8},
    number = {8},
    pages = {2078-2083},
    year = {1996},
    month = {08},
    issn = {1070-6631},
    doi = {10.1063/1.869010},
    url = {https://doi.org/10.1063/1.869010},
}

@article{TRUSCOTT_TECHET_2009, title={Water entry of spinning spheres}, volume={625}, DOI={10.1017/S0022112008005533}, journal={Journal of Fluid Mechanics}, author={Truscott, Tadd T. and Techet, Alexandra H.}, year={2009}, pages={135–165}}

@article{Yu_2023,
    author = {Yu, Y. Q. and Zong, Z. and Zhang, Q.},
    title = {Mechanism of sound emission produced by enclosed cavity upon a sphere entering the water},
    journal = {Physics of Fluids},
    volume = {35},
    number = {7},
    pages = {072107},
    year = {2023},
    month = {07},
    issn = {1070-6631},
    doi = {10.1063/5.0151851},
    url = {https://doi.org/10.1063/5.0151851},
}

@article{Grumstrup_2007,
  title = {Cavity Ripples Observed during the Impact of Solid Objects into Liquids},
  author = {Grumstrup, Torben and Keller, Joseph B. and Belmonte, Andrew},
  journal = {Phys. Rev. Lett.},
  volume = {99},
  issue = {11},
  pages = {114502},
  numpages = {4},
  year = {2007},
  month = {Sep},
  publisher = {American Physical Society},
  doi = {10.1103/PhysRevLett.99.114502},
  url = {https://link.aps.org/doi/10.1103/PhysRevLett.99.114502}
}

@article{Francoeur_2025,
    author = {Francoeur, Jeremy W and Matoza, Robin S and Ortiz, Hugo D and De Negri, Rodrigo},
    title = {Identification of transient seismo-acoustic signals from crashing ocean waves: template matching and location of discrete surf events},
    journal = {Geophysical Journal International},
    volume = {243},
    number = {2},
    pages = {ggaf317},
    year = {2025},
    month = {08},
    issn = {1365-246X},
    doi = {10.1093/gji/ggaf317},
    url = {https://doi.org/10.1093/gji/ggaf317},
    eprint = {https://academic.oup.com/gji/article-pdf/243/2/ggaf317/64062194/ggaf317.pdf},
}

@article{Nelli2025,
    author = {Nelli, Filippo and Tothova, Danica and Ooi, Andrew and Manasseh, Richard},
    title = {Sound amplitude of discrete bubbles entrained by a breaking wave},
    journal = {The Journal of the Acoustical Society of America},
    volume = {158},
    number = {2},
    pages = {1443-1450},
    year = {2025},
    month = {08},
    issn = {0001-4966},
    doi = {10.1121/10.0039053},
    url = {https://doi.org/10.1121/10.0039053},
}

@article{Deike2022,
   author = "Deike, Luc",
   title = "Mass Transfer at the Ocean–Atmosphere Interface: The Role of Wave Breaking, Droplets, and Bubbles", 
   journal= "Annual Review of Fluid Mechanics",
   year = "2022",
   volume = "54",
   number = "Volume 54, 2022",
   pages = "191-224",
   doi = "https://doi.org/10.1146/annurev-fluid-030121-014132",
   url = "https://www.annualreviews.org/content/journals/10.1146/annurev-fluid-030121-014132",
   publisher = "Annual Reviews",
   issn = "1545-4479",
   type = "Journal Article",
  }

@inproceedings{Fourtakas2017,
    author = {Fourtakas, Georgios and Stansby, Peter K. and Rogers, Benedict D. and Lind, Steven J. and Yan, Shiqiang and Ma, Qingwei W.},
    title = {On the Coupling of Incompressible SPH With a Finite Element Potential Flow Solver for Nonlinear Free Surface Flows},
    volume = {The 27th International Ocean and Polar Engineering Conference},
    series = {International Ocean and Polar Engineering Conference},
    pages = {ISOPE-I-17-074},
    year = {2017},
    month = {06},
    eprint = {https://onepetro.org/ISOPEIOPEC/proceedings-pdf/ISOPE17/ISOPE17/ISOPE-I-17-074/1263552/isope-i-17-074.pdf},
}

@article{Kersten2003,
    author = {Kersten, B. and Ohl, C. D. and Prosperetti, A.},
    title = {Transient impact of a liquid column on a miscible liquid surface},
    journal = {Physics of Fluids},
    volume = {15},
    number = {3},
    pages = {821-824},
    year = {2003},
    month = {03},
    issn = {1070-6631},
    doi = {10.1063/1.1542614},
    url = {https://doi.org/10.1063/1.1542614},
}

@article{Corrado2023,
title = {Large-scale wave breaking over a barred beach: SPH numerical simulation and comparison with experiments},
journal = {Coastal Engineering},
volume = {185},
pages = {104362},
year = {2023},
issn = {0378-3839},
doi = {https://doi.org/10.1016/j.coastaleng.2023.104362},
url = {https://www.sciencedirect.com/science/article/pii/S0378383923000868},
author = {Corrado Altomare and Pietro Scandura and Iván Cáceres and Dominic A. van der A and Giacomo Viccione},
}


\end{document}